\documentclass[final,1p,times]{elsarticle}
\usepackage{amsmath}
\usepackage{graphics}
\usepackage{graphicx}
\usepackage{float}
\usepackage{amsfonts}
\usepackage{multicol}
\usepackage{multirow}
\usepackage{enumerate}
\usepackage{url}
\usepackage{bm}
\usepackage{booktabs}
\usepackage{siunitx}
\usepackage{subcaption}
\usepackage{algorithm,algorithmic}
\usepackage{color}
\usepackage[table,xcdraw]{xcolor}
\usepackage{array}
\usepackage{placeins}
\usepackage[toc,page]{appendix}
\usepackage{amsfonts}
\usepackage{grffile}
\journal{Journal}
\definecolor{arsenic}{rgb}{0.23, 0.27, 0.29}
\definecolor{charcoal}{rgb}{0.21, 0.27, 0.31}
\definecolor{hanblue}{rgb}{0.27, 0.42, 0.81}
\definecolor{blue-ncs}{rgb}{0.0, 0.53, 0.74}
\definecolor{awesome}{rgb}{1.0, 0.13,0.32}
\definecolor{darkgreen}{rgb}{0, .4,0}

\begin{document}
	
\begin{frontmatter}
%\title{A note on the Earth testing of rovers bound for extraterrestrial missions}
\title{Using physics-based simulation towards eliminating empiricism in extraterrestrial terramechanics applications}
\author[sjtu,wisc]{Wei Hu}
\author[wisc]{Pei Li}
\author[nasa-ames]{Arno Rogg}
\author[nasa-glenn]{Alexander Schepelmann}
\author[nasa-glenn]{Colin Creager}
\author[pi]{Samuel Chandler}
\author[mit]{Ken Kamrin}
\author[wisc]{Dan Negrut\corref{cor}}
\ead{negrut@wisc.edu}
\cortext[cor]{Corresponding author}

\affiliation[sjtu]{
	organization={School of Naval Architecture, Ocean and Civil Engineering},
	addressline={Shanghai Jiao Tong University},
	city={Shanghai},
	postcode={200240}, 
	country={China}}
\affiliation[wisc]{
	organization={Department of Mechanical Engineering},
	addressline={University of Wisconsin-Madison},
	city={Madison},
	postcode={53706}, 
	state={WI},
	country={USA}}
\affiliation[nasa-ames]{
	organization={Ames Research Center},
	addressline={NASA},
	city={Mountain View},
	postcode={94035}, 
	state={CA},
	country={USA}}
\affiliation[nasa-glenn]{
	organization={NASA},
	addressline={Glenn Research Center, LMT/Mechanisms and Tribology Branch},
	city={Cleveland},
	postcode={44135}, 
	state={OH},
	country={USA}}
\affiliation[pi]{
	organization={ProtoInnovations},
	addressline={LLC},
	city={Pittsburgh},
	postcode={15201}, 
	state={PA},
	country={USA}}
\affiliation[mit]{
	organization={Department of Mechanical Engineering},
	addressline={Massachusetts Institute of Technology},
	city={Cambridge},
	postcode={02139}, 
	state={MA},
	country={USA}}

\begin{abstract}
Recently, there has been a surge of international interest in extraterrestrial exploration targeting the Moon, Mars, the moons of Mars, and various asteroids. This contribution discusses how current state-of-the-art Earth-based testing for designing rovers and landers for these missions currently leads to overly optimistic conclusions about the behavior of these devices upon deployment on the targeted celestial bodies. The key misconception is that gravitational offset is necessary during the \textit{terramechanics} testing of rover and lander prototypes on Earth. The body of evidence supporting our argument is tied to a small number of studies conducted during parabolic flights and insights derived from newly revised scaling laws. We argue that what has prevented the community from fully diagnosing the problem at hand is the absence of effective physics-based models capable of simulating terramechanics under low gravity conditions. We developed such a physics-based simulator and utilized it to gauge the mobility of early prototypes of the Volatiles Investigating Polar Exploration Rover (VIPER), which is slated to depart for the Moon in November 2024. This contribution discusses the results generated by this simulator, how they correlate with physical test results from the NASA-Glenn SLOPE lab, and the fallacy of the gravitational offset in rover and lander testing. The simulator developed is open sourced and made publicly available for unfettered use; it can support principled studies that extend beyond trafficability analysis to provide insights into in-situ resource utilization activities, e.g., digging, bulldozing, and berming in low gravity.

\end{abstract}


\end{frontmatter}

% ==========================================================================
\section{Introduction}
\label{sec:intro}
\subsection{Backdrop}
\label{subsec:backdrop}
Extraterrestrial exploration has experienced a significant uptick over the last three decades, with NASA alone rolling out several missions, e.g., Sojourner \cite{estier2000innovative}, Spirit and Opportunity \cite{morris2004mineralogy,kerr2009mars}, Curiosity \cite{voosen2018nasa}, and more recently Perseverance \cite{perseveranceMissionOverview2020mars}. Over the last decade, China has landed rovers on Mars, see Zhurong \cite{ding2022surface}, and on the Moon, see Yutu \cite{ding20222}. After the US, Russia (which landed Moon rovers in the early 1970s), and China -- India was the fourth nation to land a rover on the Moon, Pragyan, which was the first to land in the proximity of the south pole where it carried out a two week exploration program in mid-2023. Japan landed the SLIM rover with mixed success in early 2024, with two reduced-scale rovers, LEV-1 and LEV-2, tagging along \cite{japanRover2024}. 

From a high vantage point, due to the abundant presence of granular material on the Moon, Mars, and other moons or asteroids in our solar system, each mission is preceded by extensive \textit{physical testing} on Earth using granular soil conditions -- natural sands or simulants, e.g., the Mars Global Simulant MGS-1, Lunar Mare Simulant (LMS-1), Johnson Space Center 1A (JSC-A1), Minnesota Lunar Simulant-1 (MLS-1), Japanese lunar soil simulant (FJS-1). The simulants attempt to capture the soil conditions that are specific to an area on the celestial body of interest, e.g. \cite{heRegolithProperties2010,britt-regolithSimulants2016}.

A challenging aspect of any testing campaign on Earth has been handling different gravitational conditions experienced during missions on other celestial bodies. With the Moon gravitational pull at roughly 16\% that of Earth and that of Mars at 38\%, historically, three corrections have been used to factor in the different gravitational pull: gravity offload systems, e.g., \cite{han-reduceGrav2010,reducedGravSim-patent2015,skonieczny-red-excavation2016,valleGravity2017,heliumGravityOffset2022,heliumGravityOffset-bbc-2024}; reducing the mass of the rover \cite{lowMassRobot2009,heverlyCuriosityRover2013}; and use of lunar simulants designed for mobility studies, e.g., GRC-1 \cite{ORAVEC2010361}. Note that these techniques can be combined; i.e., using a lighter vehicle on GRC-1 or GRC-3. 

Using gravity offloading and simulants has been and continues to be done for the US and Chinese rovers, and likely for the Indian and Japanese missions. However, as demonstrated in this contribution, these techniques lead to misleading results airing on the side of being overoptimistic. Consider, for instance, the Curiosity test mentioned in \cite{curiosityTestingMojave2012}. The rover was stripped down of accessories, reducing its mass to 340 kg from the nominal 907 kg. Consequently, the weight of the vehicle carried by the soil was identical to the rover's weight on Mars. Then, owing to the higher gravitational pull acting on each soil particle in California's Mojave Desert, the terrain displayed higher strength and could therefore support higher shear stress without yielding, effectively providing an optimistic trafficability assessment. Reducing the mass of the rover, in isolation, is insufficient, unless the soil is changed to account for the lower gravitational pull at work on the targeted planet, moon, or asteroid. In \cite{ORAVEC2010361}, the authors proposed the GRC-1 regolith simulant to replicate on Earth the terramechanics experience encountered on the Moon. GRC-1 and GRC-3, designed at Glenn Research Center, were intended to capture terramechanics in lunar highlands and maria regions, respectively. It was noted that by adjusting the density and friction angle of the simulant, one would get a spectrum of penetrations in a cone penetration test that resembled, loosely, the values noted by the Apollo astronauts on the Moon. This was a key observation, since the assumption was that trafficability is identical under different gravitational pulls as long as the terrains' cone index gradients are identical. However, the theory that identical cone index gradients lead to similar terramechanics under different gravitational pulls has been recently debunked \cite{dacaAdrianaPhD2022,daca-coneIndex2022}. 

Fortuitously, experimental testing in low-gravity environments is set to benefit substantially from a technique that has re-emerged after remaining dormant for more than four decades: granular scaling laws (GSLs). The scaling laws enable one to understand how physical properties change with scale. A prime example from fluid dynamics is the use of the Reynolds number in wind tunnel experiments. In \cite{similitude-lunar1970,similitude-freitag1970,similitude-wismer1976}, the authors invoked scaling laws to predict the performance of full-scale off-road vehicles by focusing their attention on scale vehicles. Recently, the topic of GSLs has been revisited, formalized \cite{kenScalingLawPhysRevE2017,kamrin-gsl-expanded2020} and experimentally validated in Earth-like conditions \cite{thoesen2020granular,THOESEN2020336} as well as in reduced gravity parabolic flights \cite{dacaAdrianaPhD2022}. Although the newly formulated GSLs have not yet been used in extraterrestrial missions, they can provide a pivotal breakthrough through their ability to bridge disparate gravitational response scenarios. Herein, we employ GSLs to corroborate computer-simulated terramechanics predictions with experimental data. The GSLs are not without their limitations, see the Discussion section of this contribution.

This contribution is concerned with how rovers and landers are tested on Earth before deployment, emphasizing the value of utilizing physics-based simulation. Shifting the focus from terramechanics physical testing to computer simulation, the ``production'' approaches used to predict extraterrestrial mobility are almost exclusively rooted into the seminal work of Bekker, Wong, and Reece. The Bekker-Wong formula, $p =  \left(  \frac{K_c}{b} + K_\phi  \right)$, relates the normal pressure $p$ to the sinkage $z$ for a wheel of width $b$ using a semi-empirical, experiment-based curve fitting with parameters $K_c$, $K_\phi$, and $n$~\cite{bekker56}. The Janosi-Hanamoto formula, $\tau =  \tau_{\text{max}} ( 1 - e^{-J_s/K_s})$, or variants thereof, subsequently use the pressure $p$ to evaluate the shear stress $\tau$ between the wheel and terrain~\cite{janosi61}. Specifically, $\tau$ depends on $\tau_{\text{max}} =  c + p \tan \varphi$, the accumulated shear displacement $J_s$, the cohesion coefficient $c$, internal friction angle $\varphi$, and the so-called Janosi parameter or slip modulus $K_s$. 
%Each wheel circumference point in contact with the terrain has a particular sinkage $z$ and therefore a particular value of the normal pressure $p$ and shear $\tau$. By summing up these small quantities over all the points on the wheel that come in contact with the soil, one gets a resultant normal force and tractive force, along with a torque that opposes the rotation of the wheel. 
This phenomenological approach has its origins in work done in conjunction with military vehicles \cite{bekker56,hegedusBulldozing1960,janosi61,wong67a,wong67b,bekker69}. In planetary exploration, a broad family of terramechanics models have built off the Bekker-Wong model \cite{iagnemma2004,shibly05,bauer2005experimental,ishigami2007,Krenn2008Rover,Krenn2008SCM,Krenn2011,chinaLunarSoilModel2020,chronoSCM2022}. 

For most simulations, a Bekker-Wong/Janosi-Hanamoto (BWJH) type model runs at a real time factor (RTF) of 1 and below, which indicates faster than real time operation (the RTF is defined as the amount of time a computer has to work to simulate one second of system evolution). Thus, the BWJH models are suitable for expeditious simulations aimed at testing autonomy algorithms, e.g., state estimators, path planners, control policies \cite{chiang2010human,dedonato2015human}. The BWJH results are satisfactory under three main assumptions: the wheel sinkage is small, slip ratio is low, and the wheel geometry is close to a cylinder without lugs or grousers \cite{Smith2014,meirion2011modified}. However, there are several problems with employing the BWJH class of models for \textit{predictive} extraterrestrial terramechanics studies; i.e., simulating scenarios that would predict mobility on the Moon, for instance. To start with, low-gravity terramechanics is poorly understood and subject to ongoing research \cite{kobayashi-mobility-lowG-2010,dacaAdrianaPhD2022}. The BWJH model abstraction, a phenomenological/semi-empirical framework, has been established in conjunction with mobility in Earth gravitation and the community discarded the role of gravity \cite{dingIagnemma2015}. When it became apparent that the BWJH class should factor in gravity aspects, corrections have been attempted \cite{wong-Accounting4Gravity2012}, yet the ensuing methodology called for yet additional empirical parameters that were hard to produce. Thus, amending the BWJH-class of models needed additional calibration, which went beyond the bevameter test employed to produce the stock BWJH parameters. It is also noted that the bevameter test is involved, not standardized, and calls for a heavy and bulky apparatus. Moreover, results associated with bevameter tests carried on Earth, e.g., \cite{Apfelbeck2011Systematic,edwards2017bevameter}, have not been correlated to low-gravity BWJH model parameters. 

In lieu of bevameter tests, there have been a-posterirori BWJH parameter identification efforts done while the rover operated in-situ, see, for instance, the recent ChangE-4 mission that deployed the Yutu-2 rover to the Moon \cite{dingScienceRobotics2022,dingNatureCommunications2024}. Similar BWJH identification efforts, Earth-bound though, are reported in \cite{Iagnemma2002online,ojeda2006terrain}. The nature of being a-posteriori, i.e., the rover operates at the time when model parameters are calibrated, curtails the effectiveness of the BWJH insofar as the mission preparation and rover \textit{design} are concerned. The BWJH can be employed, upon meeting the three aforementioned assumptions, to ground-control an ongoing mission. However, changing the rover wheel or celestial body of destination would require a new BWJH model that would need to be calibrated from scratch yet again. 

Being semi-empirical, additional adjustments need to be made to the BWJH model to account for attributes such as nontrivial grousers \cite{irani-updatedBekkerWong2011}, steering \cite{ishigami2007}, light weight and/or small size \cite{meirion2011modified}, etc. The BWJH models are documented as lacking in handling of irregular terrain for which the equivalent geometric factor $b$ is hard to gauge since the interaction with irregular terrain is complex and non-stationary. Finally, the terrain in BWJH lacks any dynamic response -- there is no material and therefore mass movement associated with soil and its deformation; the terrain is simply a force element that prevents the sinking of the wheel and yields a tractive force. As such, dig-in and material ejection cannot be captured. For a list of other limitations and mitigating approaches, see \cite{rodriguezHighSpeedGranMatMobility2019}. 

In a broader context, beyond the class of BWJH models, there are two other terramechanics simulation options: approaches that embrace a continuum representation model (CRM) \cite{sulsky1994particle,bardenhagen2000material,bui2008lagrangian,chenSPH3DgranMat2012,chauchat2014three,ionescu2015viscoplastic,dunatunga2017continuum}; and fully-resolved approaches, in which the motion of the particles that constitute soil is tracked forward in time using the so called discrete element method (DEM)~\cite{cundall1979discrete,iwashita1999mechanics,jensen1999simulation}. The DEM approach is slow but accurate; the CRM lies in between BWJH and DEM, both in terms of speed and accuracy. 

When a wheel operating on granular soils features complex lugs or grouser geometries, or experiences very high slip ratios, the DEM can be relied upon for accurate numerical results~\cite{iagnema2015,ucgul-EDEM-tillage2015,zhao-FEM-DEMtire-terrain2017,antonioVehicleTireGranMatSim2017}. However, since many engineering problems can involve billions of discrete grains, the computational cost of a fully resolved DEM simulation can become prohibitively high. The RTF of DEM terramechanics simulations can be in the range \num{3000} to \num{15000}, see, for instance, \cite{antonioVehicleTireGranMatSim2017}. By comparison, the real time factor for CRM terramechanics simulations can be as low as 30-150 \cite{weiTracCtrl2022}. Another strength of CRM is that it is a physics-based approach in which the input parameters, e.g., density, friction angle, stiffness, shear modulus, cohesion can be easily obtained, see, for instance, \cite{heRegolithProperties2010}, or estimated. Consequently, little to no parameter calibration is needed before running the simulations. The three attractive attributes of CRM -- speed, accuracy, and setup convenience, come at the price of a more involved solution methodology. Indeed, being the solution of a time-dependent set of partial differential equations, the continuum problem is spatially discretized using either a finite element method (FEM) \cite{chauchat2014three,ionescu2015viscoplastic}; or a meshless solution, e.g., the material point method (MPM) \cite{sulsky1994particle,bardenhagen2000material,kenichiMPM2016,kamrin2019}, or the smoothed particle hydrodynamics (SPH) method \cite{bui2008lagrangian,chenSPH3DgranMat2012,nguyenSPHgranFlows2017,hurley2017continuum,xu2019analysis,chen2020gpu,weiGranularSPH2021}. Since the soil is subject to plastic flow with large deformation at high slip ratio, ill-shaped FEM elements can lead to numerical instabilities or require costly re-meshing operations, which places meshless methods at an advantage. 

In this contribution, we report on a new simulation-anchored framework for designing rover and lander missions. The terramechanics modeling methodology adopted is based on the Continuous Representation Model (CRM) \cite{weiGranularSPH2021} due to its accuracy and efficiency traits, and employs the SPH spatial discretization of the equations of motion. Our contribution is fourfold -- specifically, we: established CRM as a viable approach for terramechanics simulation; implemented an open-source publicly available CRM simulator validated against VIPER-related experimental data; demonstrated that the physics-based simulator is predictive -- it produces results that match experimental test results and obey the scaling predicted by GSL; and, most importantly, demonstrated that the simulator reveals misconceptions in the way the physical testing of rovers is carried out today.

\begin{figure}[h]
	\centering
	\includegraphics[width=6in]{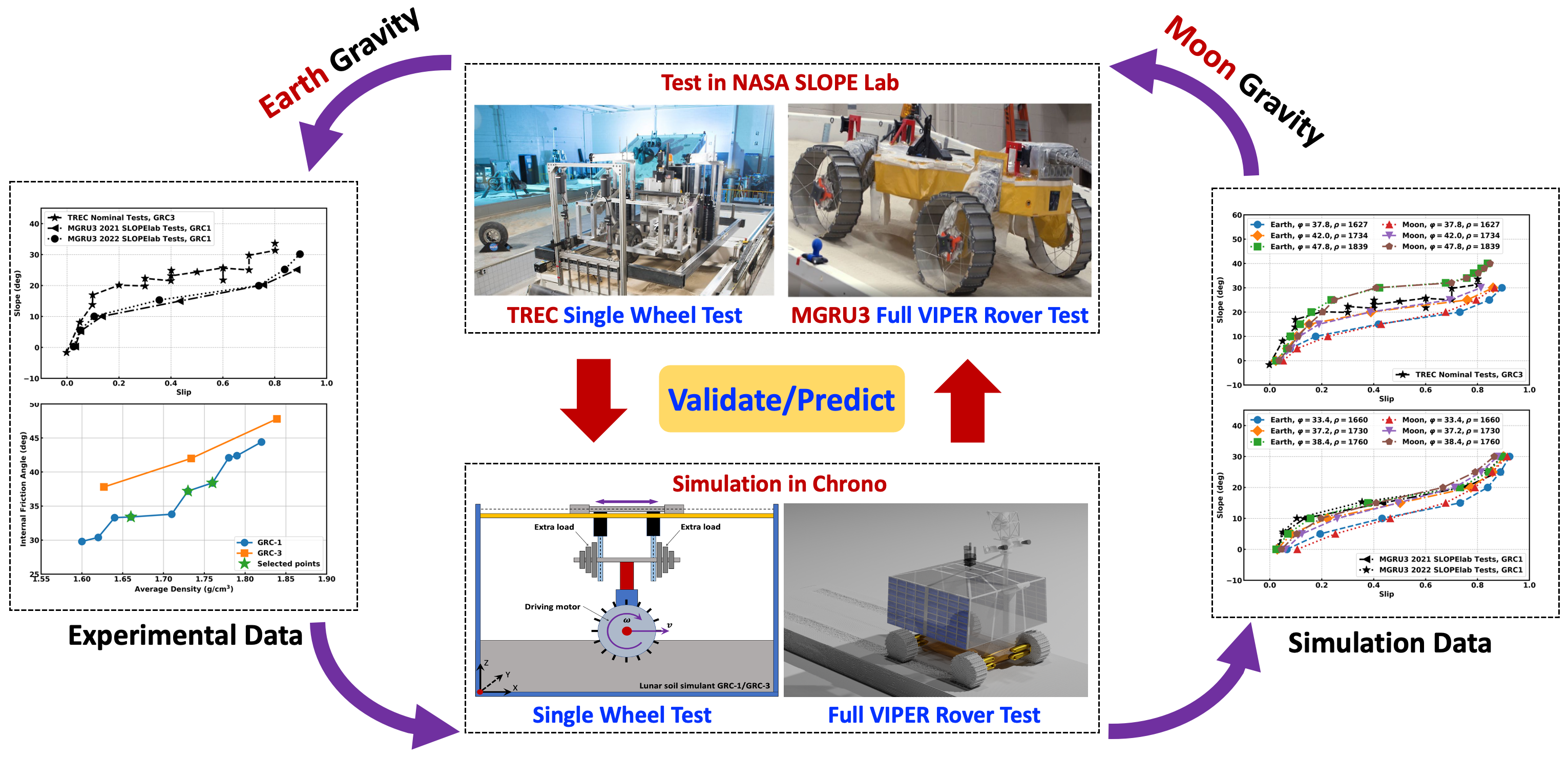}
	\caption{Schematic view of the workflow: experimental test results are used to validate the simulator, which is subsequently used to predict the VIPER rover's performance on the Moon. The experimental data was generated at NASA's SLOPE lab with tests performed under Earth gravity. The same tests were conducted in the physics-based Chrono simulator, to judge its predictive traits before using it to produce results under Moon gravity.} 
	\label{fig:schematic}
\end{figure}

\subsection{Experimental setup}
\label{subsec:exSetup}
The study discussed in this contribution is summarized in Fig.~\ref{fig:schematic}. We present results for both single wheel and full-rover tests; the rover used is a 1/6 mass replica of VIPER. The validation experimental test data was collected at NASA's SLOPE lab. The deformable terrain was modeled using the CRM approach; details can be found in the supplementary materials. Being a physics-based methodology, the CRM model parameters are identical to the material parameters associated with the GRC-1 \cite{ORAVEC2010361} and GRC-3 \cite{he2013geotechnical} lunar soil simulant used in NASA's SLOPE lab. In other words, compared to the semi-empirical BWJH approach, the parameter calibration needs are significantly reduced as the parameters needed are friction angle, bulk density, etc.; i.e., parameters with immediate physical meaning. The single wheel simulations were run in ``VV''-mode, when both the translational ``V''elocity and angular ``V''elocity of the wheel were controlled to yield a certain wheel slip; and then in ``slope''-mode, where the angular velocity of the wheel was constant, and the translational velocity was measured once the wheel reached a steady state on a terrain with a fixed slope. For the full rover, all simulations were in slope-mode. The single wheel test rig and the rover were modeled as multibody systems, thus capturing the full nonlinear dynamics of the system. All simulations were conducted in a co-simulation framework with the multibody dynamics solved using a multi-core CPU and the CRM terramechanics solved on a GPU. The slope/slip and power/slip relationships obtained in simulation were validated against experimental data.

\begin{figure}[h]
	\centering
	\includegraphics[width=4in]{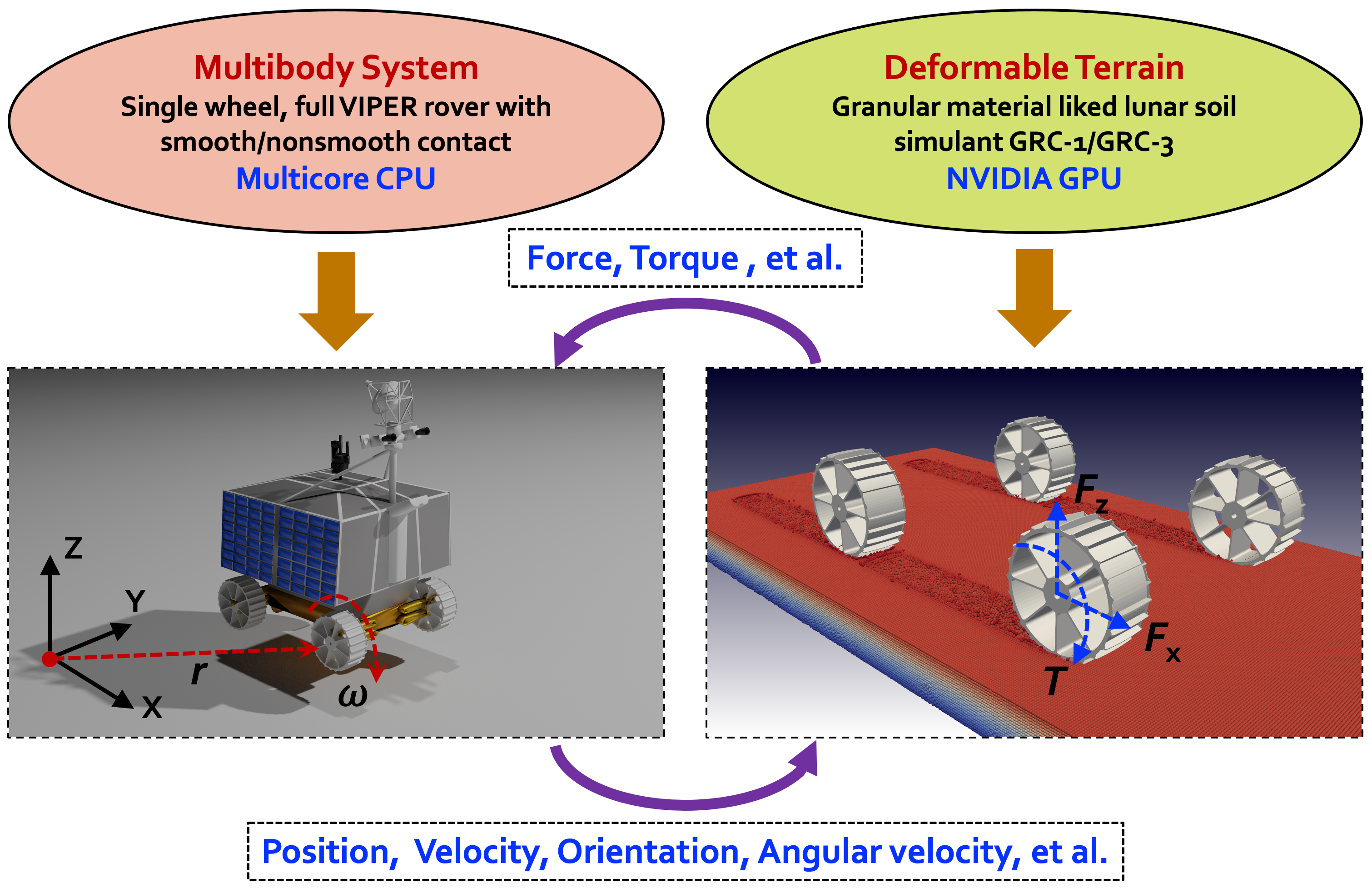}
	\caption{The co-simulation framework employed: the rover was modeled as a multibody system whose dynamics was solved using a multicore CPU. The deformable terrain was modeled as a continuum whose dynamics was solved using GPU computing. The two modules communicated passing force and torque information from the terrain to the rover; and position, velocity, orientation, and angular velocity coming from each wheel and going to the terrain CRM solver.}
	\label{fig:co_sim_setup}
\end{figure}

In the simulations performed for both single wheel and full rover, two modules come into play in the co-simulation framework implemented in the open source software Chrono \cite{chronoOverview2016,projectChronoGithub}. One is the multibody dynamics simulation engine, which is used to propagate forward in time the motion of the solid bodies, e.g., the dynamics of the single wheel or the full rover. The frictional contact between the rigid bodies is handled therein using a differential variational inequality (DVI) approach  \cite{armanDEMP-DEMC2017,negrutSerbanTasoraJCND2017}. The second module handles the dynamics of the granular lunar terrain, which is accomplished using the CRM approach. Since the SPH particles used to discretize the CRM simulation domain are usually much larger than the actual terrain grains, the degree of freedom count is significantly reduced, which explains the major CRM simulation speed gains over DEM simulation. The dynamics of the SPH particles was integrated forward in time using GPU acceleration. Since the dynamics of the rover and the terrain systems was solved separately in two different simulation engines, one running on the CPU and one on the GPU, a communication was required between these two hardware assets to enforce the coupling effect, i.e. the wheel-soil interaction.  Figure \ref{fig:co_sim_setup} illustrates the developed co-simulation framework. At each time step, the dynamics of terrain was solved first, hence the force and torque that was applied from the soil to the wheel can be calculated and passed from the GPU memory to the CPU memory. Then the dynamics of the rover system was solved with the external force applied from the terrain side. Once the new position, velocity, orientation, and angular velocity of each wheel were updated, they were passed back to the GPU memory to advance in time the state of the terrain.% It is noted that the wheel-soil interaction cannot be directly accomplished between the SPH particle and the mesh that was used to discretize the wheel. It was actually calculated using the boundary condition enforcing (BCE) particles which were attached to the wheel according to the geometry of it, as shown in Fig. \ref{fig:BCE} in the supplementary materials. The BCE particles can be automatically generated using an integrated mesh-to-particle tool in Chrono. The BCE particles only need to be generated once at the beginning of the simulation. Since we only pass small mount of data between CPU memory and GPU memory at each time step, the communication between them is super efficient which does not affect the performance of the simulation a lot.

\subsection{Granular Scaling Laws}
\label{subsec:scalingLaws}
Two important points highlighted in this contribution are as follows: (a) by careful experimental design, terramechanics in low gravity environments can be correlated with terramechanics under Earth's gravity; and (b) CRM serves as a predictive method for understanding rover terramechanics over a range of gravitational accelerations. These claims are supported by results and observations that align with the expectations spelled out by the Granular Scaling Laws, which are summarized below. Several accounts are available for these laws, herein the discussion is anchored by recent work reported in \cite{kamrin-gsl-expanded2020}. The two laws of interest pertain to the scaling of ($i$) the power necessary to produce a certain motion of an artifact, e.g., a wheel; and ($ii$) the translational velocity experienced by the implement. Specifically, the scaling laws assert the existence of a function ${\boldsymbol{\Psi}}$ of five inputs that produces two outputs, the latter being the scaling of the power $P$ and longitudinal velocity $V$ associated with the terramechanics of an implement (here a wheel) \cite{kamrin-gsl-expanded2020}:
\begin{equation}
	\label{eq:scalingLaws}
	\begin{bmatrix}
		\frac{P}{Mg\sqrt{Lg}} \\
		%\\
		\frac{V}{\sqrt{Lg}}
	\end{bmatrix}
	={\boldsymbol{\Psi}}\left(\sqrt{\frac{g}{L}} t,f,\frac{g}{L\omega^2},\frac{\rho_0 D \: L^2}{M}, \theta \right) \;.
\end{equation}
Above, it is assumed that the granular material is cohesionless (see \cite{kamrin-gsl-expanded2020} for cohesive terrain scenarios); $L$ is a characteristic length (for a wheel, its effective radius); $M$ is the total mass; $D$ is the width; $\omega$ is the the angular velocity; $f$ is a shape factor; $\rho_0$ is the bulk density in dense state; $g$ is the gravitational pull; and $\theta$ is the tilt of the slope negotiated by the wheel. The laws in Eq.~(\ref{eq:scalingLaws}) state that if two experiments are carried out, with a set of parameters indexed by subscript 1 and 2, respectively, and if $\frac{\rho_{01} D_1 \: L_1^2}{M_1g_1} = \frac{\rho_{02} D_2 \: L_2^2}{M_2g_2}$, $\frac{g_1}{L_1\omega^2_1} = \frac{g_2}{L_2\omega^2_2}$, $f_1=f_2$, $\sqrt{\frac{g_1}{L_1}} t_1 = \sqrt{\frac{g_2}{L_2}} t_2$, and $\theta_1 = \theta_2$, then the corresponding quantities with ``1'' and ``2'' on the left side of Eq.~(\ref{eq:scalingLaws}) are identical for the two experiments. Consequently, if $P_1$ and $V_1$ are measured in experiment ``1,'' they can be used to estimate the power and velocities in the experiment indexed by the subscript ``2.'' In subsection \S\ref{subsec:actual_slope}, the subscript ``1'' will be associated with Earth tests, while ``2'' with Moon tests.

\FloatBarrier
\section{Results}\label{sec:results}
The results reported are organized in two subsections. The first concentrates on single-wheel tests and speaks to the predictive attributes of the simulator, a topic also addressed in \cite{weiGranularSPH2021,weiVirtualBevameter2023}. The second subsection, which is the linchpin of this contribution, compares single wheel and rover physical testing results obtained in the SLOPE lab with simulation data produced in Earth and Moon gravitational pull conditions. The testing and the simulations were conducted using both GRC-1 and GRC-3. VIPER is slated for operation at the lunar South pole, a highlands area that has a soil containing mostly finely crystalline anorthosite, which GRC-1 attempts to capture. As such, subsection \S\ref{subsec:actual_slope} will mostly but not entirely report GRC-1 results, with GRC-3 results provided in the supplementary material. Qualitatively, there is no remarkable difference between the GRC-1 and GRC-3 results, be it for single wheel or full rover. Finally, for both GRC-1 and GRC-3, it is noted that the terrain can exhibit a spectrum of friction angles and bulk densities, see Fig.~\ref{fig:densityVSfricangle} for a range of values for the friction angle and bulk density.% Whether GRC-1 is a good simulant as far as terramechanics studies are concerned is a topic that is not addressed in this contribution. 

\subsection{VV-Mode: single wheel experiments}\label{subsec:single_vv}
Several single-wheel VV-mode physical and numerical tests were conducted on flat terrain to two ends: produce a plot that relates the DrawBar-Pull (DBP) force to the wheel slip; and generate traction slope vs. wheel slip plots -- see \cite{wongTerramechanics2009} for a discussion of DBP and these plots. The traction slope associated with a specific wheel slip is calculated as $arctan(\text{DBP}/\text{N})$, where N is the load impressed by the rover wheel on the deformable terrain under Earth gravity. The goal of this exercise was to show that the physics-based simulator is predictive and captures well how key model parameters that have a clear physical meaning, e.g., bulk density and friction angle, reflect in the simulated response of the wheel performance. In all tests, physical and numerical, the material was assumed cohesionless.

\begin{figure}[h]
	\centering
	\begin{subfigure}{0.38\textwidth}
		\centering
		\includegraphics[width=2.4in]{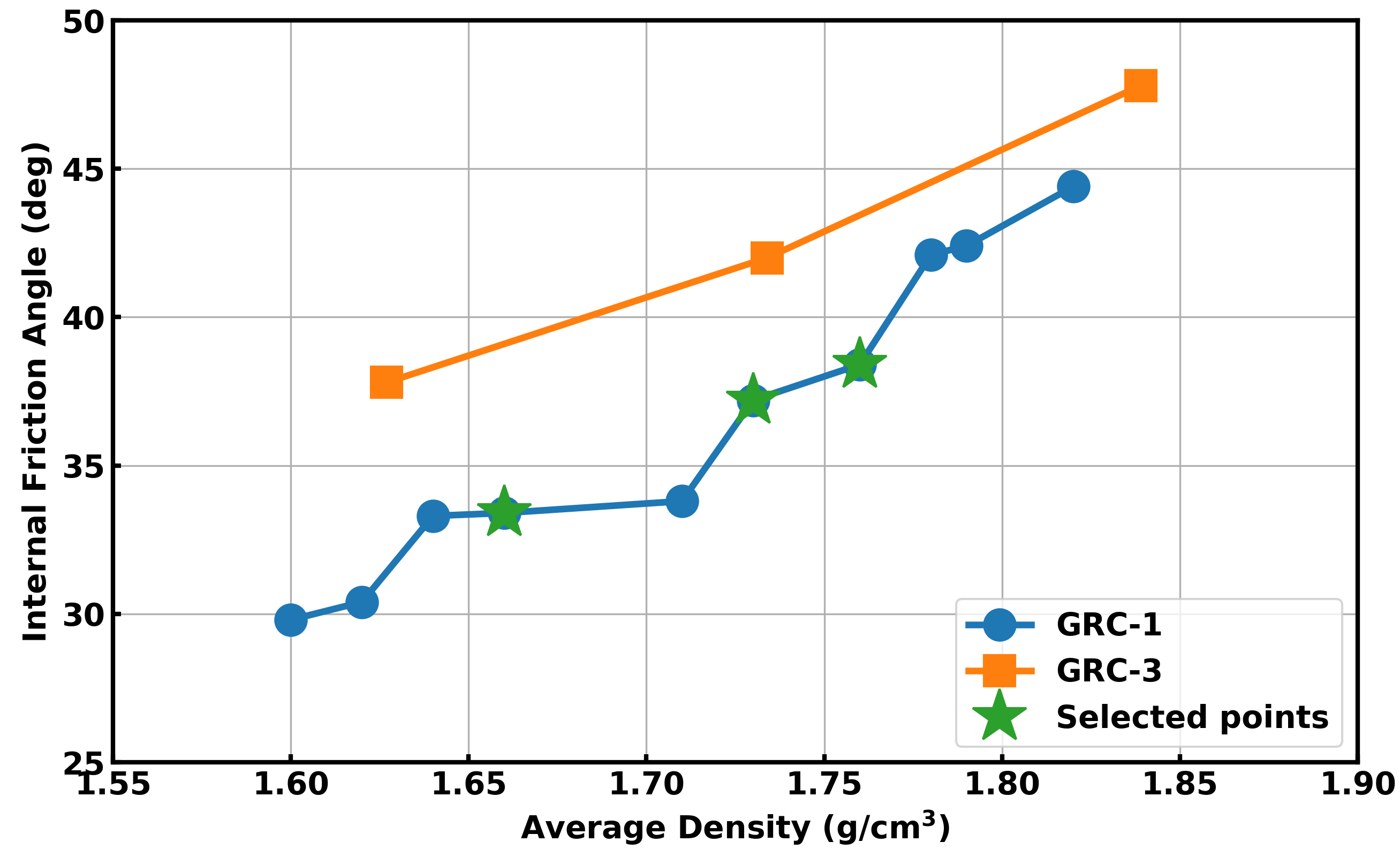}
		\caption{Properties of GRC-1 and GRC-3 lunar soil simulant} 
		\label{fig:densityVSfricangle}
	\end{subfigure}
	\begin{subfigure}{0.58\textwidth}
		\includegraphics[width=3.6in]{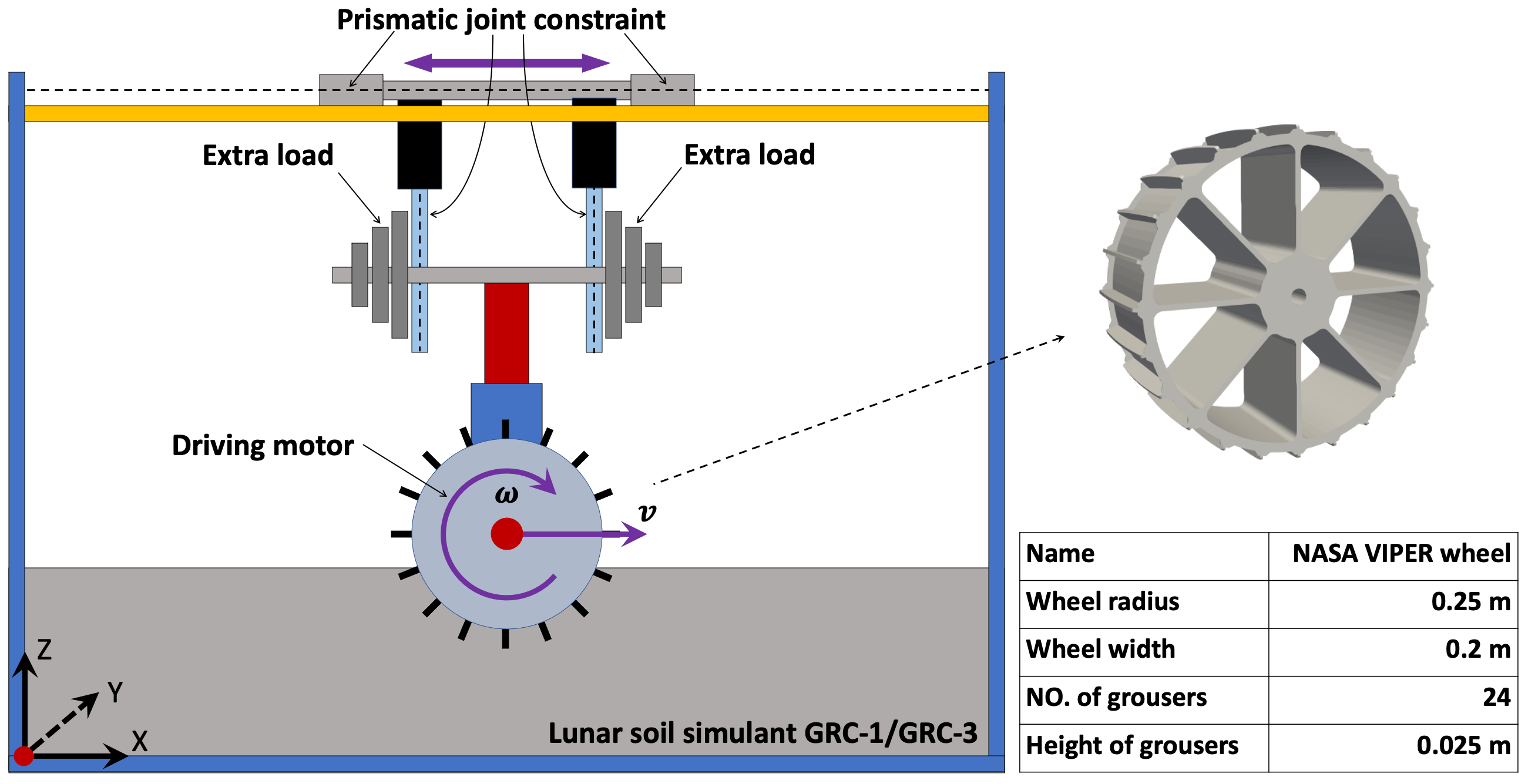}
		\caption{Schematic of single-wheel test rig.} 
		\label{fig:single_schematic}
	\end{subfigure}
	\caption{Schematic view of the single wheel test under velocity control mode (VV-mode). Both translational velocity and angular velocity of the wheel can be controlled using the test rig modeled in Chrono. Excess mass can be added on the wheel assembly to model wheel-soil interaction under various loads. The wheel used in the simulation shown here has the geometry used in NASA's SLOPE lab -- the radius is 0.25m, the width is 0.2m, and there are 24 grousers. The height of each grouser is 0.025m. } 
	\label{fig:setup}
\end{figure}

\begin{figure}[h]
	\centering
	\begin{subfigure}{0.49\textwidth}
		\centering
		\includegraphics[width=3.2in]{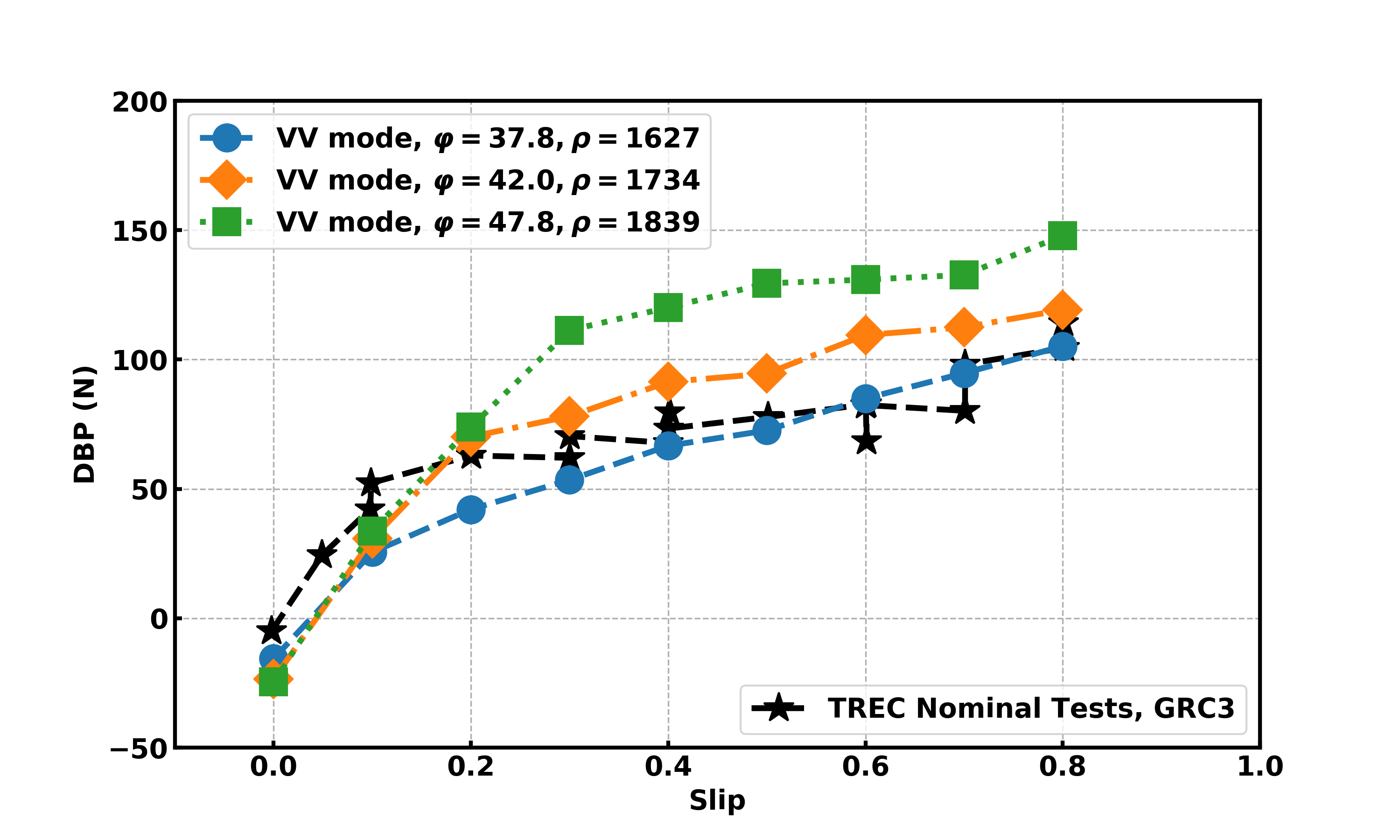}
		\caption{DBP vs. slip.}
	\end{subfigure}
	\begin{subfigure}{0.49\textwidth}
		\centering
		\includegraphics[width=3.2in]{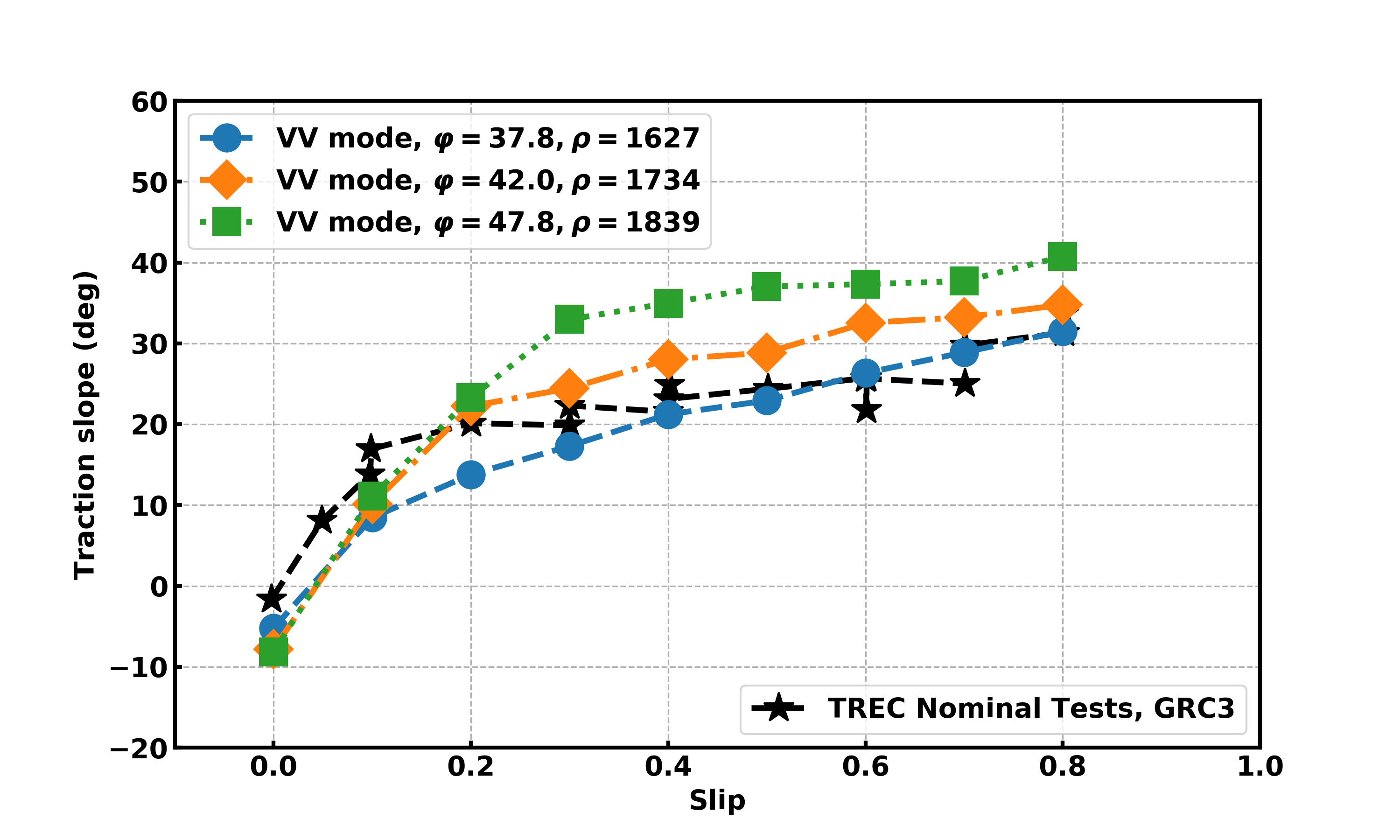}
		\caption{Traction slope vs. slip.} 
	\end{subfigure}
	\caption{Single wheel physical testing \& simulation results on GRC-3 \cite{he2013geotechnical} lunar soil simulant using the 17.5kg wheel. Tests were performed under VV-mode. In simulation, three different sets of GRC-3 material proprieties associated with the lunar soil simulant were chosen -- with bulk densities 1627, 1734, and 1839 $\si{kg/m^3}$, and internal friction angles $37.8^{\circ}$, $42.0^{\circ}$, and $47.8^{\circ}$, respectively. In this, and all subsequent images, curves listed with dotted lines correspond to \textit{experimental} data. The experimental data is connected with a dotted line to emphasize the physical measurements, which can at times be hard to discern against the simulation results. Note that the experimental measurements might list multiple results for the same experimental setup, reflecting the uncertainty in physical measurements.} 
	\label{fig:single_grc3}
\end{figure}

The physical test was performed at NASA Glenn's SLOPE lab using the Traction \& Excavation Capabilities (TREC) Rig, see Fig. \ref{fig:schematic}. The results obtained in the lab are shown in Fig. \ref{fig:single_grc3} with black star markers. The corresponding digital twin was built according to the rig information shown in Fig. \ref{fig:single_schematic}. The total load acting onto the deformable lunar soil simulant was as induced by a 17.5 kg mass, of which half came from the wheel, the other half coming from extra non-wheel mass added to account for a part of the chassis. The wheel was driven with a constant translational velocity $v = 0.2~\si{m/s}$ on the bed of lunar simulant. The angular velocity was controlled to yield a predefined slip value $s=1-\frac{v}{\omega r}$, where $\omega > 0$ is the angular velocity and $r$ is the effective radius of the wheel. Given a slip ratio $s$, in VV-mode, the wheel angular velocity was set to $\omega = \frac{v}{r(1-s)}$. 

\begin{figure}[h]
	\centering
	\includegraphics[width=2.in]{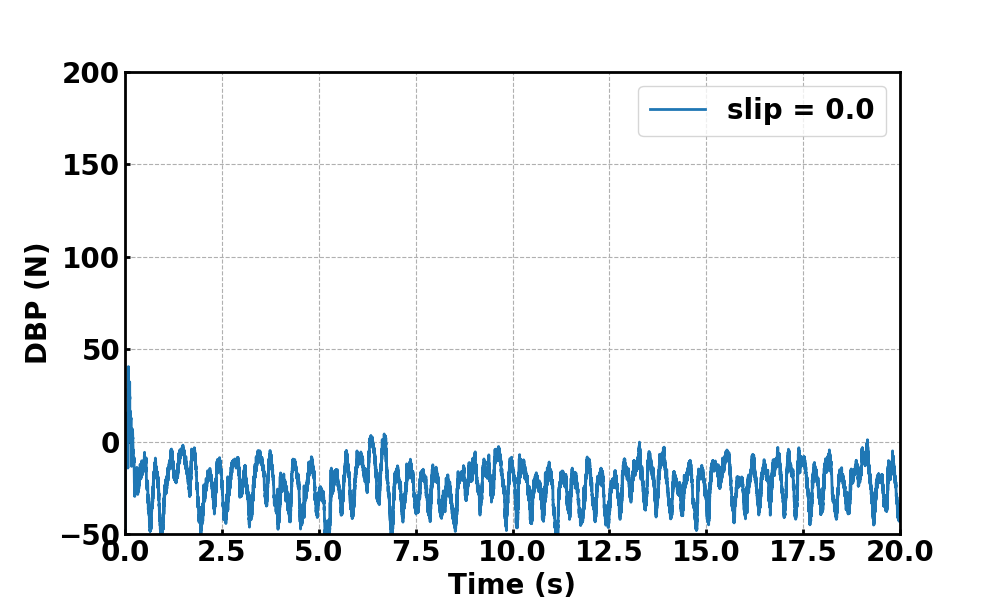}
	\includegraphics[width=2.in]{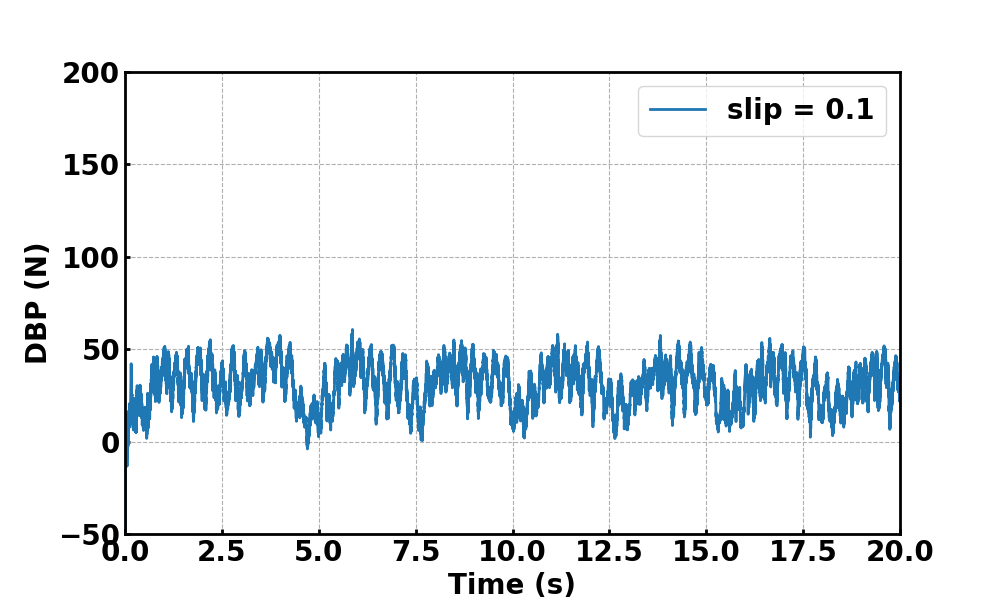}
	\includegraphics[width=2.in]{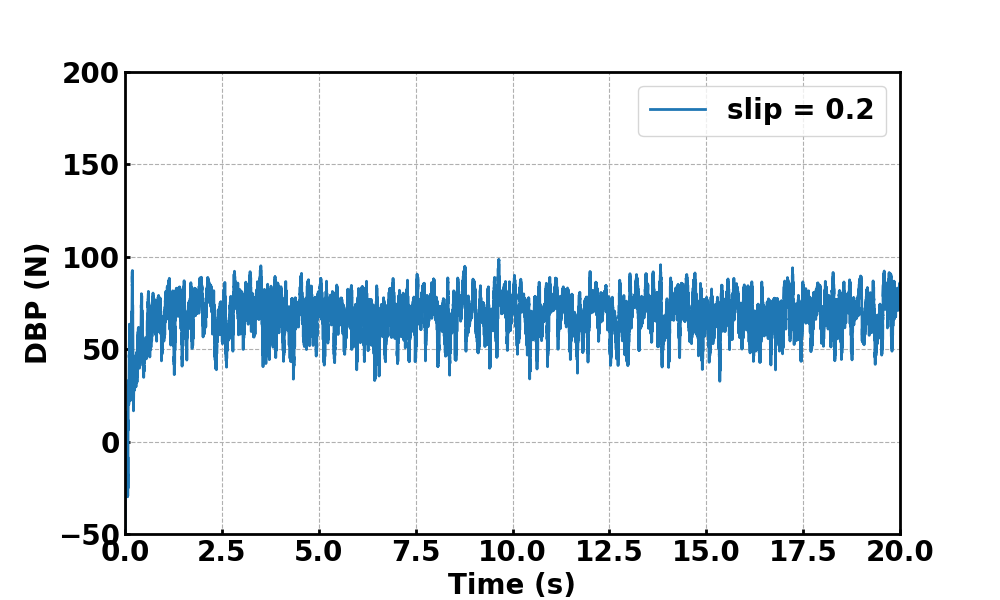}\\
	\includegraphics[width=2.in]{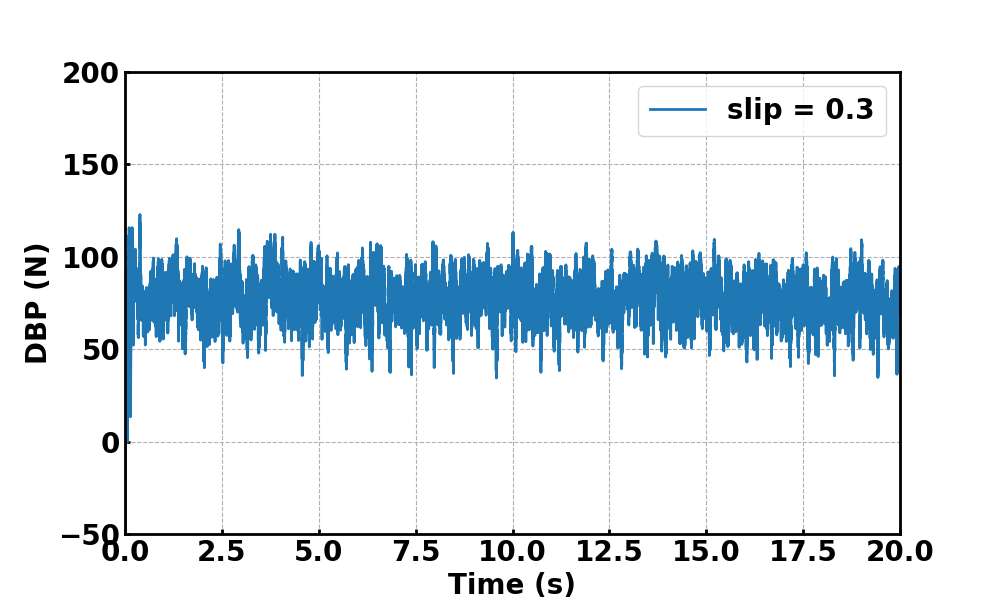}
	\includegraphics[width=2.in]{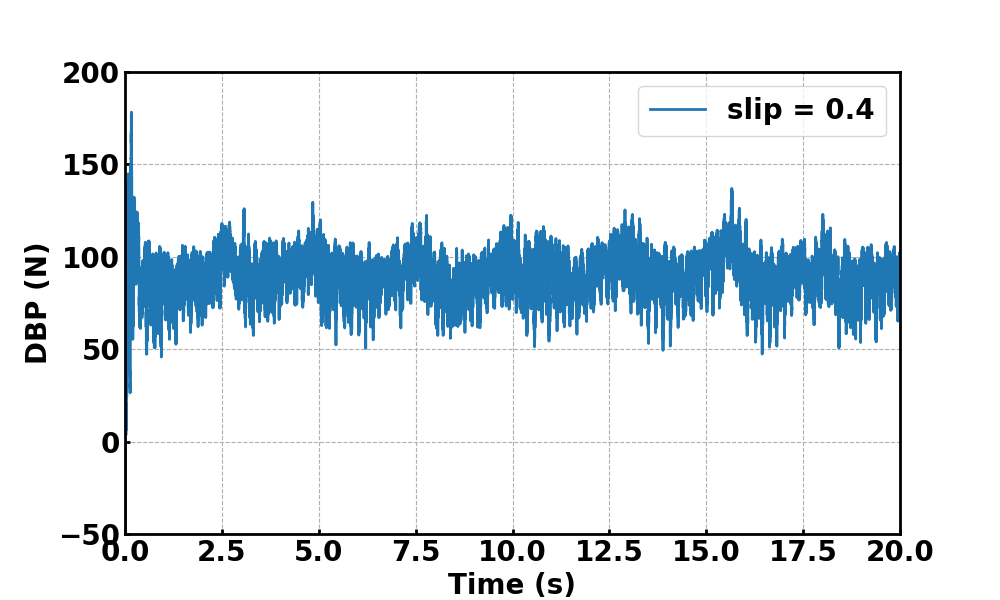}
	\includegraphics[width=2.in]{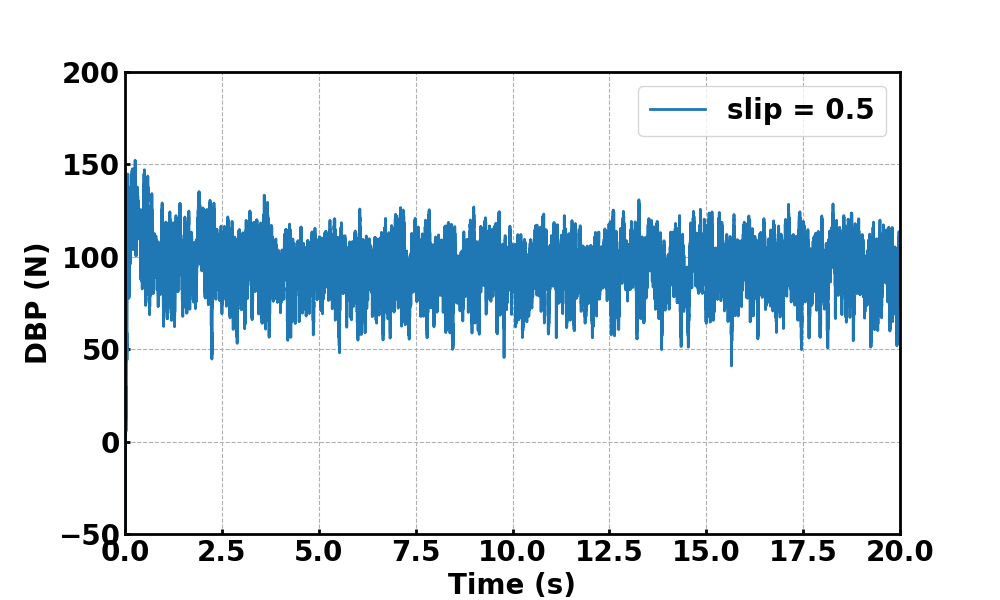}\\
	\includegraphics[width=2.in]{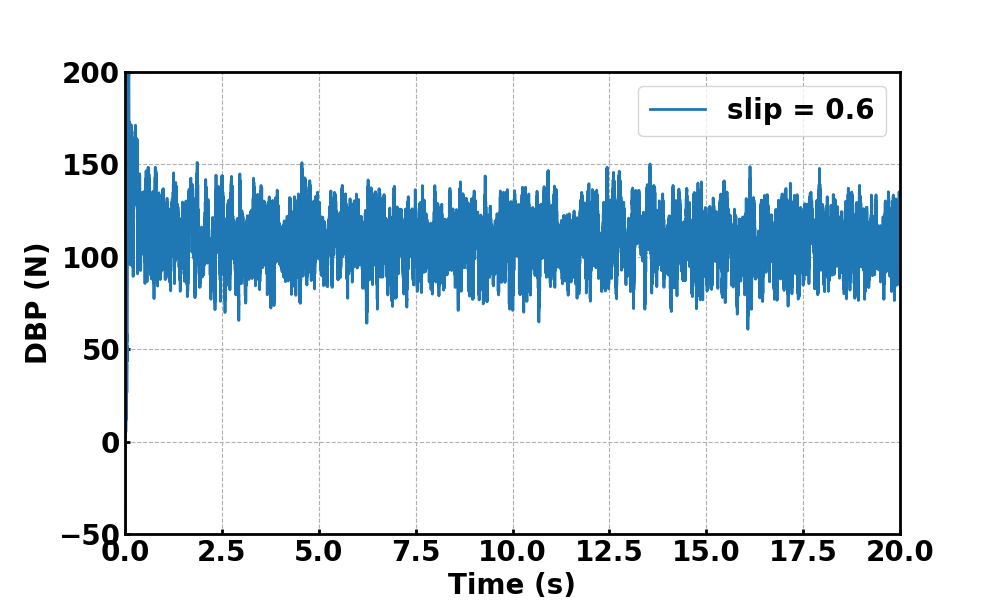}
	\includegraphics[width=2.in]{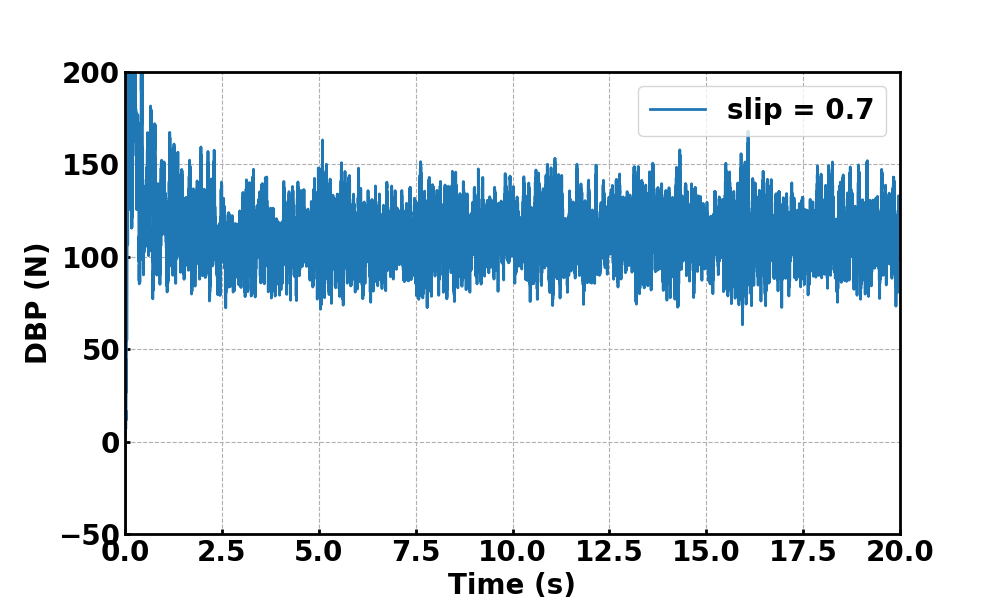}
	\includegraphics[width=2.in]{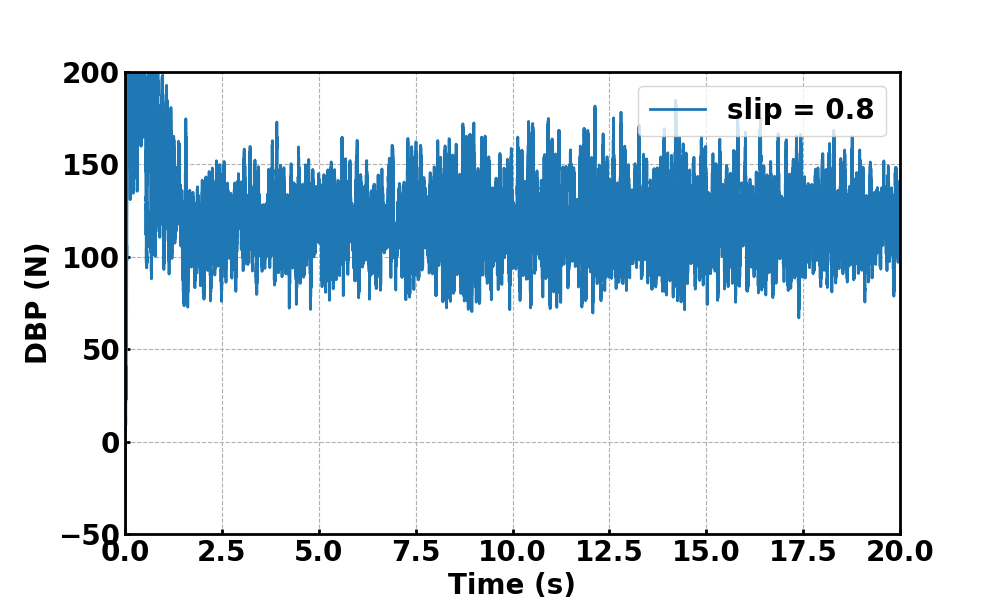}\\
	\caption{Time history of DBP force measured on the wheel at each slip ratio. The simulations were performed on GRC-3 lunar soil simulant in VV-mode. Steady state can be observed in each of the experiments. The density and internal friction angle were 1734 $\si{kg/m^3}$ and $42.0^{\circ}$, respectively.} 
	\label{fig:vv_dbp_time}
\end{figure}

Each simulation was run for approximately 20 s with a slip $s$ fixed at a predefined value in the 0 to 0.8 range. The ensuing average DBP force was measured as the force needed to be impressed at the wheel center to achieve this controlled VV-mode wheel movement. The simulation results are given in Fig.~\ref{fig:single_grc3}, and when compared with the TREC experimental data they show good agreement for both the DBP vs. slip and traction slope vs. slip relationships. It is noted that each of the markers shown in the plot requires one complete 20 s simulation since the DBP force is an averaged value. The time histories of the DBP force for each slip ratio considered are shown in Fig.~\ref{fig:vv_dbp_time}. The value of the DBP force was the averaged value of each simulation at its steady state regime. Note that at zero slip, the force is negative. In these simulations, three different sets of GRC-3 material proprieties associated with the lunar soil simulant were chosen -- with bulk densities 1627, 1734, and 1839 $\si{kg/m^3}$, and internal friction angles $37.8^{\circ}$, $42.0^{\circ}$, and $47.8^{\circ}$, respectively, see Fig. \ref{fig:densityVSfricangle} to place these values in context.

Two salient points associated with this simulator validation test are as follows: the physics-based simulator produces results in line with physical test results. Second, it is more convenient to use a physics-based simulator, compared to a BWJH-class model. For the latter, one cannot use intuitive and relatively readily available gravity-independent parameters such as bulk density and friction angle; rather, a bevameter test is required to identify the model parameters. Note that the outcomes of the bevameter test are gravity dependent -- Moon parameters would require testing in Moon conditions.

\FloatBarrier

\begin{figure}[h]
	\centering
	\begin{subfigure}{0.58\textwidth}
		\includegraphics[width=3.6in]{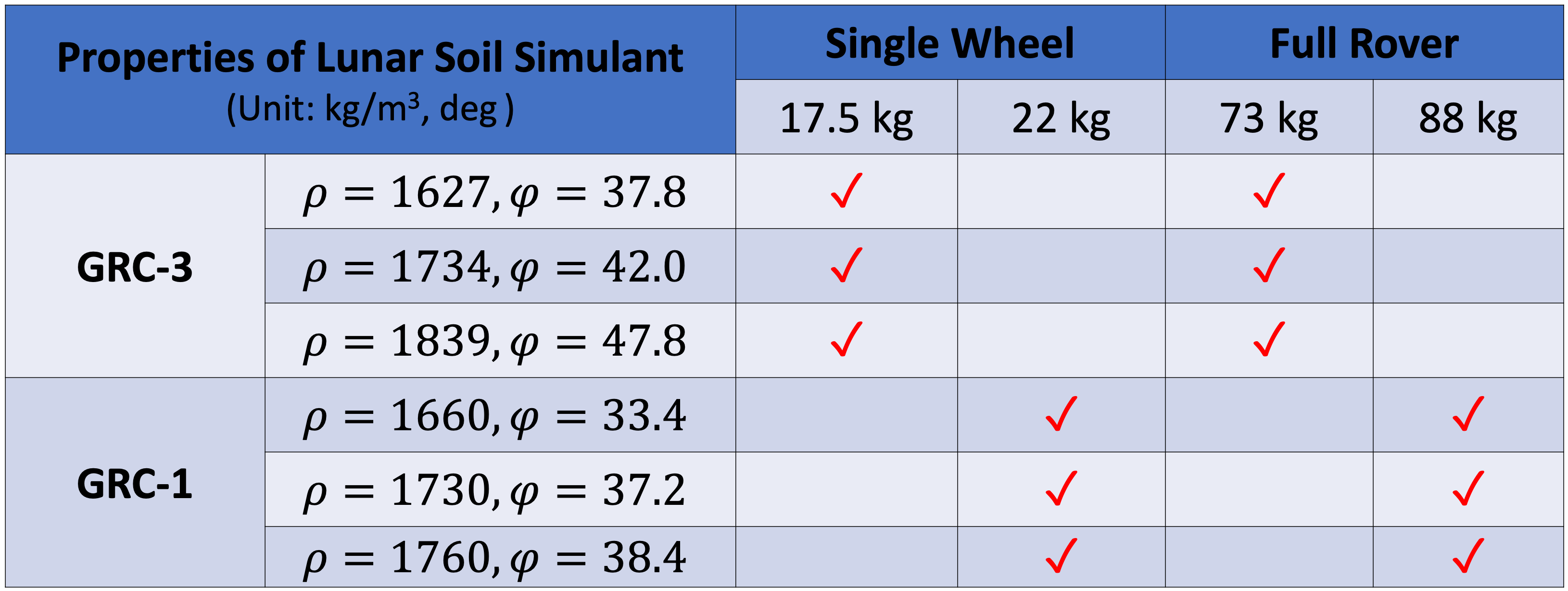}
		\caption{} 
		\label{fig:sim_summary}
	\end{subfigure}
	\begin{subfigure}{0.38\textwidth}
		\centering
		\includegraphics[width=2.4in]{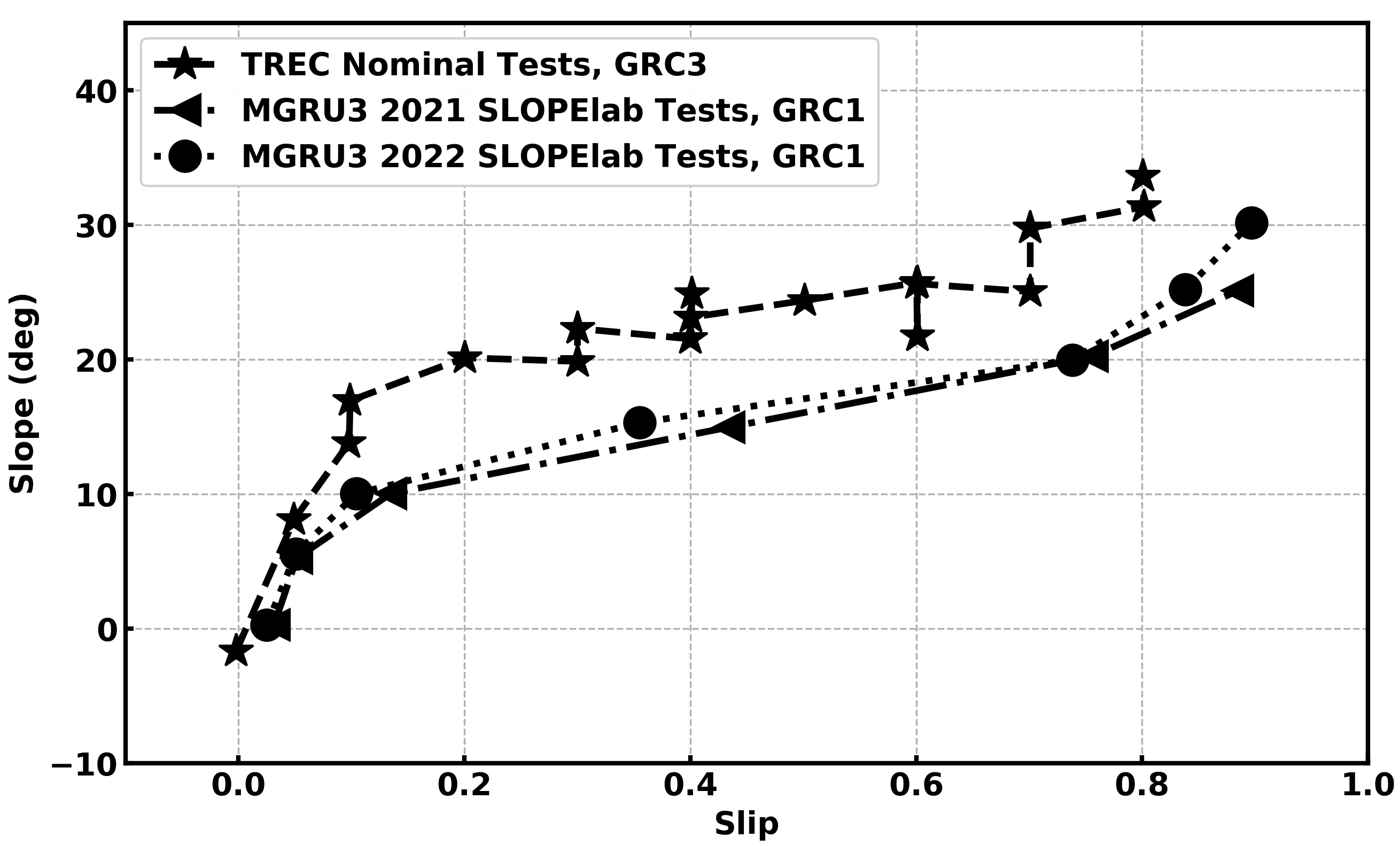}
		\caption{Experimental results in NASA's SLOPE lab.} 
		\label{fig:nasaSLOPEresults}
	\end{subfigure}
	\caption{The summary of ``slope-mode'' simulation cases. Three different sets of material properties (density and friction angle) of GRC-3 and GRC-1 were used. The simulations were run under both Earth gravity and Moon gravity. Two different angular velocities ($0.8~\si{rad/s}$ and $0.33~\si{rad/s}$) were used in the simulations under Moon gravity to assess the extent to which the simulation results come in line with the granular scaling laws.} 
	\label{fig:inputDataExperiments}
\end{figure}

\subsection{Slope-mode: VIPER and corresponding single wheel tests}\label{subsec:actual_slope}
This subsection presents results that highlight the following two aspects demonstrated via CRM terramechanics: there is no need to modify the mass or topology of the rover to predict through Earth tests the slope vs. slip map or the power draw experienced by the rover while operating on the Moon in steady state conditions; and results obtained for a single-wheel test are indicative of full rover behavior.

In slope-mode, the rover was placed on a tilted terrain with an actual slope varied from $\theta=0^{\circ}$ to $\theta = 30^{\circ}$. As such, the gravitational pull might not be perpendicular to the terrain surface, which has implications in relation to the strength attributes of the soil. In these experiments, the wheels of the rover were driven with a constant angular velocity $\omega = 0.8~\si{rad/s}$; the translational velocity  up the incline was not controlled -- it was an outcome of the experiment. The MGRU3 results are shown in Fig. \ref{fig:nasaSLOPEresults} with circle and triangle markers. 

Information about the slope-mode experiments is provided in Fig. \ref{fig:sim_summary}. There were $12 \times 3 \times 7 = 252$ simulations run: 12 red ``check marks'' in Fig.~\ref{fig:sim_summary}; three angular velocities -- $\omega = 0.8~\si{rad/s}$ for Earth, and $\omega = 0.33~\si{rad/s}$ and $\omega = 0.8~\si{rad/s}$ for the Moon; and 7 slopes, $\theta=0^{\circ}, 5^{\circ}, 10^{\circ}, 15^{\circ}, 20^{\circ}, 25^{\circ}, 30^{\circ}$. Each simulation ran for approximately 20 s to ensure that a steady state was reached. At steady state, we measured the average rover translational velocity $v$ and subsequently calculated the associated slip ratio $s$. To investigate whether single-wheel results are indicative of full rover dynamics, we also ran the single wheel simulation in the same slope-mode with approximately 1/4 of the mass of the rover. The densities of GRC-1 used in the CRM simulation were 1660, 1730, and 1760 $\si{kg/m^3}$. The internal friction angles were $33.4^{\circ}$, $37.2^{\circ}$, and $38.4^{\circ}$, respectively; see Fig. \ref{fig:densityVSfricangle} for placing these values in context. To be consistent with the experimental data obtained from the NASA's TREC test rig for single wheel and the Moon Gravitation Representative Unit 3 (MGRU3) for the full VIPER, a 73 kg digital twin was built for the scenarios with GRC-3 lunar simulant, while a 88 kg digital twin was used in the GRC-1 scenarios. In the single wheel test, 17.5 kg (which is close to 73/4) and 22 kg (which is 88/4) wheels were used, respectively, in accordance to how MGRU3's mass changed during the design phase of VIPER. Under Moon gravity, we considered an angular velocity $\omega = 0.33~\si{rad/s}$ since this is roughly $\frac{1}{\sqrt{6}}$ of the value used under Earth gravity. This ratio is dictated by the scaling law outlined in \cite{kamrin-gsl-expanded2020}) as the one necessary to predict the rover's performance on the Moon.

\begin{figure}[h]
	\centering
	\begin{subfigure}{0.49\textwidth}
		\centering
		\includegraphics[width=3.2in]{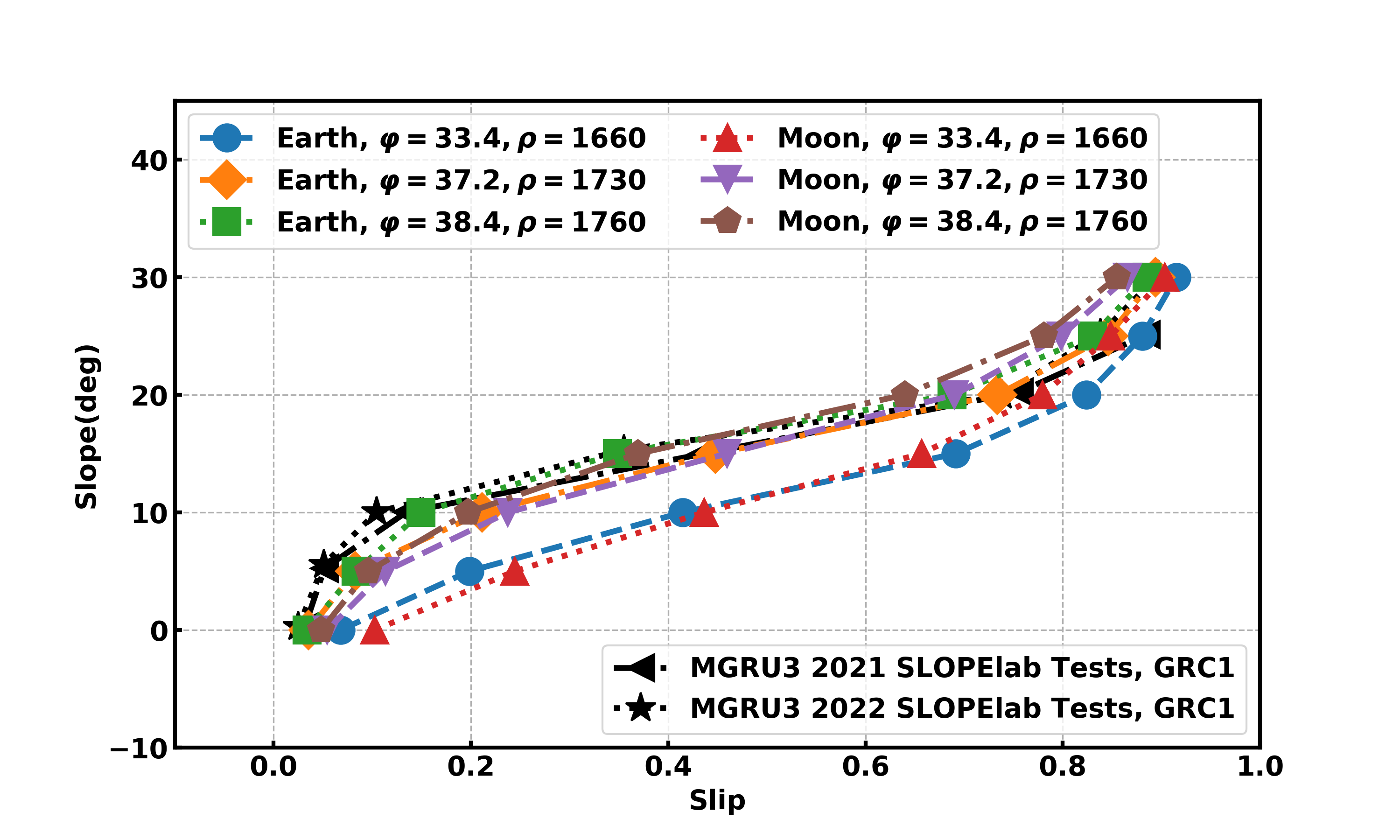}
		\caption{Single wheel on GRC-1.}
		\label{fig:single_wheel_earth_vs_moon}
	\end{subfigure}
	\begin{subfigure}{0.49\textwidth}
		\centering
		\includegraphics[width=3.2in]{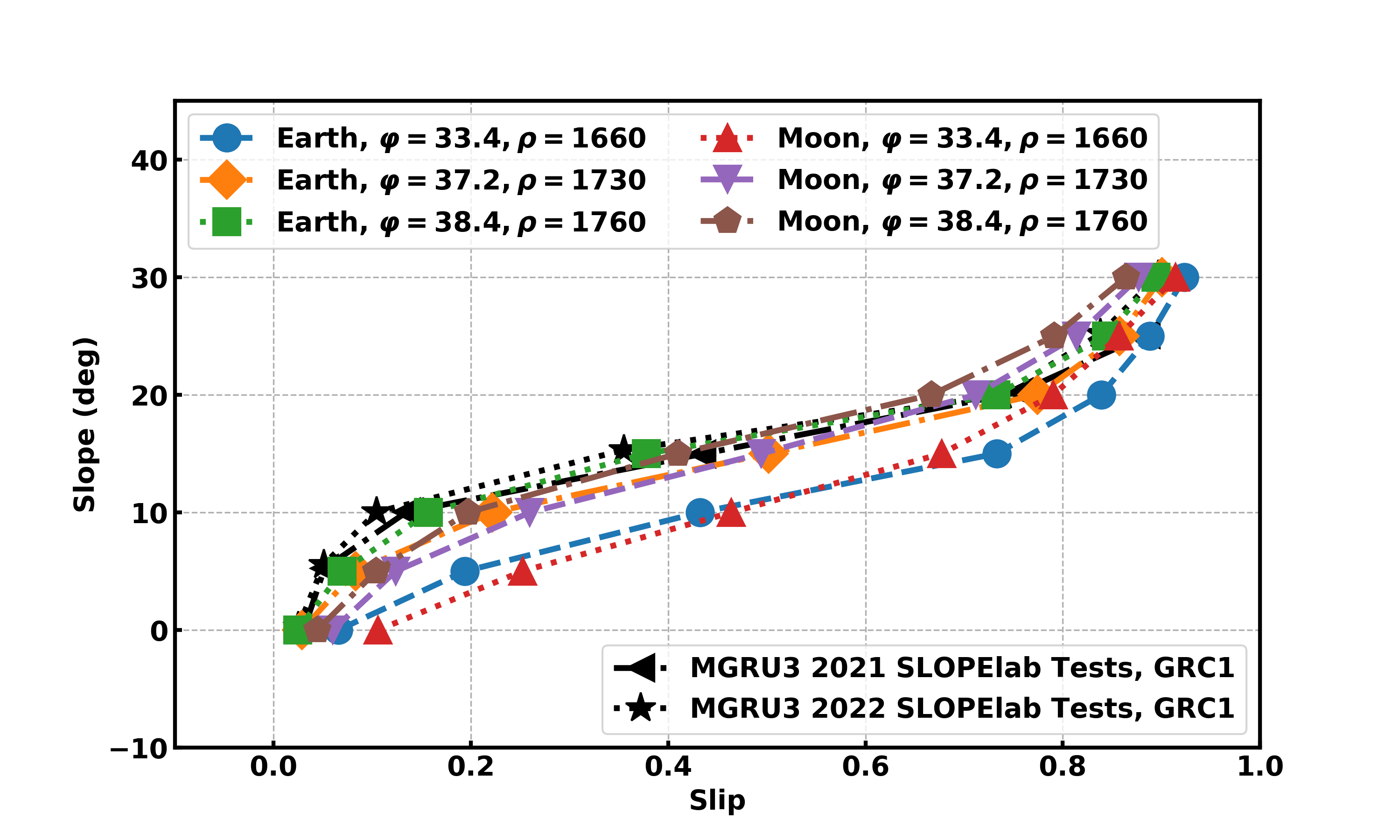}
		\caption{Full rover on GRC-1.} 
		\label{fig:full_earth_vs_moon}
	\end{subfigure}
	\caption{Single wheel and full rover simulation using CRM on GRC-1 lunar soil simulant. A 22 kg single wheel and 88 kg MGRU3 rover were used in the simulations. The wheel angular velocity was set to $0.8~\si{rad/s}$ in the simulations with Earth gravity, and $0.33~\si{rad/s}$ for Moon gravity according to the granular scaling laws discussed in \cite{kamrin-gsl-expanded2020}. All the values used to generated the markers in the figures were averaged out in steady state for each slope $\theta$ scenario. In the slope-mode plots, one would start with a slope on the $y$ axis and note the slip it led to on the $x$ axis. The single wheel results were obtained in the Traction and Excavation Capabilities (TREC) Rig at Glenn Research Center. Similar results for GRC-3 are provided in the Supplementary Material component; the GRC-3 results are qualitatively identical to the ones presented here.} 
	\label{fig:single_full_earh_vs_moon}
\end{figure}

Figure \ref{fig:single_full_earh_vs_moon} compares simulation results and the experimental data under both Earth and Moon gravity. The salient points are as follows: (i) the single wheel and full rover simulations performed under Earth gravity match well the physical test results obtained at SLOPE lab; and the rover's performance on the Moon is consistent with that observed on Earth in terms of the slope/slip relationship if the wheel driving angular velocity meets the requirement according to scaling law reported in \cite{kamrin-gsl-expanded2020}. By inspecting data provided in the Supplementary Material, it is noted that the performance of the rover on GRC-3 is slightly better than that on GRC-1 due to the higher friction coefficient and higher soil density.

\begin{figure}[h]
	\centering
	\begin{subfigure}{0.49\textwidth}
		\centering
		\includegraphics[width=3.2in]{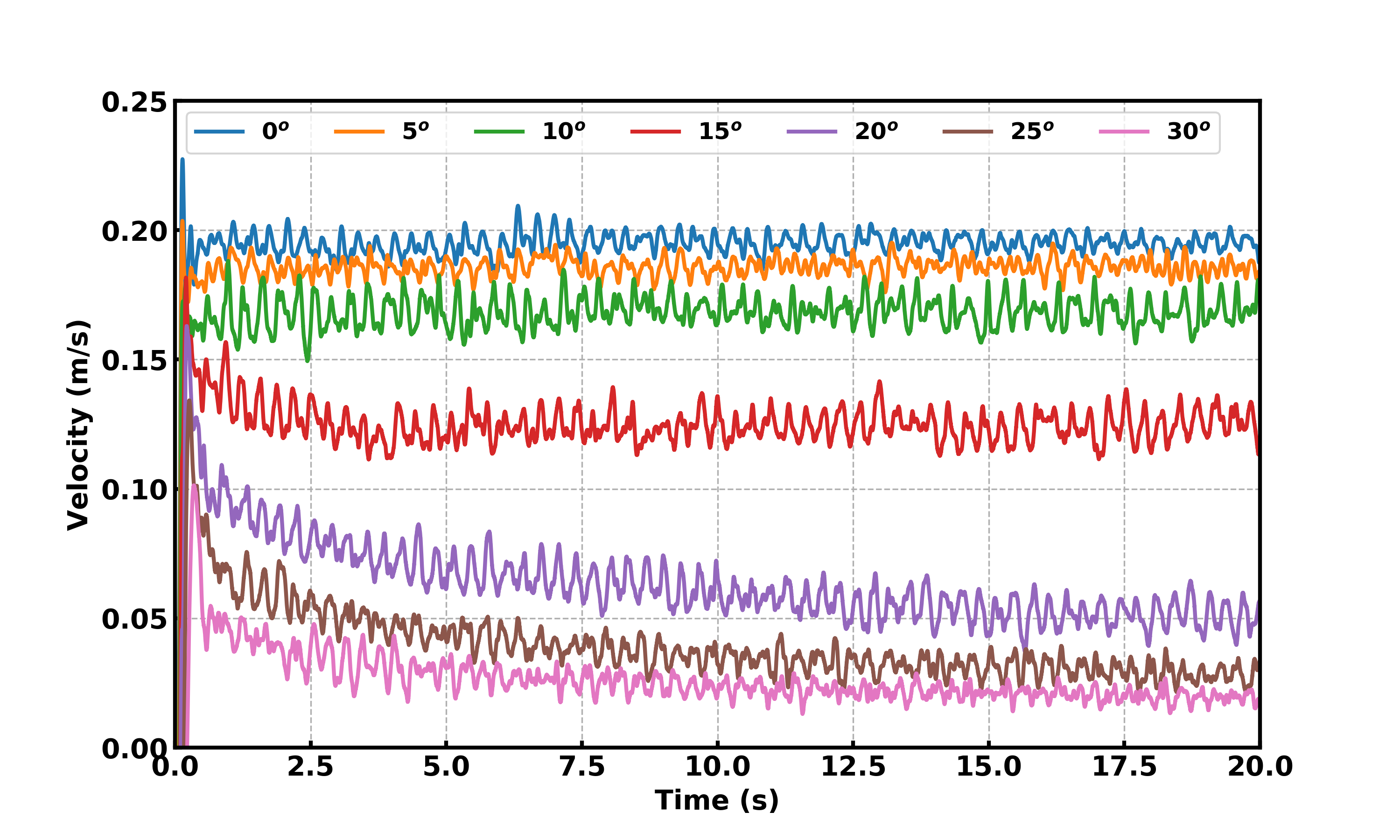}
		\caption{Earth gravity, angular velocity $\omega=0.8~\si{rad/s}$.}
		\label{fig:details-multipleslopes-fullrover-earth}
	\end{subfigure}
	\begin{subfigure}{0.49\textwidth}
		\centering
		\includegraphics[width=3.2in]{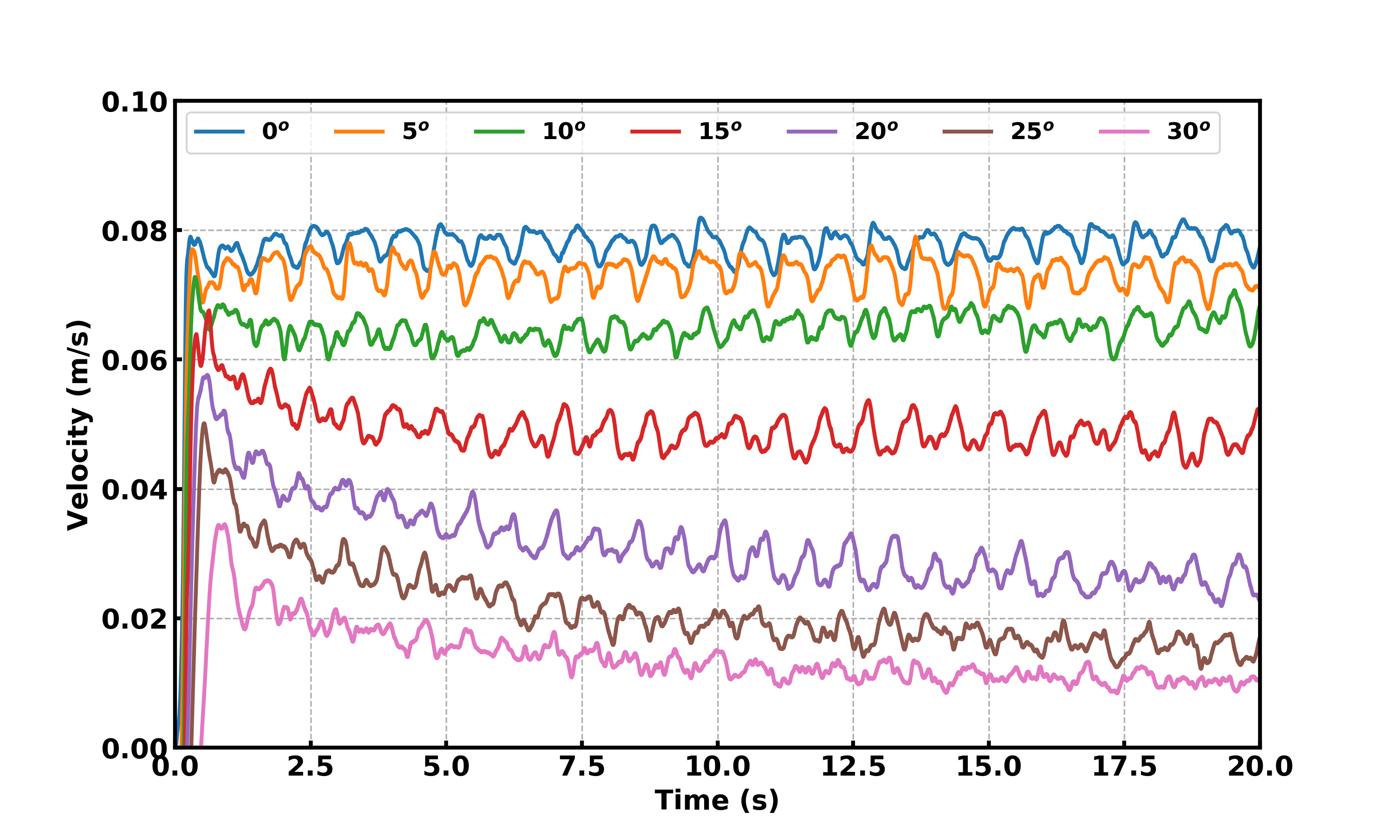}
		\caption{Moon gravity, angular velocity $\omega=0.33~\si{rad/s}$.} 
		\label{fig:details-multipleslopes-fullrover-moon}
	\end{subfigure}
	\caption{Time history for MGRU3's velocity simulated on GRC-1. Tests were done for $\theta$ between $0^{\circ}$ and $30^{\circ}$ in increments of $5^{\circ}$ with bulk density of 1760 $\si{kg/m^3}$ and friction angle $38.4^{\circ}$. The information in these two plots was used to generate the green and brown curves in Fig.~\ref{fig:full_earth_vs_moon}.} 
	\label{fig:rover_vel_his_grc1}
\end{figure}

The results illustrated in Fig.~\ref{fig:rover_vel_his_grc1} were used to generate the green and brown curves in Fig.~\ref{fig:full_earth_vs_moon}. Specifically, GRC-1 simulations with bulk density of 1760 $\si{kg/m^3}$ and friction angle $38.4^{\circ}$ were run for $\theta$ between $0^{\circ}$ and $30^{\circ}$ in increments of $5^{\circ}$ -- under Earth gravity in Fig.~\ref{fig:details-multipleslopes-fullrover-earth} and Moon gravity in Fig.~\ref{fig:details-multipleslopes-fullrover-moon}. Each slope $\theta$ leads to a velocity profile in Fig.~\ref{fig:details-multipleslopes-fullrover-earth}, and that velocity profile, when averaged out at steady state leads to a slip value, which represents one dot on the green line in Fig.~\ref{fig:full_earth_vs_moon}. Likewise, each $\theta$ leads to a lunar velocity profile in Fig.~\ref{fig:details-multipleslopes-fullrover-moon}, and that velocity profile, when averaged out at steady state leads to a slip value, which represents one dot on the brown line in Fig.~\ref{fig:full_earth_vs_moon}. Note how the scaling law emerges from the results reported in Fig.~\ref{fig:rover_vel_his_grc1}: for instance, when the slope of the terrain was $15^{\circ}$ (red lines in the plots), on the left, the rover average velocity on Earth was approximately 0.125 m/s; in Moon gravity, the speed averaged at 0.05 m/s. The ratio between these numbers works out to be approximately $\sqrt{6}$. The same $\sqrt{6}$ ratio holds if one compares any two curves of identical color in the left and right plots in Fig.~\ref{fig:rover_vel_his_grc1}.

\begin{figure}[h]
	\centering
	\begin{subfigure}{0.49\textwidth}
		\centering
		\includegraphics[width=3.2in]{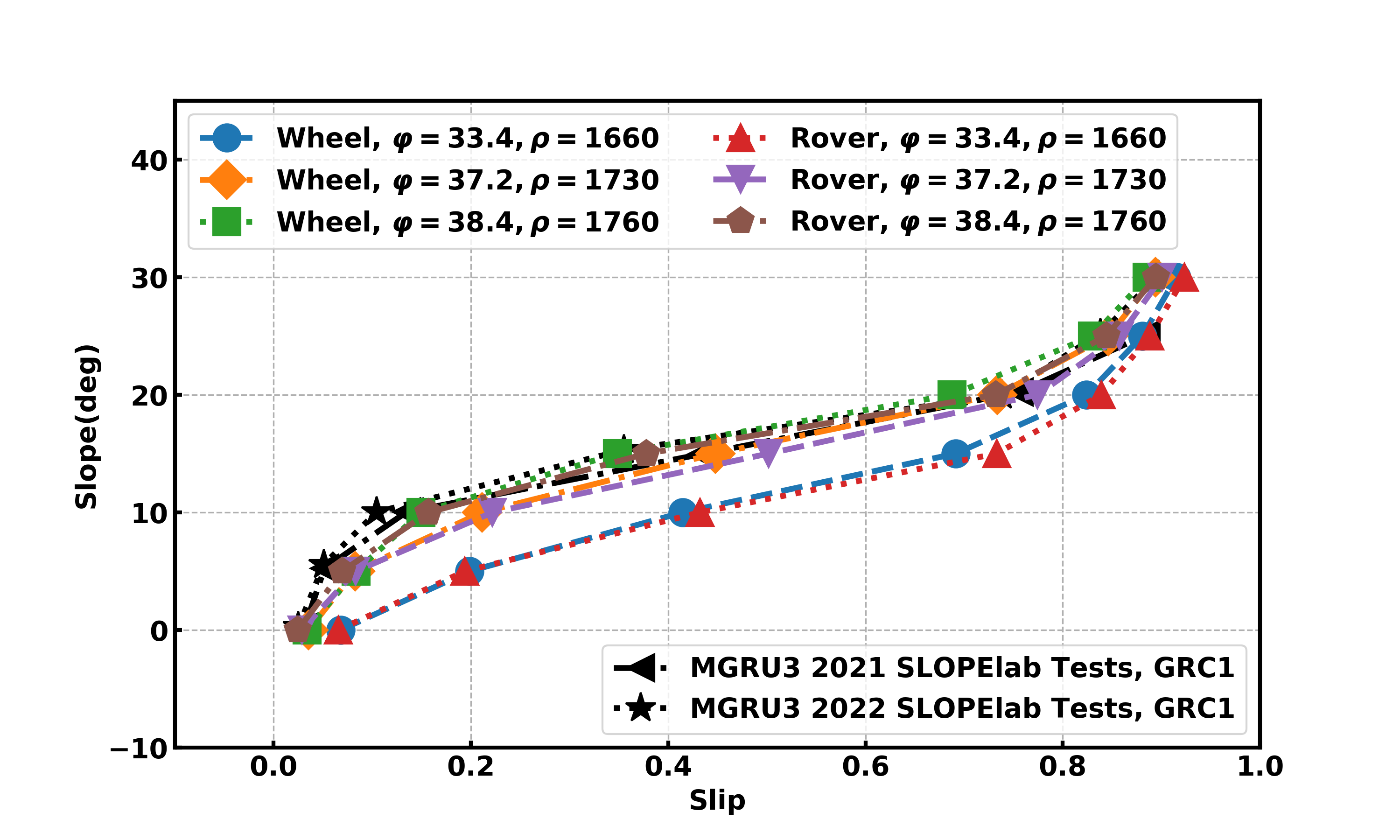}
		\caption{Single wheel vs. full rover results -- Earth gravity on GRC-1; $\omega=0.8$ rad/s}
	\end{subfigure}
	\begin{subfigure}{0.49\textwidth}
		\centering
		\includegraphics[width=3.2in]{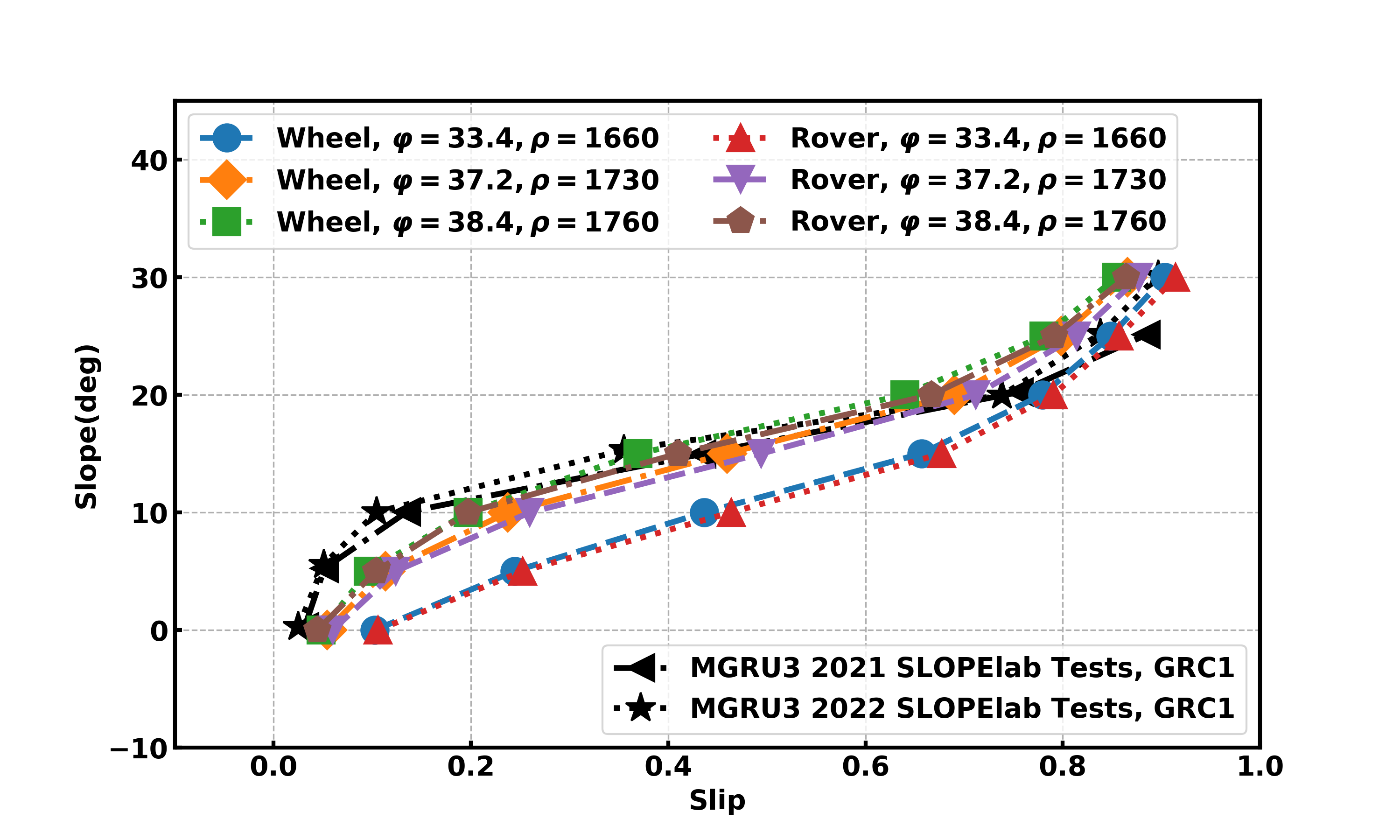}
		\caption{Single wheel vs. full rover results -- Moon gravity on GRC-1; $\omega=0.33$ rad/s} 
	\end{subfigure}
	\caption{A comparison between single wheel and MGRU3 simulations using GRC-1 lunar soil simulant with both Earth and Moon gravity. Note that single wheel performance is indicative of full rover performance. Similar results for GRC-3 are provided in the Supplementary Material component; the GRC-3 results are qualitatively identical to the ones presented here.} 
	\label{fig:wheel_vs_rover_earth_moon}
\end{figure}

Figure \ref{fig:wheel_vs_rover_earth_moon} reports single wheel and full rover simulation results on GRC-1 simulant under Earth and Moon gravity. The performance of a single wheel exhibits a slope/slip relationship comparable to that of the entire rover, indicating that simulations using just a single wheel are generally sufficient to predict the rover's overall performance in this respect. Note that the single wheel simulation is approximately four times faster due to the fewer SPH particles that participate in the CRM simulation. This is accomplished by using ``active domains'' -- only the dynamics of the material in the proximity of the implements that come in contact with the soil, i.e., the active domain, is simulated rather than the terramechanics of the entire mass of regolith. 

\begin{figure}[h]
	\centering
	\begin{subfigure}{0.49\textwidth}
		\centering
		\includegraphics[width=3.2in]{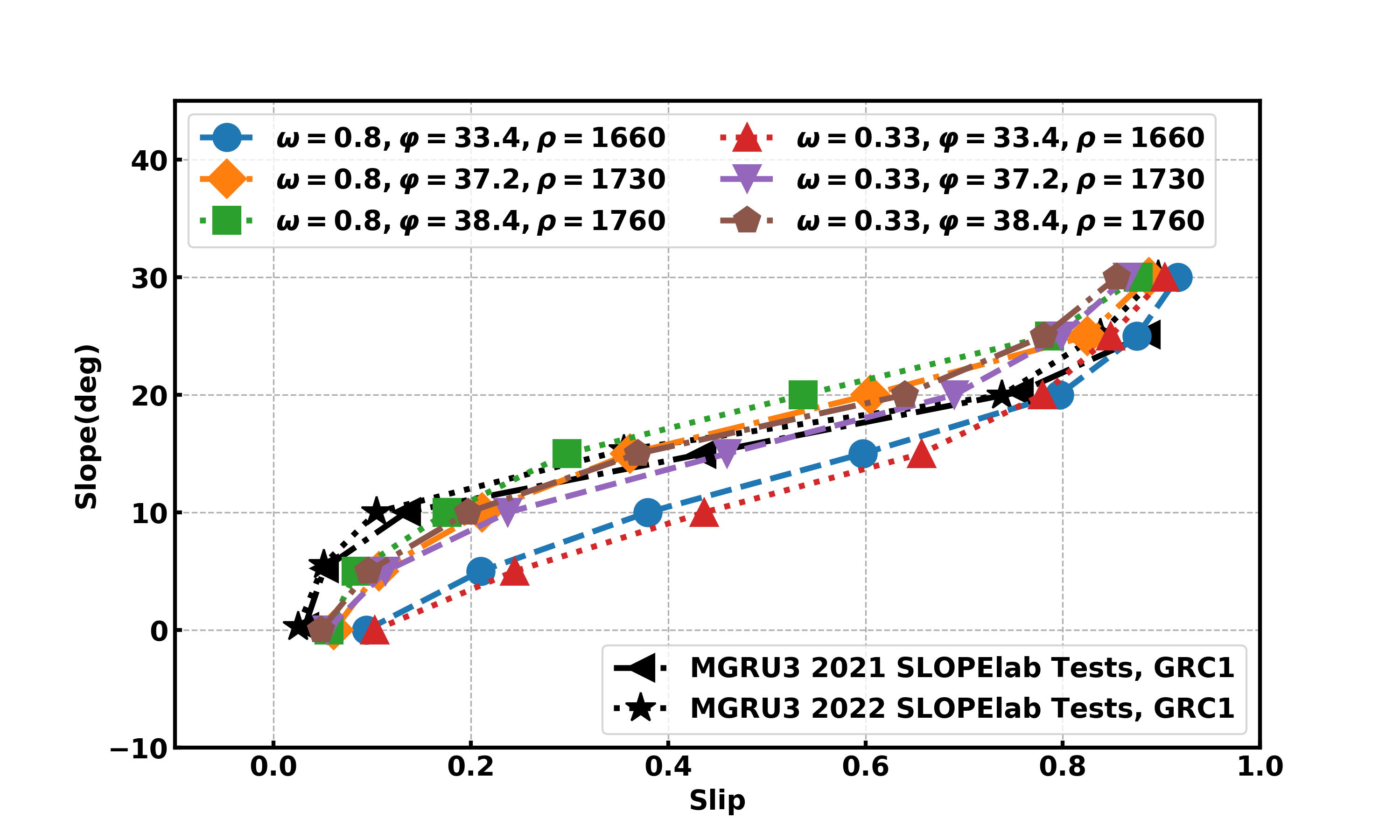}
		\caption{Single wheel on GRC-1.}
		\label{fig:2AngVelMoonGravSingleWheel}
	\end{subfigure}
	\begin{subfigure}{0.49\textwidth}
		\centering
		\includegraphics[width=3.2in]{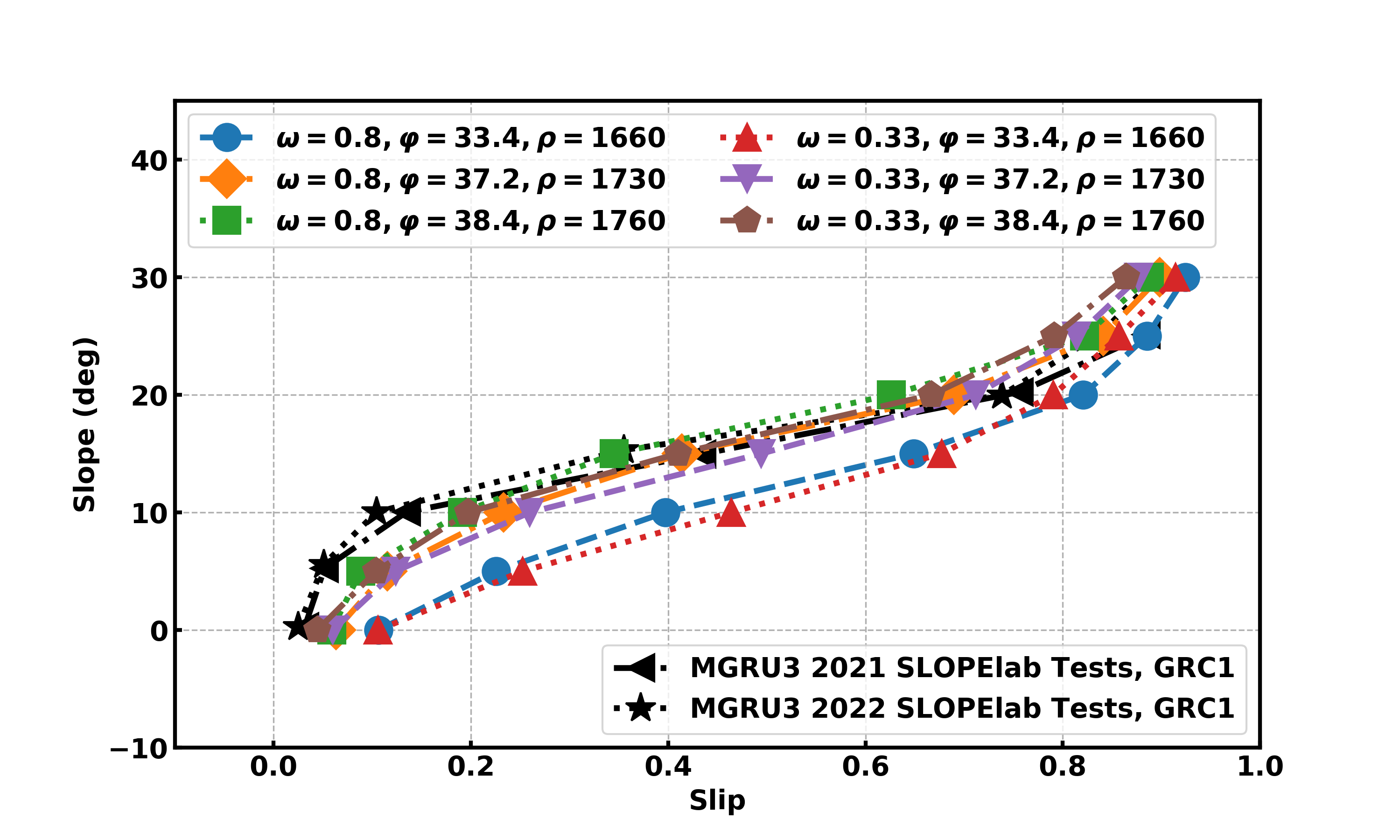}
		\caption{Full rover on GRC-1.} 
		\label{fig:2AngVelMoonGravFullRover}
	\end{subfigure}
	\caption{Single wheel and full rover simulation using on GRC-1 lunar soil simulant \textit{under Moon gravity}. Two different angular velocity were used -- $0.33~\si{rad/s}$  and $0.8~\si{rad/s}$, yet the slope vs. slip curves are identical. Similar results for GRC-3 are provided in the Supplementary Material component; the GRC-3 results are qualitatively identical to the ones presented here.} 
	\label{fig:diff_angW}
\end{figure}

To gain insights into how the mobility attributes change with the angular velocity, a set of single wheel and full rover simulations were run under Moon gravity with a higher angular velocity $\omega = 0.8~\si{rad/s}$; a comparison with results obtained for $\omega = 0.33~\si{rad/s}$ are provided in Fig. \ref{fig:diff_angW}. The results are almost identical, as up to a critical value of the angular velocity, the slope/slip relationship is not sensitive to angular velocity, see Fig. 3A in \cite{kenGoldman2021}. Note, however, that should one look at the translational velocity of the lunar rover at $\omega = 0.8~\si{rad/s}$, the scaling law would not be able to correlate that translational velocity to the one of the rover moving on Earth when the wheels are driven at $\omega = 0.8~\si{rad/s}$.

\FloatBarrier

\begin{figure}[h]
	\centering
	\begin{subfigure}{0.49\textwidth}
		\centering
		\includegraphics[width=3.2in]{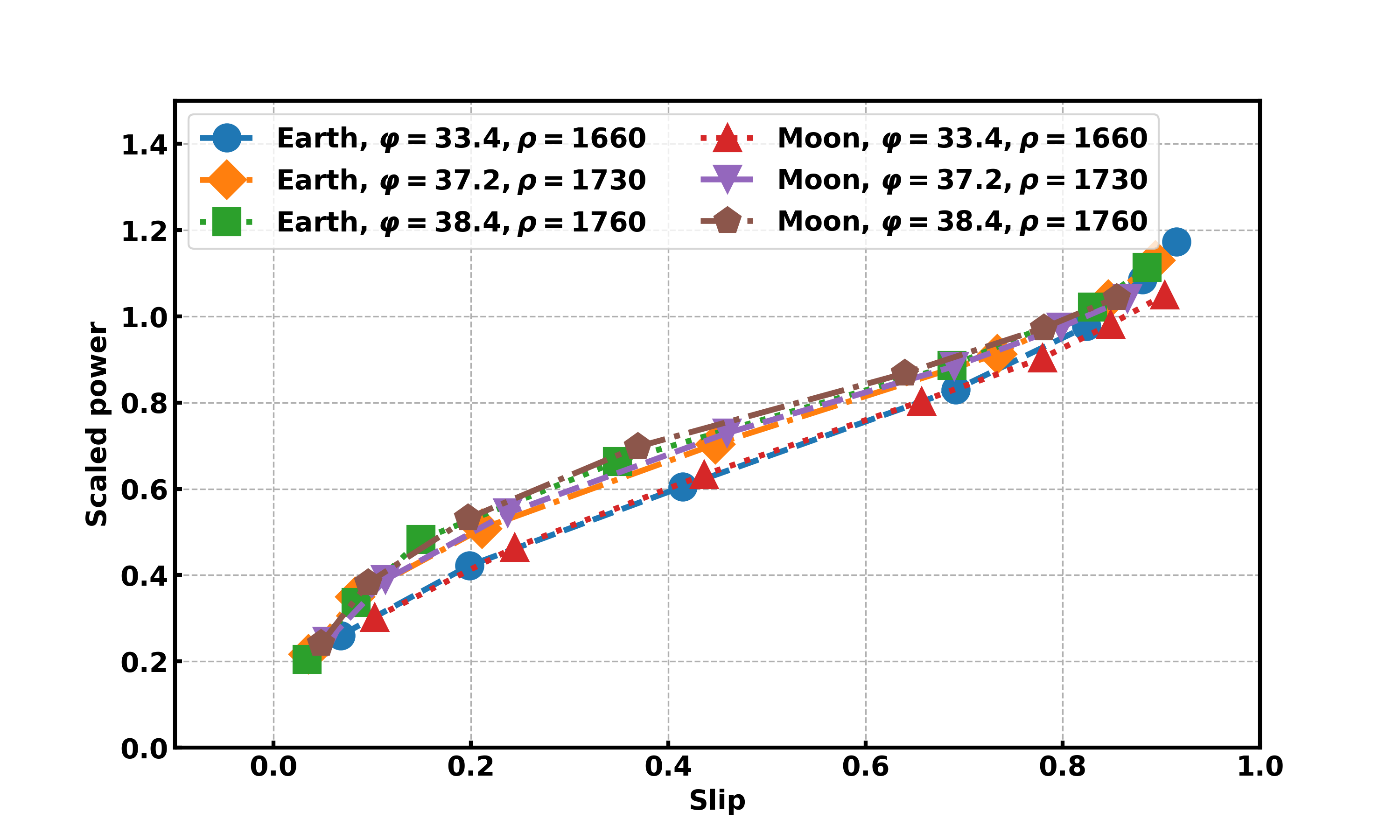}
		\caption{Single wheel: $\omega$=0.8~\si{rad/s} on Earth, $\omega$=0.33~\si{rad/s} on Moon}
	\end{subfigure}
	\begin{subfigure}{0.49\textwidth}
		\centering
		\includegraphics[width=3.2in]{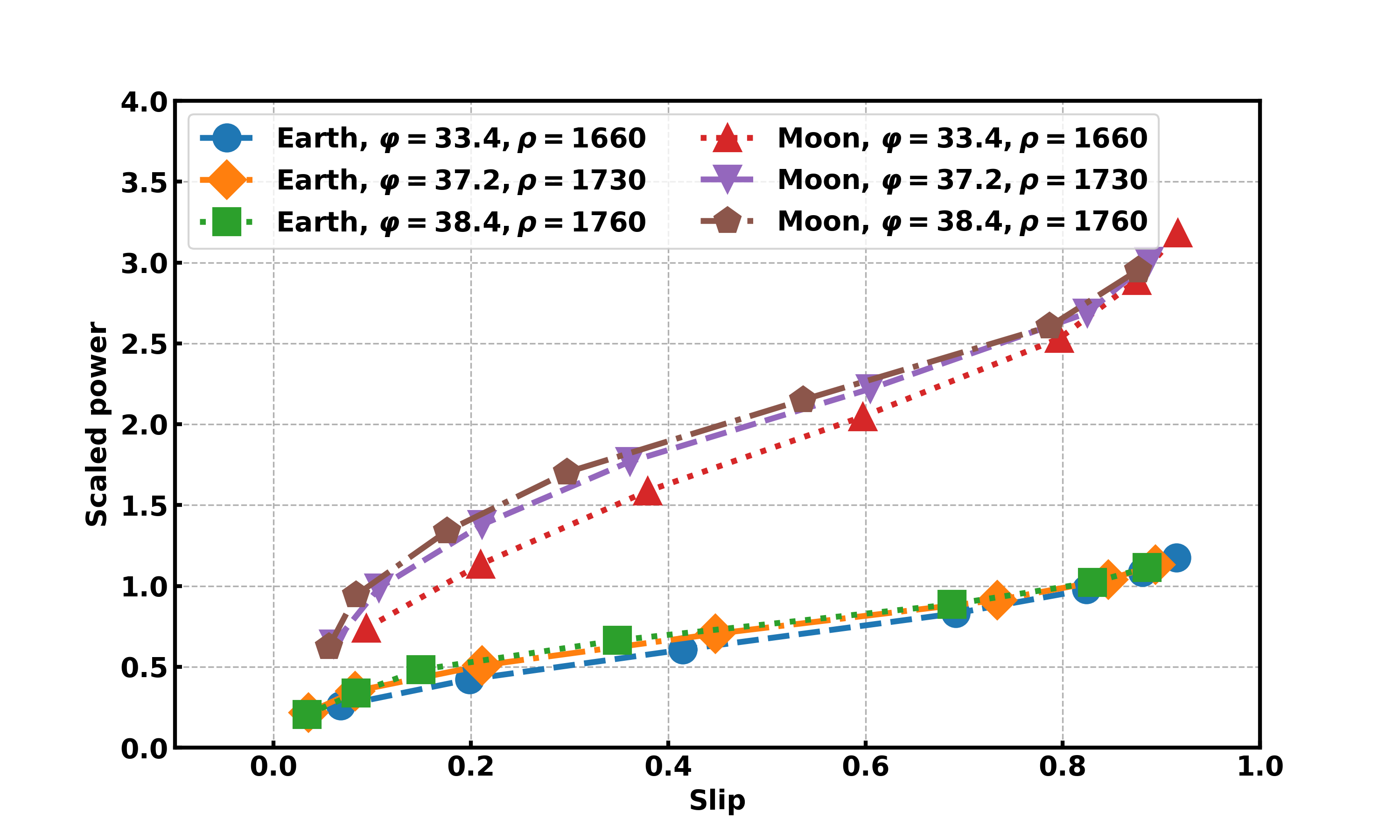}
		\caption{Single wheel: $\omega$=0.8~\si{rad/s} on Earth, $\omega$=0.8~\si{rad/s} on Moon} 
	\end{subfigure}
	\begin{subfigure}{0.49\textwidth}
		\centering
		\includegraphics[width=3.2in]{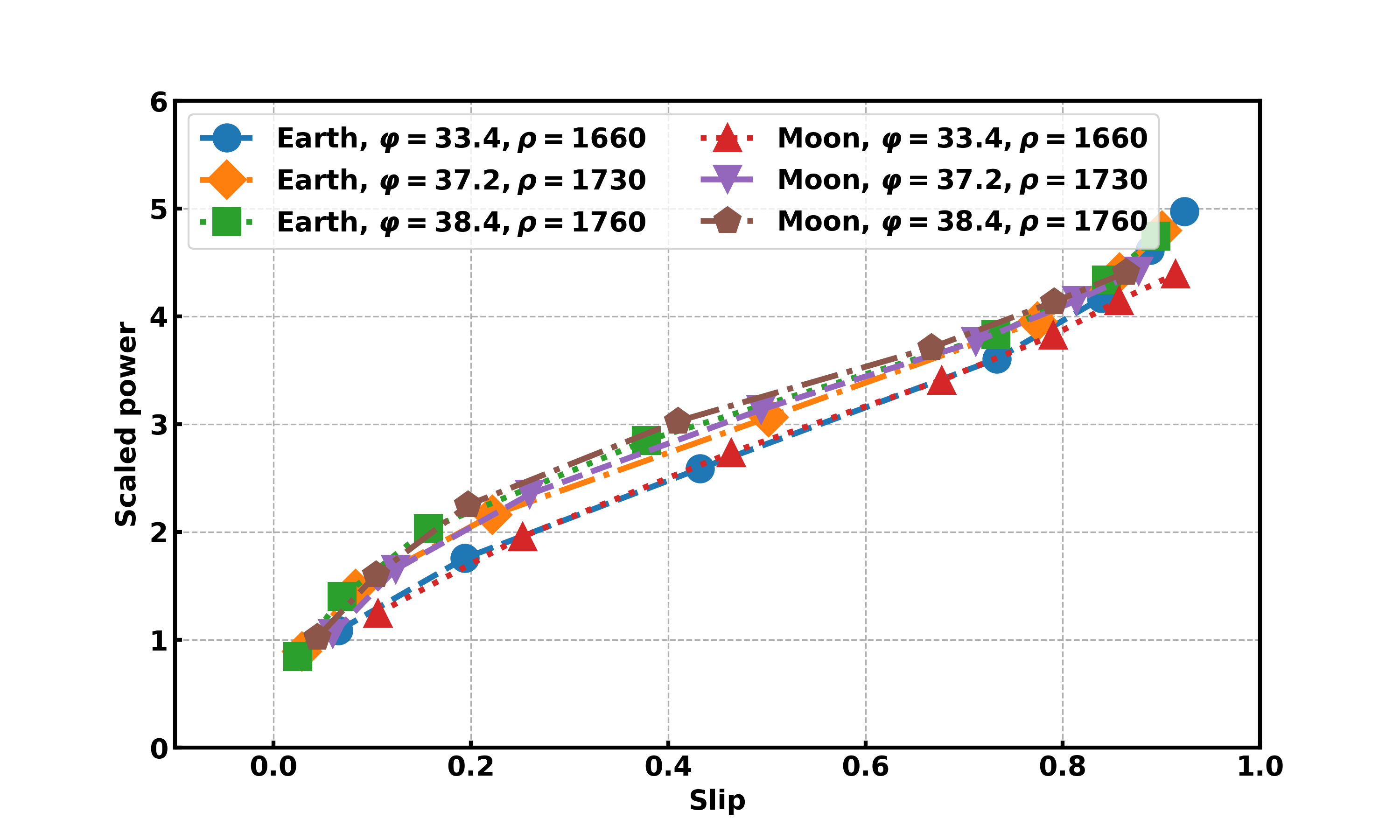}
		\caption{Full rover: $\omega$=0.8~\si{rad/s} on Earth, $\omega$=0.33~\si{rad/s} on Moon}
	\end{subfigure}
	\begin{subfigure}{0.49\textwidth}
		\centering
		\includegraphics[width=3.2in]{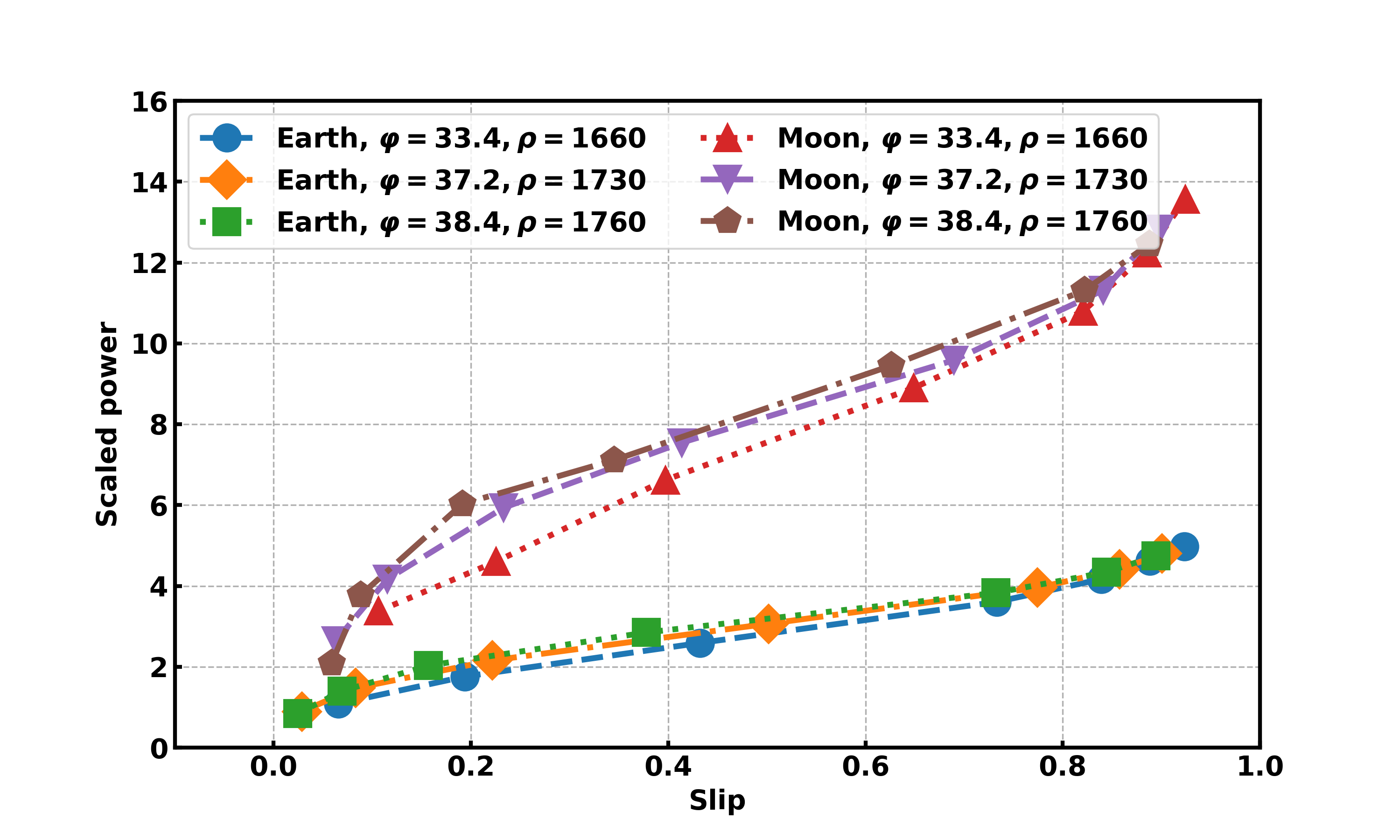}
		\caption{Full rover: $\omega$=0.8~\si{rad/s} on Earth, $\omega$=0.8~\si{rad/s} on Moon} 
	\end{subfigure}
	\caption{Scaled wheel power at steady state of single wheel/full rover simulation using CRM on GRC-1 lunar soil simulant.} 
	\label{fig:rover_power_grc1}
\end{figure}

Figure \ref{fig:rover_power_grc1} gives the \textit{scaled} power/slip relationship for two scenarios: single wheel and full rover on GRC-1 lunar soil simulant, where the \textit{scaled} power $\frac{P}{Mg\sqrt{Lg}}$ is the term in the left side of Eq.~(\ref{eq:scalingLaws}). One angular velocity $0.8~\si{rad/s}$ was used in the simulations under Earth gravity, while two different angular velocities, $0.33~\si{rad/s}$ and $0.8~\si{rad/s}$, were used in the simulations under Moon gravity. Note that if the angular velocity in the Earth experiment is roughly $\sqrt{6}$ times larger than that used under Moon gravity, the \textit{scaled} powers are identical. If we use same angular velocity to do the tests on both Moon and Earth, there will be a gap between these two sets of simulations for both single wheel and full rover, as shown in Fig. \ref{fig:rover_power_grc1} b and d. Similar results can be noted for GRC-3, see the Supplementary Material section. The results indicate that the \textit{scaled} power/slip relationship obtained in simulation produces results predicted by the granular scaling law \cite{kamrin-gsl-expanded2020}. 

\begin{figure}[h]
	\centering
	\begin{subfigure}{0.49\textwidth}
		\centering
		\includegraphics[width=3.0in]{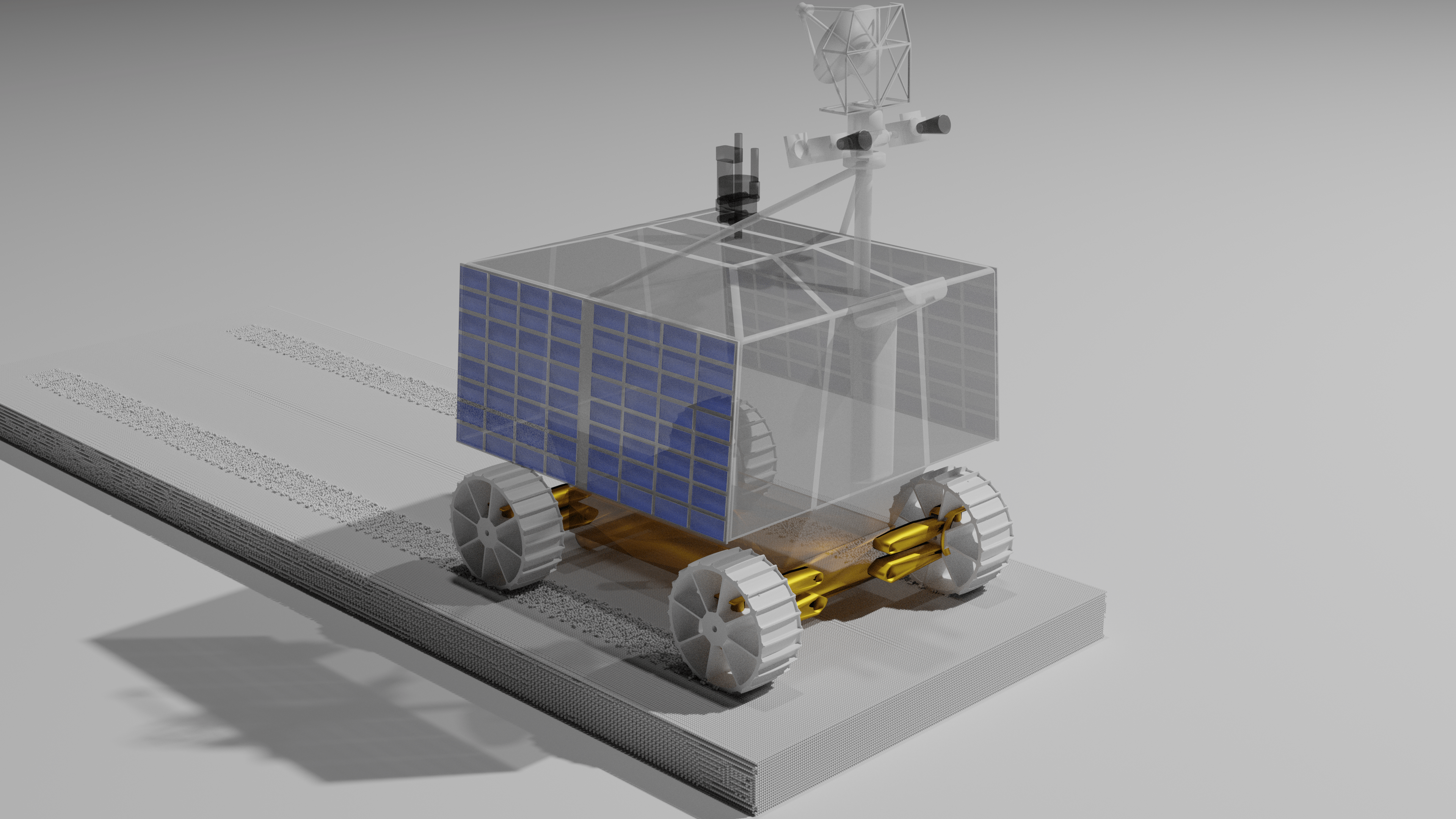}
		\caption{slope = $0^{\circ}$}
	\end{subfigure}
	\begin{subfigure}{0.49\textwidth}
		\centering
		\includegraphics[width=3.0in]{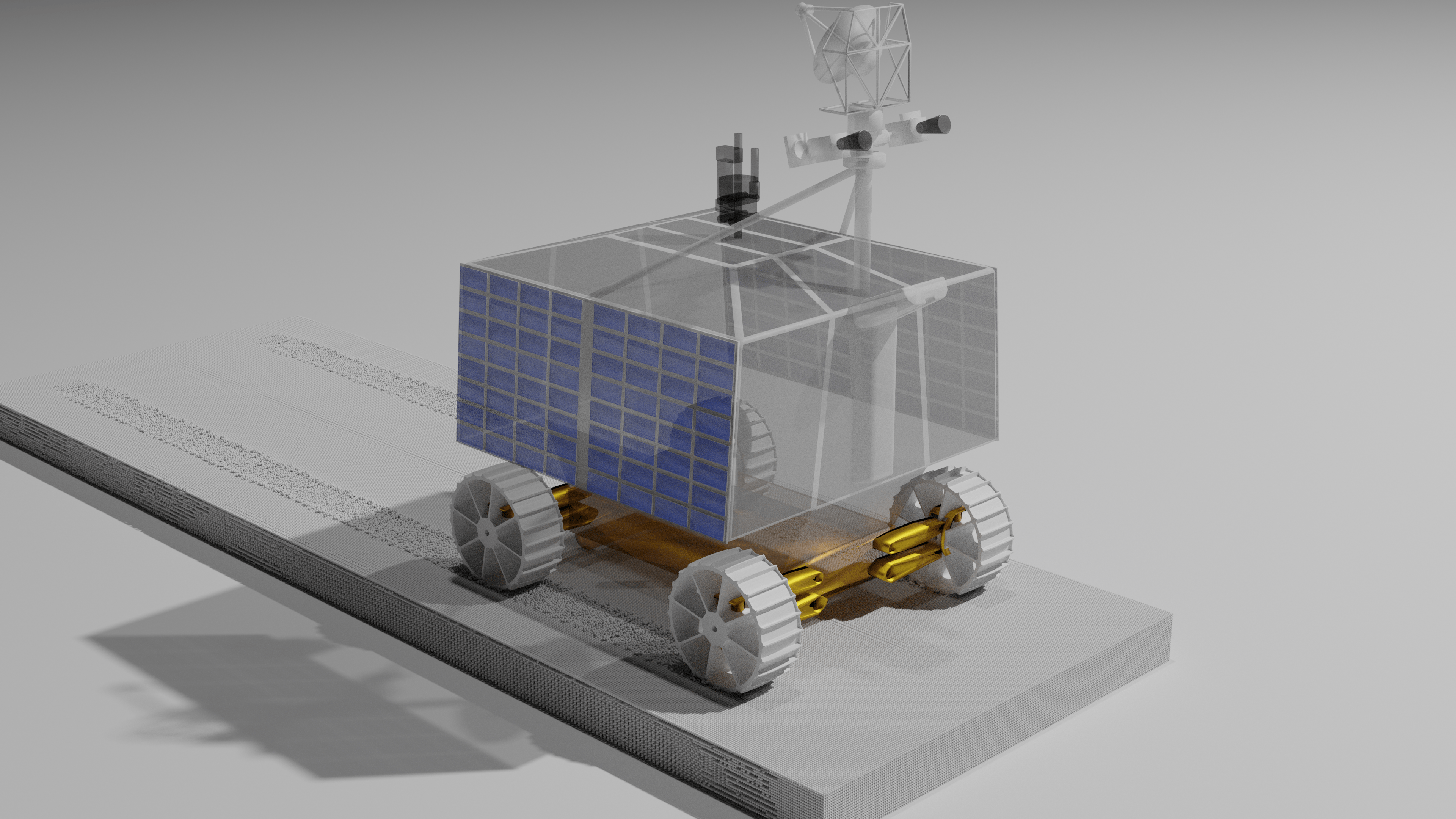}
		\caption{slope = $10^{\circ}$} 
	\end{subfigure}
	\begin{subfigure}{0.49\textwidth}
		\centering
		\includegraphics[width=3.0in]{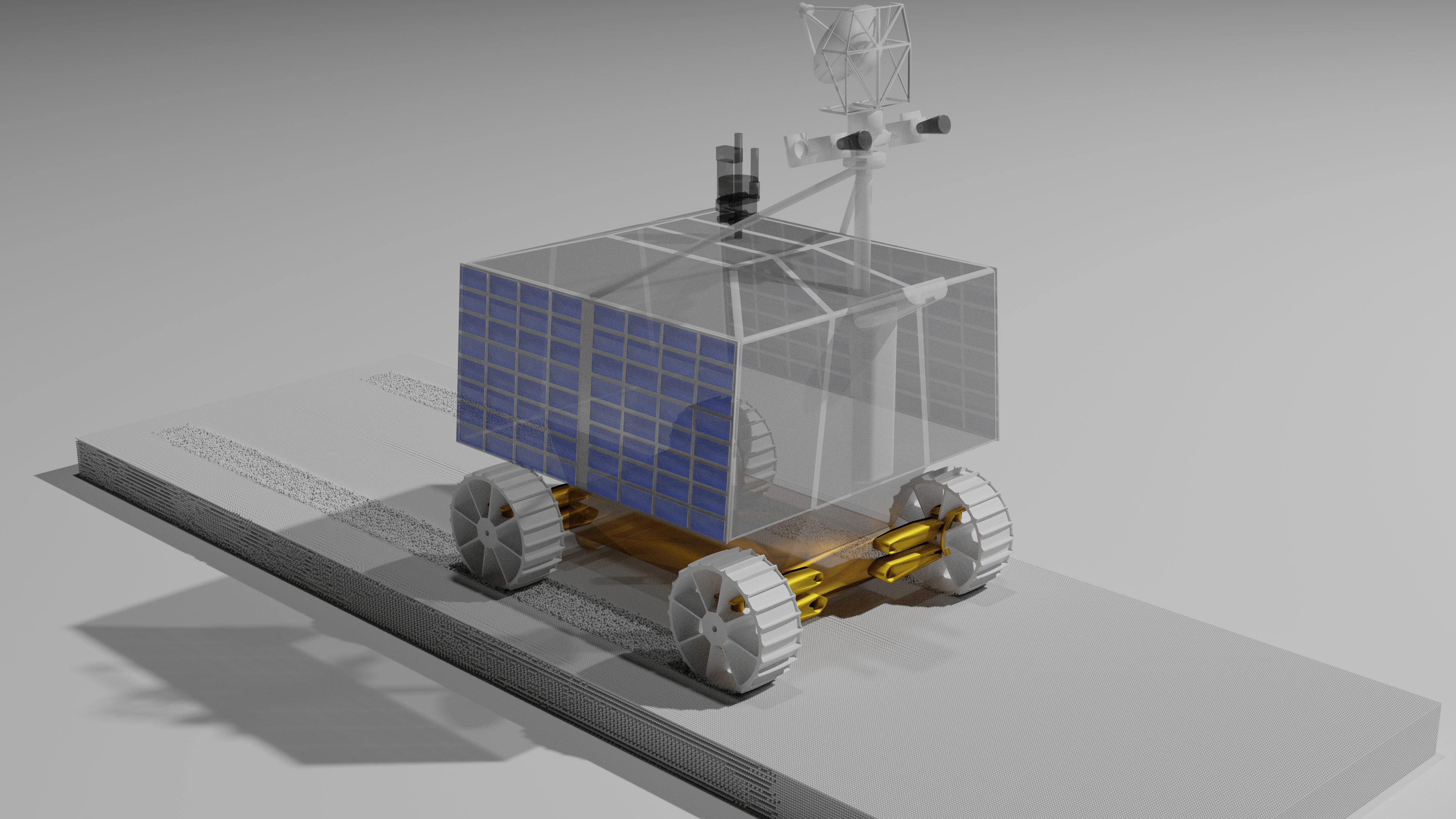}
		\caption{slope = $20^{\circ}$}
	\end{subfigure}
	\begin{subfigure}{0.49\textwidth}
		\centering
		\includegraphics[width=3.0in]{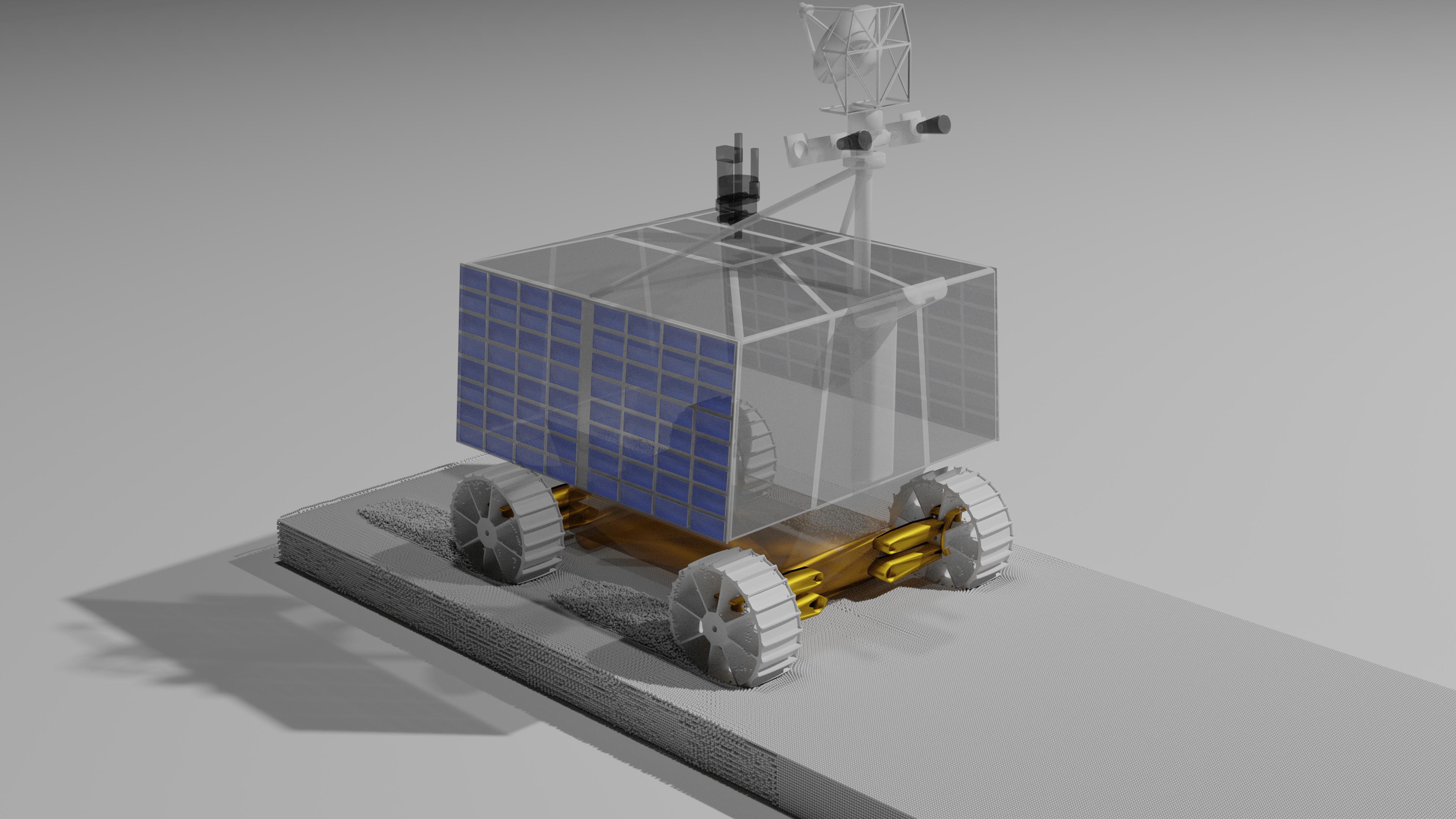}
		\caption{slope = $30^{\circ}$} 
	\end{subfigure}
	\caption{Screenshots of the 73 kg rover simulated on GRC-3 under Earth gravity, 20 seconds into the motion of the rover. The terrain slope is accounted for by changing the direction of the gravitational pull.} 
	\label{fig:full_grc3_screen}
\end{figure}

Figure \ref{fig:full_grc3_screen} shows screen shots of the 73 kg rover moving over GRC-3 lunar soil simulant at several terrain slopes $\theta$, between 0 to $30^\circ$. The angular velocity at the wheel was $\omega$=0.8~\si{rad/s}; the GRC-3 density and internal friction angle were 1734 $\si{kg/m^3}$ and $42.0^{\circ}$, respectively. To compare the images, all terrains were rotated back with the terrain to be shown as horizontal. As expected, the higher the terrain slope $\theta$, the shorter the distance the rover can move up the incline in a given amount of time, and the higher the wheel soil sinkage. 

\FloatBarrier
\section{Discussion}\label{sec:discussion}
This manuscript summarizes lessons learned in a simulation campaign undertaken during the design phase of VIPER. Assessing the trafficability worthiness of the rover design was anchored by the  methodology in use at NASA and currently embraced by other space agencies. Given that (a) at the onset of this study VIPER's mass was approximately 440 kg (during the design phase, its mass increased from 440 kg to 520 kg due to instrumentation decisions); and (b) the rover is Moon bound -- the Moon Gravitation Representative Unit 3 (MGRU3) rover was built at a mass roughly 1/6 of the nominal rover's mass. This gravity-offload decision explains why the 2021-2022 physical testing results reported in the previous section are for a \textit{rover} with masses of 73 kg and 88 kg. These MGRU3 masses were used in the experimental campaign at the SLOPE lab to collect data and assess the mobility traits of the roughly 500 kg VIPER. In retrospect, there was no need to keep the geometry of the rover yet reduce its mass. In doing so, a light rover is placed on granular material that is acted upon by the Earth's gravitational pull, which yields overoptimistic performance for the nominal rover when deployed on a celestial body of lower gravity. One might argue that the use of GRC-1 and GRC-3 counterbalance this, but this theory was proven incorrect \cite{dacaAdrianaPhD2022}.  

\begin{figure}[h]
	\centering
	\includegraphics[width=4in]{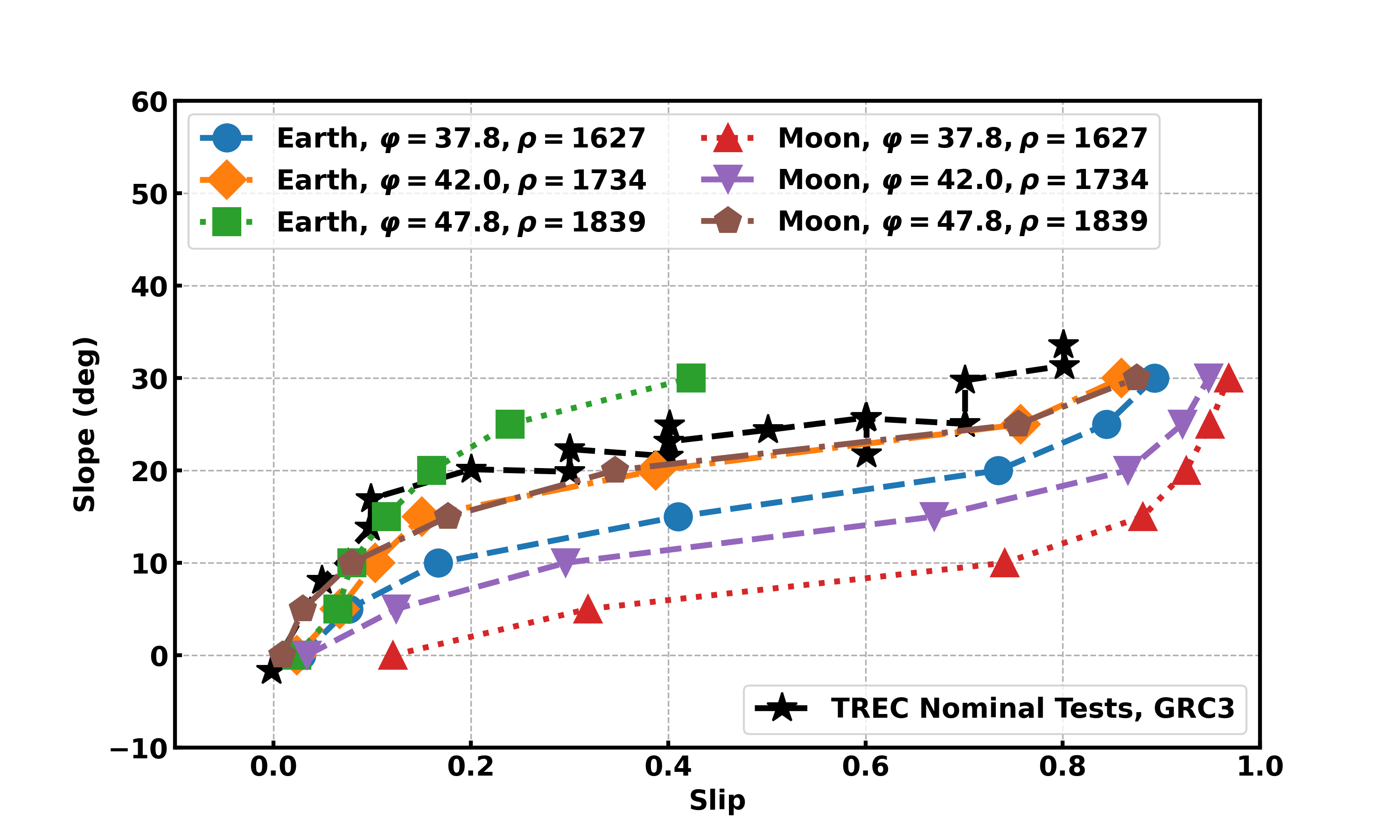}
	\caption{Slope-mode analysis: physical testing data vs. CRM simulation results for the MGRU3 rover on Earth, and VIPER rover on the Moon. The former rover has a mass of 73 kg, a reflection of the misconception that gravitational offset is necessary on Earth. The VIPER rover has a mass of 440 kg and is simulated in Moon gravity. The simulation results confirm that placing a light vehicle on terrain that has Earth-gravity induced higher strength leads to misleading results.} 
	\label{fig:gravitationalOffset}
\end{figure}

From a high vantage point, if the results in the previous section argued that ($i$) the physics-based simulator is predictive, and ($ii$) it produces results that obey the granular scaling laws, the information synthesized in Fig.~\ref{fig:gravitationalOffset} confirms that gravitational offload should not be used to test on Earth rovers that will operate in lower gravitational environments. The information captured in the figure is associated with slope-mode testing. The plot reports physical testing data vs. CRM simulation results for the MGRU3 rover on Earth, and VIPER rover on the Moon. The former rover has a mass of 73 kg -- a low mass value reflecting the common belief that gravitational offset is necessary on Earth. The VIPER rover has a mass of 440 kg and is simulated in Moon gravity. The simulation results confirm that placing a light vehicle on terrain that has Earth gravity-induced higher strength leads to overly optimistic results. Consider, for instance, the situation when the GRC-3 terrain has a friction angle $\phi=47.8$ and bulk density $\rho = 1839$. On Earth, the results are associated with the line with green squares; on the Moon, this would be the line with brown pentagons. The results indicate that MGRU3 climbs a $\theta = 30^{\circ}$ slope and it can do so at a slip value of approximately 42\%. However, if the VIPER rover was to climb on the Moon a $\theta = 30^{\circ}$ slope, it would experience significantly higher slip, approximately 85\%. This is an example of over-optimistic results produced by Earth-testing, slip values above 80\% place the rover in a situation that increases the propensity for dig-in. The same overly optimistic behavior is inferred for other friction angle \& bulk density values. For instance, consider the $\phi=37.8$ and $\rho = 1627$ GRC-3 case -- blue circles on Earth for MGRU3, red triangles on the Moon for VIPER. MGRU3 would climb a $\theta = 10^{\circ}$ slope at 18\% slip, while the VIPER on the Moon would experience approximately 75\% slip. Note that the data in Fig.~\ref{fig:gravitationalOffset} is obtained by running a suite of experiments like the ones reported in Fig.~\ref{fig:rover_vel_his_grc1}.

\FloatBarrier

The main insights drawn from this study are as follows. The rover that will undertake an extraterrestrial mission can be used on Earth for tests whose results can be extrapolated, via granular scaling laws, to predict steady-state macro-behavior of the rover in yet to be visited scenarios. Time and money can be saved if single wheel tests are used instead of full rover tests. CRM simulation can be used to predict specific scenarios of interest and micro-scale and/or transitory behavior. Even if one does not have a good idea about the parameters defining terrain models for other celestial bodies, being physics-based, CRM, and for that matter DEM, simulation can be used to perform parameter sweeps to reveal average behavior and worst case scenarios. Finally, one unexpected result was that the slope vs. slip curves obtained in slope-mode test are invariant under a change of angular velocity, see Fig.~\ref{fig:diff_angW}.

Considering the body of evidence obtained in parabolic flights and via the revamped scaling laws, we posit that a paradigm shift is justified when judging trafficability in different gravity fields. This contribution adds to this body of evidence, and it does so by employing a validated simulator that strays away from the BWJH model. The latter is empirical and cannot capture important factors that shape the performance of the rover, e.g., nontrivial grouser geometries, impact of the gravity, non-flat terrain, dynamic effects (soil ejection, wheel sinking process). We posit that the wide adoption of the BWJH model, which is semi-empirical, has prevented the community from understanding the fallacy of using gravitational offset since BWJH, in its common use, does not factor the gravitational acceleration in the terrain model. Peeking into the future provides even more impetus to move away from BWJH models. Indeed, NASA's lunar habitation plans are anchored by in-situ resource utilization that will require terramechanics studies of vehicles that dig into terrain, bulldoze it, etc. These operations call for physics that the BWJH class of models cannot capture.

%This latter aspect is key, since NASA's lunar plans that are anchored by in situ resource utilization will require terramechanics studies of vehicles that dig into terrain, bulldoze it, etc. 

\medskip

Ultimately, the results obtained in this effort make a strong case for relying heavily on physics-based terramechanics models when designing rovers and landers for extraterrestrial exploration. Terramechanics simulation is presently coming of age for two reasons. First, leveraging GPU computing, as pursued here, results in substantial gains in simulation speed. This choice opens the door for the use of CRM and DEM, two models previously dismissed as too slow to be relevant for large-scale terramechanics studies. The strength of both CRM and DEM is that they are physics-based. Therefore, (a) by comparison with the BWJH class of models, the parameters used to set up the CRM or DEM digital-twin terrain are intuitive and more easily accessible; and (b) the spectrum of applications in which CRM and DEM can come into play is richer than that of semi-empirical methods of BWJH class, which was set up to address mobility only and thus lacks the context necessary to handle other physics, e.g., digging, bulldozing, or change in gravity. Second, and more importantly, using physics-based simulation provides insights that are otherwise difficult or impossible to obtain. How would these insights be otherwise obtained? As pointed out, using helium balloons or gantry-type systems for gravitational offset leads to overly optimistic results regarding trafficability outcomes. Then, the options left are parabolic flights and scaling laws. The former are challenging to set up for two reasons: the duration of an experiment is necessarily short; and collecting relevant information is challenging, e.g., it is difficult in a short parabolic flight to gauge how the material shears under a wheel rover since imaging these phenomena, while not impossible, is costly and cumbersome. Inexpensive solutions that employ transparent walls sometimes impact the very dynamics of the phenomena that are of interest and are by necessity providing only a 2D snapshot of the relevant physics. As for the granular scaling laws, while elegant and insightful, they are limited to macro-scale information tied to \textit{steady-state} regimes. Moreover, the insights provided pertain to the macro-scale performance of the vehicle and not that of the terrain. How the terrain is disturbed and how its dynamics is coupled with that of the implement is averaged out of the conversation. In other words, when using GSLs, one cannot tell much about how the terrain will be disturbed, which is an issue, e.g., for the rear wheels moving over the ruts of the forward wheel, or when the ego rover or a companion one revisits the perturbed terrain. 

\FloatBarrier
\section{Materials and methods}\label{sec:mat_meth}
To resolve the dynamics of the deformable terrain in its two-way coupling with the rover's wheels movement, we employed a homogenization of the granular-like lunar soil and used an elasto-plastic CRM approach~\cite{dunatunga2017continuum}. The CRM solution is obtained using the SPH method, which belongs to the class of meshless, Lagrangian particle-based approaches~\cite{Lucy1977,Monaghan1977}. The state information is advected with the SPH particles, and the dynamics equations are enforced at the location of these particles. Each particle moves based on its interactions with neighbor particles and moving boundaries (e.g., the rover wheel), and the presence of external forces (e.g., gravity). The SPH method has proved effective in the modeling and simulating of granular material problems with large deformation \cite{bui2008lagrangian,pengSPH-geomechanics2015,nguyenSPHgranFlows2017,hurley2017continuum}. The two-way coupling between terrain and implements is discussed in \cite{weiGranularSPH2021,weiTracCtrl2022}; the approach therein captures large deformation of the granular material terrain and large overall 3D motion of the solid bodies. The interaction between implements and terrain is posed and solved as a fluid-solid interaction (FSI) problem using so-called boundary-conditions enforcing particles rigidly attached to the boundary of the solid bodies. To connect the dynamics of the granular material with the update in the stress field, we employ the constitutive law proposed in \cite{KamrinFluidMechanics2015}. More details about our SPH-based terramechanics solution can be found in the supplementary materials. Note that in CRM, one can replace the SPH-based spatial discretization of the equations of motion with an alternative one anchored by the material point method, likely yielding equally good results \cite{KamrinFluidMechanics2015,haeriMPM-terramechanics2022}.

The physical results reported were obtained in NASA's SLOPE lab with tests performed under Earth gravity. In lieu of lunar regolith, the soil simulants used were GRC-3 and GRC-1. First, single wheel and full rover physical testing was carried out to obtain the slope/slip and power/slip maps reported herein. Subsequently, digital twins were built and the simulations were carried out in Chrono; for the rover test, a full multi-body system was set up to match MGRU3. For the wheel test rig or full rover simulations, the soil parameters used were those of the actual GRC-3 and GRC-1 simulant. The simulator uses a co-simulation framework in which the wheel/rover dynamics was solved in a multicore CPU chip while the terrain dynamics was solved at the same time using an NVIDIA GPU. A small amount of data was CPU-GPU exchanged at each numerical integration time step to enforce the coupling between wheel and soil. Changing from Earth conditions to Moon conditions was as simple as changing one line of code, from Earth's gravitational acceleration to that of the Moon. Due to the lack of lunar physical test data, we could not validate the accuracy of the Moon simulation results directly. However, the Moon gravity simulation results matched well the results obtained using Earth gravity if one analyzes the data through the lens of the scaling law theory. 

In relation to the materials and methods used, one caveat is that the results reported were obtained in conjunction with the regolith simulants GRC-1 and GRC-3. As pointed out, there is an ongoing debate in the community about the suitability of using these simulants. Providing a comprehensive answer to this question falls outside the scope of this contribution. However, there are two relevant and salient points relevant in this context. First, the actual results reported, e.g., the slope/slip curves, might not be identical to the results that will be noted on the Moon. This is because the terrains (the one used in simulation and the real one on the Moon) are likely different. The second salient point is that one can nonetheless rely on a \textit{physics-based} simulator to conduct a battery of simulations sweeping over ranges of likely terrain properties (bulk densities, friction angles, etc.) to obtain a comprehensive image of the possible performance of the rover. In time, once the geomechanics attributes of the lunar soil become available, the physics-based simulator will produce results of lesser uncertainty.

% ==========================================================================
\section*{Supplementary materials} 
Additional simulation results and an account of the CRM method used are available in the supplementary information. The open-source code to reproduce the results reported in this study is available at \url{https://github.com/sjtumsd/crm_sim_nasa_exp_scripts}. The simulations can be run under Linux or Windows by using the scripts provided. The Chrono simulator is publicly available on GitHub for unfettered use and distribution owing to its BSD3 license \cite{projectChronoGithub}.

\section*{Acknowledgments} 
Support for part of this work was provided by NASA STTR \#80NSSC20C0252, National Science Foundation grant OAC2209791 and US Army Research Office, under grants W911NF1910431 and W911NF1810476. 

%\clearpage
\bibliographystyle{elsarticle-num-names}
\biboptions{numbers}
\bibliography{main-dn}

\newpage
\setcounter{page}{1}
\section*{Supplementary materials}\label{sec:supp_mat}
\subsection*{Overview of the CRM method}\label{subsec:crm_methods}
For the CRM approach used in this work, we employ a homogenization of the granular material and use an elasto-plastic continuum model to capture the dynamics of the deformable lunar soil terrain \cite{KamrinFluidMechanics2015}. Herein, the CRM solution is obtained using the SPH method, which is a Lagrangian particle-based solution that requires no background grid \cite{Lucy1977,Monaghan1977}. The state information is advected with the SPH particles, and the dynamics equations are enforced at the location of the SPH particles. The particles move based on the interactions among neighbor particles and the external forces, e.g. gravity. The SPH method has proven effective and efficient in simulating granular material problems with large deformation \cite{nguyenSPHgranFlows2017,hurley2017continuum,weiGranularSPH2021,weiTracCtrl2022}.

In CRM, the problem unknowns, i.e., field velocity vector $\mathbf{u}$ and the Cauchy stress tensor $\boldsymbol{\sigma}$, enter the mass and momentum balance equations as: 
\begin{equation}\label{equ:NS}
	\begin{cases}
		\frac{\text{d} \mathbf{u}}{\text{d} t} = \frac{\nabla \boldsymbol{\sigma}}{\rho}  + \mathbf{f}_b  \vspace{5pt} \\
		\frac{\text{d}\rho}{\text{d} t}=-\rho \nabla \cdot \mathbf{u} 
	\end{cases}~ \; ,
\end{equation}
where $\rho$ is the density of the deformable terrain, and $\mathbf{f}_b$ represents external forces, e.g., the gravity force. The total stress tensor $\boldsymbol{\sigma} \in {\mathbb{R}}^{3\times 3}$ is split in two components expressed as $\boldsymbol{\sigma} \equiv -p{\bf I}+\boldsymbol{\tau}$, where $\boldsymbol{\tau}$ is the deviatoric component of the total stress tensor and $p$ is the pressure which can be calculated from the trace of the total stress tensor as $p=-\frac{1}{3}\mathrm{tr}(\boldsymbol{\sigma}) = -\frac{1}{3}(\sigma_{xx}+\sigma_{yy}+\sigma_{zz})$.  For closure, a stress rate tensor formula is employed. We use Hooke's law as well as the work described in \cite{KamrinFluidMechanics2015,Monaghan2000,Gray2001sph,yue2015continuum} to express the objective total stress rate tensor as: 
\begin{equation}\label{equ:stressRate}
	\frac{\text{d} \boldsymbol{\sigma}}{\text{d} t} = \dot{\boldsymbol{\phi }}\cdot{\boldsymbol{\sigma}}-{\boldsymbol{\sigma}}\cdot\dot{\boldsymbol{\phi }} + 2G[\dot{\boldsymbol{\varepsilon}}-\frac{1}{3}\mathrm{tr}(\dot{\boldsymbol{\varepsilon}}){\bf I}] + \frac{1}{3}K\mathrm{tr}(\dot{\boldsymbol{\varepsilon}}){\bf I} \; .
\end{equation}
 
In Eq.~(\ref{equ:stressRate}), when the material is not subject to plastic flow, the elastic strain rate tensor $\dot{\boldsymbol{\varepsilon}}$ of the granular material is defined as $\dot{\boldsymbol{\varepsilon}} =\frac{1}{2}[\nabla\textbf{u} + (\nabla\textbf{u})^\intercal]$; the rotation rate tensor is expressed as $\dot{\boldsymbol{\phi }} =\frac{1}{2}[\nabla\textbf{u} - (\nabla\textbf{u})^\intercal]$. Herein, $G$ and $K$ denote the shear modulus and bulk modulus of the granular material-like deformable terrain, respectively, and $\bf I$ is the identity matrix. It is noted that the expression of the elastic strain rate tensor given above only works in cases without a plastic flow. Once the granular material starts to flow, the elastic strain rate tensor is defined as: 
\begin{equation}\label{equ:strRate}
	\dot{\boldsymbol{\varepsilon}} =\frac{1}{2}[\nabla\textbf{u} + (\nabla\textbf{u})^\intercal] - \frac{\dot{\lambda}}{\sqrt{2}} \frac{\boldsymbol{\tau}}{\bar{\tau}} \; ,
\end{equation}
in which the second term on the right-hand side comes from the contribution of the plastic flow of the continuum representation of the granular material. Therein, $\dot{\lambda}$ and $\bar{\tau}$ are the plastic strain rate and equivalent shear stress, respectively \cite{KamrinFluidMechanics2015}.

We use the SPH method to spatially discretize the mass and momentum balance equations in Eq.~(\ref{equ:NS}) and the expression of total stress rate tensor in Eq.~(\ref{equ:stressRate}). In SPH, the simulation domain (including the deformable granular material terrain, solid bodies, and wall boundaries) is discretized using SPH and BCE particles. The former are used in conjunction with the deformable granular material terrain, with which they advect. The motion of the SPH particles is obtained by solving the governing equations, see Eqs.~(\ref{equ:NS}) and~(\ref{equ:stressRate}). Conversely, the motion of the BCE particles is tied to that of the solid bodies, to which they are rigidly attached. Their role is to couple the motion of the SPH particles to the motion of the solid bodies \cite{weiGranularSPH2021}.

According to the SPH method, the value of a function $f$ at the position of particle $i$ can be approximated as \cite{Monaghan2005a}: 
\begin{equation}\label{equ:sphFun}
	f_i = \sum\nolimits_j f_j W_{ij}\mathcal{V}_j \; ,
\end{equation}
where $W_{ij}$ is a kernel function, and $\mathcal{V}_i$ is the volume of particle $i$, defined as $\mathcal{V}_i = (\sum\nolimits_{j}W_{ij})^{-1}$. Thus, the mass associated with particle $i$ can be obtained as $m_i = \rho_i \mathcal{V}_i$. Herein, we use a cubic spline kernel function:
 
\begin{equation}\label{equ:sphKer}
	W_{ij} = W(\textbf{r}_{ij}) =  \alpha_d \cdot \:
	\begin{cases}
		\frac{2}{3} - R^2 + \frac{1}{2} R^3, & 0 \leq R < 1 \\
		\frac{1}{6}(2-R)^3, & 1 \leq R < 2 \\
		0, & R \geq 2  \\
	\end{cases} \; ,
\end{equation}
for which the relative position between particles $i$ and $j$ is defined as $\textbf{r}_{ij} = \textbf{x}_i - \textbf{x}_j$, with $\textbf{x}_i$ and $\textbf{x}_j$ being the positions of particle $i$ and $j$, respectively. For a three-dimensional problem, $\alpha_d = 3/(2\pi h^3)$. The scaled length parameter $R$ is defined as $R=r_{ij}/{h}$, where $r_{ij}$ is the length of the vector $\textbf{r}_{ij}$, and $h$ the characteristic smoothing length (one to two times the initial particle spacing $\Delta x$). In the light of Eq.~(\ref{equ:sphKer}), a field variable (e.g., velocity $\mathbf{u}$ or density $\rho$) at the position of particle $ i $ receives contributions from the values at all neighbor particles $j$ according to Eq.~(\ref{equ:sphFun}) as long as $j \in {\cal{N}}_{h,i} \equiv \left\{ \textbf{x}_j :~ r_{ij} < 2h \right\}$. 

For the gradient $\nabla f$ evaluated at the position of SPH particle $i$, both consistent and inconsistent discretizations are available \cite{fatehi2011}. While computationally slightly more expensive, the consistent SPH discretization 
 
\begin{equation}\label{equ:graOpe}
	\nabla f_{i}=\sum\nolimits_j (f_{j}-f_{i})\left(\textbf{G}_{i}\cdot \nabla_i W_{ij}\right)\mathcal{V}_{j} \; ,
\end{equation}
gives higher accuracy and is used herein. The gradient of the kernel function $W_{ij}$ with respect to the position of particle $i$ is expressed as: 
 
\begin{equation*}
	\nabla_i W_{ij} = \frac{\alpha_d}{h} \frac{\textbf{r}_{ij}}{r_{ij}}
	\begin{cases}
		-2R + \frac{3}{2}R^2,  & 0 \leq R < 1 \\
		-\frac{1}{2}(2-R)^2, & 1 \leq R < 2 \\
		~~~0,  & R \geq 2 \\
	\end{cases} .
\end{equation*}
 
In Eq.~(\ref{equ:graOpe}), $\textbf{G}_i \equiv -\left[ \sum\nolimits_j \textbf{r}_{ij} \otimes\nabla_{i}W_{ij}\mathcal{V}_{j}\right]^{-1} \in {\mathbb{R}}^{3\times 3}$ is a symmetric correction matrix associated with particle $i$. With $\textbf{G}_i$ being involved in the discretization of the gradient operator, an exact gradient for a linear function $ f $ can be guaranteed regardless of the ratio of $h/\Delta x$~\cite{fatehi2011}, where $\Delta x$ is the initial SPH discretization spacing. This higher accuracy allows one to use a relatively smaller $h$, thus saving computational cost, see, for instance, \cite{huConsistent2019}.

Hence, the consistent discretizations of the momentum balance and continuity equations are obtained by substituting Eq. \eqref{equ:NS} into Eq. \eqref{equ:graOpe}, which yields 
 
\begin{equation}\label{equ:momentum_dis}
	\frac{{\text{d}}\mathbf{u}_i}{{\text{d}}t}= \frac{1}{\rho_i} \sum\limits_j (\boldsymbol{\sigma} _{j} - \boldsymbol{\sigma} _{i}) \cdot {\textbf{b}}_{ij} + \mathbf{f}_{b,i} \; , 
\end{equation}
 
\begin{equation}\label{equ:continuity_dis}
	\frac{{\text{d}}\rho_i}{{\text{d}}t}= -\rho_i  \sum\limits_j (\textbf{u} _{j}- \textbf{u}_{i}) \cdot {\textbf{b}}_{ij} \; .
\end{equation}
Similarly, the consistent discretization of the rotation rate and strain rate tensors assume the expression
\begin{equation}\label{equ:rotation_rate_dis}
	\dot{\boldsymbol{\phi }}_i =
	\frac{1}{2} \sum\limits_j \left [
	\textbf{u} _{ji}  {\textbf{b}}_{ij}^\intercal - \left ( \textbf{u}_{ji}  {\textbf{b}}_{ij}^\intercal \right )^\intercal \right ]
\end{equation}
 
\begin{equation}\label{equ:strain_rate_dis}
	\dot{\boldsymbol{\varepsilon}}_i =
	\frac{1}{2} \sum\limits_j \left [
	\textbf{u} _{ji}  {\textbf{b}}_{ij}^\intercal + \left ( \textbf{u}_{ji}  {\textbf{b}}_{ij}^\intercal \right )^\intercal \right ],
\end{equation}
where $ {\textbf{b}}_{ij} \equiv  \textbf{G}_{i} \cdot \nabla_i W_{ij} \mathcal{V}_{j}$. Finally, the consistent discretization of the total stress rate tensor is obtained by substituting Eqs. \eqref{equ:rotation_rate_dis} and \eqref{equ:strain_rate_dis} into Eq. \eqref{equ:stressRate}, which yields
 
\begin{align}\nonumber\label{equ:stress_rate_dis}
	\frac{{\text{d}}\boldsymbol{\sigma}_i}{{\text{d}}t}  
	&=\frac{1}{2}\left\{
	\sum\limits_j \left [
	\textbf{u} _{ji}  {\textbf{b}}_{ij}^\intercal - \left ( \textbf{u}_{ji}  {\textbf{b}}_{ij}^\intercal \right )^\intercal\right ]\boldsymbol{\sigma}_i
	-\boldsymbol{\sigma}_i  \sum\limits_j \left [
	\textbf{u} _{ji}  {\textbf{b}}_{ij}^\intercal - \left ( \textbf{u}_{ji}  {\textbf{b}}_{ij}^\intercal \right )^\intercal\right ]
	\right\} \\
	&+G\left\{
	\sum\limits_j \left [ 
	\textbf{u} _{ji}  {\textbf{b}}_{ij}^\intercal + \left ( \textbf{u}_{ji}  {\textbf{b}}_{ij}^\intercal \right )^\intercal\right ]-\frac{1}{3}\textrm{tr}\left(\sum\limits_j \left [
	\textbf{u} _{ji}  {\textbf{b}}_{ij}^\intercal + \left ( \textbf{u}_{ji}  {\textbf{b}}_{ij}^\intercal \right )^\intercal\right ]\right)\textbf{I} 
	\right\} \tag{12}\\ 
	&+\frac{1}{6}K\left\{
	\textrm{tr}\left( \sum\limits_j \left [
	\textbf{u} _{ji}  {\textbf{b}}_{ij}^\intercal + \left ( \textbf{u}_{ji}  {\textbf{b}}_{ij}^\intercal \right )^\intercal\right ] \right)\textbf{I}
	\right\} \nonumber \; .
\end{align}

\setcounter{figure}{0}  
\makeatletter 
\renewcommand{\thefigure}{S\@arabic\c@figure}
\makeatother

\subsection*{Wheel soil interaction}\label{subsec:Bou_Cons}  
\begin{figure}[h]
	\centering
	\includegraphics[width=2.5in]{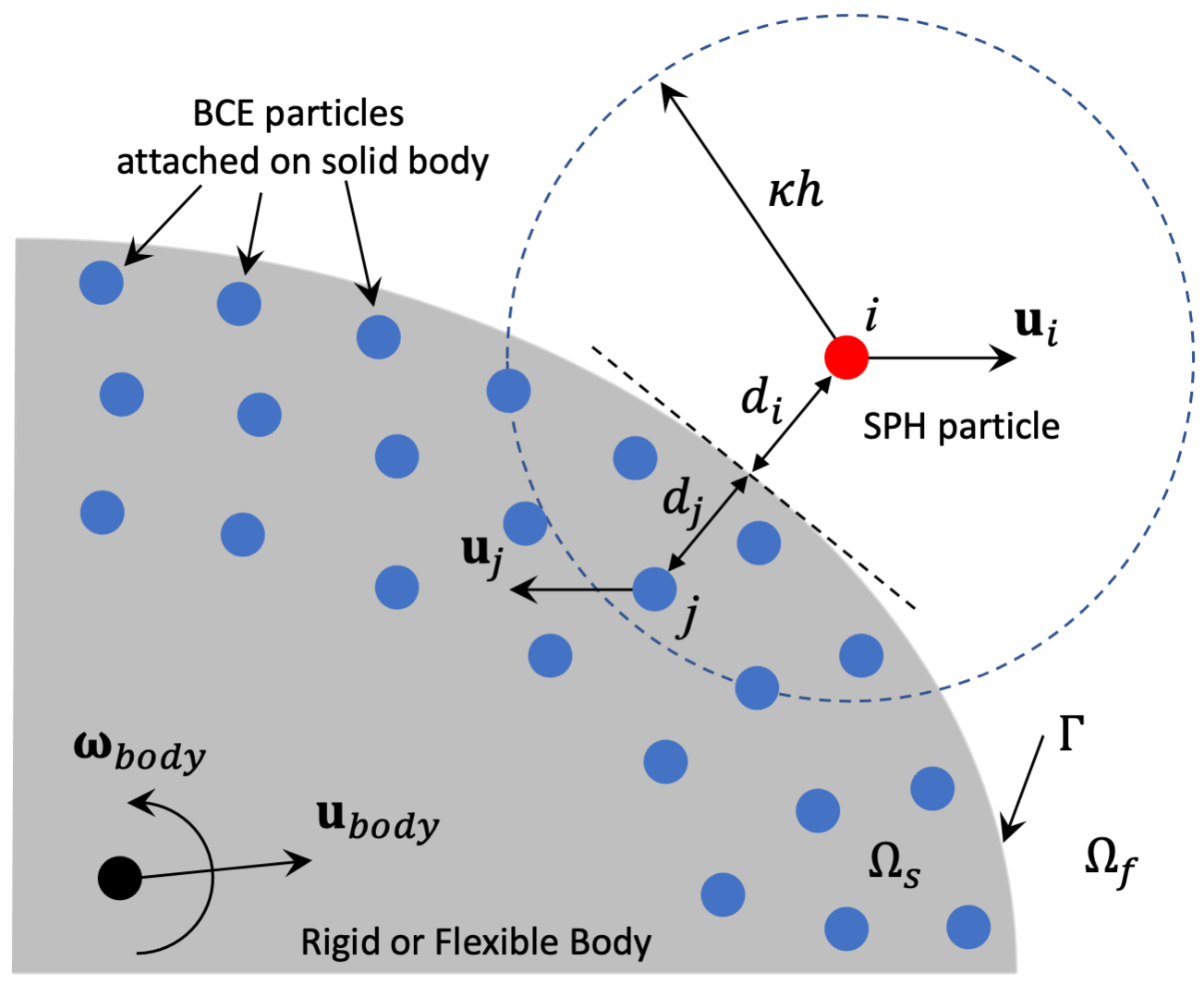}
	\caption{SPH particles and BCE particles close to the solid body or wall boundary. The Dirichlet boundary condition is imposed by extrapolating the velocities to the BCE particles.}
	\label{fig:BCE}
\end{figure}

In this work, a two-way coupling approach is modeled by imposing a Dirichlet (no-slip and no-penetration) boundary condition (BC) for the deformable granular material terrain at the solid boundary (moving wheels or fixed wall). To accurately impose a Dirichlet BC for the granular material's velocity, a full support domain contained in ($\Omega_f \cup \Omega_s$) should be attained to guarantee accurate SPH approximation for particles close to the boundary $\Gamma$. To this end, as shown in Fig.~S\ref{fig:BCE} in the supplementary materials, we follow the strategy proposed in \cite{TakedaSPHBC1994,Morris1997,Holmes2011Smooth,Pan2017Modeling,hu2017consistent} to generate several layers of BCE particles in the solid area $\Omega_s$ close to the boundary $\Gamma$. The velocities of the BCE particles can be linearly extrapolated from the SPH particles' velocities close to the boundary, i.e.,
\begin{equation} \label{equ:linear_extrapol}
	\textbf{u}_j = \frac{d_{j}}{d_{i}}(\textbf{u}_B - \textbf{u}_i) + \textbf{u}_{B} \; ,
\end{equation}
where $d_{i}$ and $d_{j}$ represent the perpendicular distances to the solid boundary $\Gamma$ for an SPH particle $i$ and a BCE particle $j$, respectively. Here, $\textbf{u}_{B}$ denotes the velocity at the solid boundary, which is expressed as
\begin{equation}\label{equ:noslipBC}
	\textbf{u}_B = \mathbf{u}_{body} + \boldsymbol{\omega}_{body} \times \mathbf{r}_c(\bf{x})  \;,
\end{equation}
where $\mathbf{u}_{body}$ and $\boldsymbol{\omega}_{body}$ are the translational and angular velocities of the solid body (e.g. the moving rover wheel), respectively; and $\mathbf{r}_c(\bf{x})$ denotes the vector from the center of mass (e.g. the wheel center) of the solid body to the location ${\bf x}$ at the boundary $\Gamma$. Note that the velocities of the BCE particles extrapolated from that of the SPH particles and the solid boundary are only used to enforce the Dirichlet BC; these velocities cannot be used to advect the BCE particles since they will move along with the solid body to which they are rigidly attached. 

For the total stress tensor $\boldsymbol{\sigma}_{j}$ at the position of a BCE particle $j$, we follow the approach in \cite{zhan2019three} to extrapolate it from the SPH particles' total stress tensor close to the boundary $\Gamma$, i.e.,
\begin{equation}\label{equ:stress_boundary}
	\boldsymbol{\sigma}_{j} =  \frac{\sum\limits_{i\in \Omega_f} \boldsymbol{\sigma} _{i} W_{ji} + [diag(\textbf{f}_b - \textbf{f}_j)] \sum\limits_{i\in \Omega_f}\rho_i [diag(\textbf{r}_{ji})] W_{ji}}
	{\sum\limits_{i\in \Omega_f} W_{ji}}  \; ,
\end{equation}
where $\mathbf{r}_{ji} = \mathbf{x}_j - \mathbf{x}_i$; the function $diag(\textbf{f}_b - \textbf{f}_j)$ creates a diagonal matrix from the vector $\textbf{f}_b - \textbf{f}_j$ and so does the function $diag(\textbf{r}_{ji})$; $\textbf{f}_b$ is the body force of the granular material (e.g., the gravity); and $\textbf{f}_j$ is the inertial force associated with the BCE particle $j$ and can be evaluated as:
\begin{equation}
	\textbf{f}_{j} =  \dot{\mathbf{u}}_{body}  + \dot{\boldsymbol{\omega}}_{body} \times \mathbf{r}_{jc} + \boldsymbol{\omega}_{body} \times (\boldsymbol{\omega}_{body} \times \mathbf{r}_{jc}) \; ,
\end{equation}
where $\mathbf{r}_{jc}$ is the vector from the solid body's center of mass to the position of the BCE particle $j$. The total force $\textbf{F}_{body}$ and torque $\mathbf{T}_{body}$ exerted by the deformable terrain upon the solid body is then calculated by summing the forces contributed by the SPH particles onto the BCE particles as described in the conservative SPH method \cite{bian2014splitting}, i.e., 

\begin{equation}
	\textbf{F}_{body} = \sum\limits_{j\in \Omega_s} m_j \:  {\dot {\mathbf{u}}}_j ~~~\text{and}~~~ \mathbf{T}_{body} = \sum\limits_{j\in \Omega_s} \mathbf{r}_{jc} \times (m_j \: {\dot {\mathbf{u}}}_j) \; .
\end{equation}

\subsection*{Update of field variables}\label{subsec:Time_Integ}  
In this work, the field variables (e.g. the velocity, position, and total stress tensor) of the SPH particles are updated using an explicit predictor-corrector time integration scheme with second-order accuracy \cite{Monaghan1989,Monaghan1994}. There are two half steps involved in this integration scheme for each time step. In the first half step, an intermediate value of velocity $\bar{\textbf{u}}_{i}$, position $\bar{\textbf{x}}_{i}$, and total stress tensor $\bar{\boldsymbol{\sigma}}_i$ are first predicted at $t+\frac{\Delta t}{2}$. Using predicted values, Eqs. \eqref{equ:momentum_dis}, \eqref{equ:stress_rate_dis} and \eqref{equ:XSPH} are evaluated again to update the velocity, position, and total stress tensor to the corrected values. Finally, the field variables of the SPH particles are updated based on the initial and corrected values at $t+\Delta t$. More details about the interaction scheme for granular material dynamics can be found in \cite{weiGranularSPH2021}. 

To enforce the condition that the particles advect at a velocity close to an average velocity of their neighboring particles, the so-called XSPH technique \cite{Monaghan1989} is used in this study. According to the XSPH method, the relationship between the displacement of the SPH particle $i$ and its velocity is expressed as: 
\begin{equation}\label{equ:XSPH}
	\frac{{\text{d}}\textbf{x}_i}{{\text{d}}t} = \textbf{u}_i-\xi\sum_j\textbf{u}_{ij}W_{ij}\mathcal{V}_{j} \; ,
\end{equation}
where the second term is a correction term with the coefficient $\xi$ in this work being set to 0.5. More details about the XSPH method used in granular material dynamics can be found in \cite{weiGranularSPH2021}. Therein, a comprehensive study about how to choose the value of the coefficient $\xi$ was performed by gauging the influence of $\xi$ onto the kinetic energy of the dynamic system. 

To accurately update the value of the total stress tensor $\boldsymbol{\sigma}$ of the deformable granular material terrain, we employ an approach originally proposed in MPM \cite{KamrinFluidMechanics2015} and apply it within the framework of SPH. The total stress tensor of each SPH particles is first updated explicitly from $t$ to $t+\Delta t$ according to the predictor-corrector scheme described in \cite{Monaghan1989,Monaghan1994,weiGranularSPH2021}. Once the update is done, at the end of this time step, the total stress tensor is then further corrected based on a four-step post-processing strategy expressed as: $(i)$ Calculate the value of $p^*$ and $\boldsymbol{\tau}^*$ according to the value of total stress tensor $\boldsymbol{\sigma}^*$ that is already obtained through the predictor-corrector scheme using Eq.~(\ref{equ:stressRate}); $(ii)$ If $p^*<0$, then simply set $\boldsymbol{\sigma} = \mathbf{0}$ at $t+\Delta t$ and start a new integration time step; $(iii)$ If $p^*>0$, set $ p=p^* $, compute the double inner product of $\boldsymbol{\tau}^*$ as $\bar{\tau}^* = \sqrt{\frac{1}{2}(\tau^{*}_{\alpha\beta}):(\tau^{*}_{\alpha\beta})}$, and compute $S_0$ as $S_0 = \mu_s p^*$, {here, $\alpha$ and $\beta$ are indices for the stress components}; $(iv)$ If $\bar{\tau}^*<S_0$, simply set $\boldsymbol{\tau}=\boldsymbol{\tau}^*$ as the deviatoric component of $\boldsymbol{\sigma}$ at $t+\Delta t$ since no plastic flow occurs at this moment; else, use the Drucker-Prager yield criterion to scale the deviatoric component of $\boldsymbol{\sigma}$ back to the yield surface as $\boldsymbol{\tau} = \frac{\mu p^*}{\bar{\tau}^*}\boldsymbol{\tau}^*$. Here, the friction coefficient used in step $(iii)$ is defined as $\mu = \mu_s+\frac{\mu_2-\mu_s}{I_0/I + 1}$ \cite{KamrinFluidMechanics2015}, where $\mu_s$ is the static friction coefficient, and $\mu_2$ is the limiting value of $\mu$ when $I \to \infty$; $I_0$ is a material constant which is set to 0.03 in this work; $I$ is the inertial number. More details about the four-step strategy and the parameters' calculation can be found in \cite{weiGranularSPH2021}.

\subsection*{Supplementary data and results}\label{subsec:supp_fig}

\begin{figure}[h]
	\centering
	\begin{subfigure}{0.49\textwidth}
		\centering
		\includegraphics[width=3.2in]{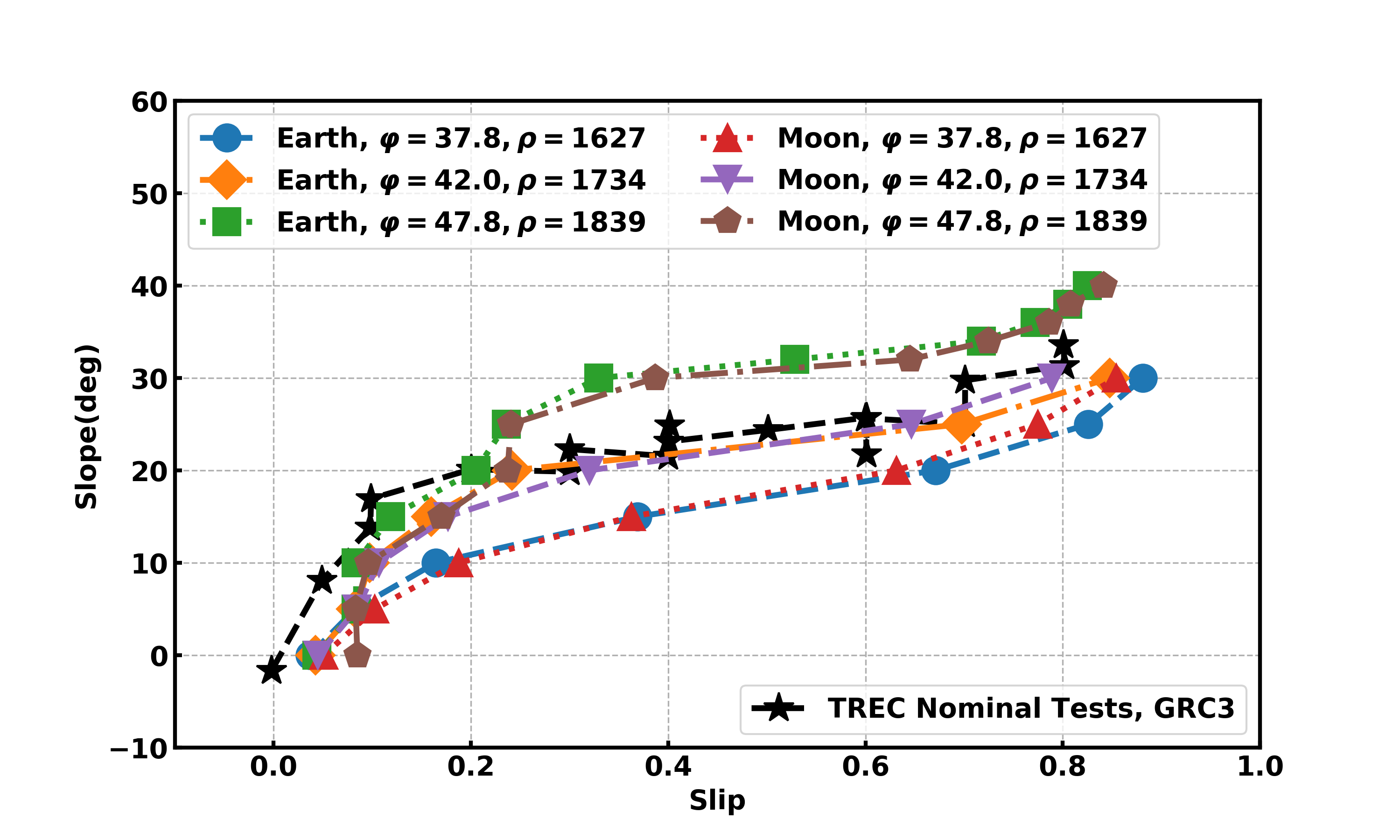}
		\caption{Single wheel on GRC-3}
	\end{subfigure}
	\begin{subfigure}{0.49\textwidth}
		\centering
		\includegraphics[width=3.2in]{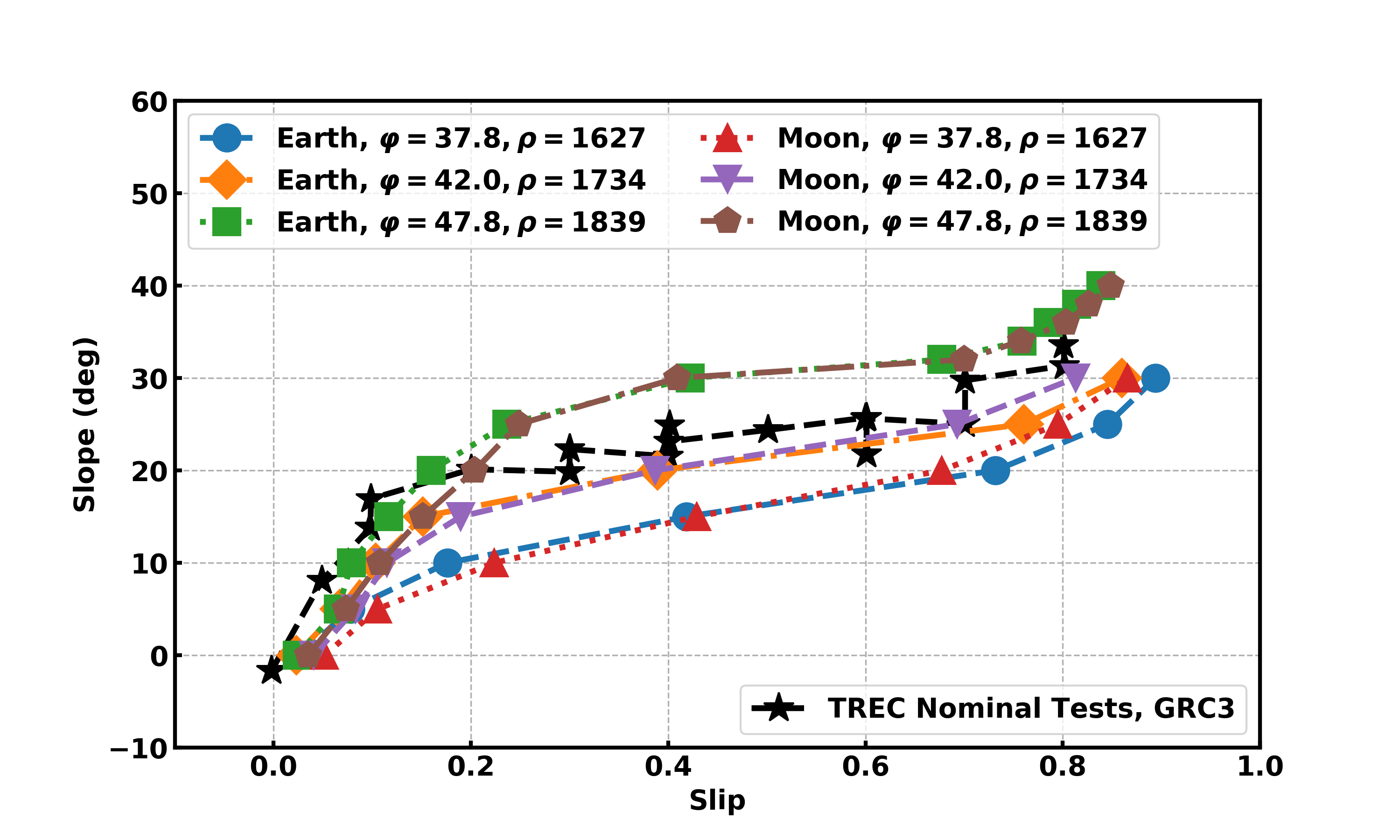}
		\caption{Full rover on GRC-3} 
	\end{subfigure}
	\caption{Single wheel and full rover simulation using CRM on GRC-3 lunar soil simulant. A 17.5 kg single wheel and 73 kg MGRU3 rover were used in the simulations on GRC-3. The wheel angular velocity were fixed to $0.8~\si{rad/s}$ in the simulations with Earth gravity, while fixed to $0.33~\si{rad/s}$ in the ones with Moon gravity according to the scaling law \cite{kamrin-gsl-expanded2020}. All the values used to generated the markers in the figures were obtained in steady state for each individual slip scenario. The single wheel results were obtained in the Traction and Excavation Capabilities (TREC) Rig at Glenn Research Center.} 
	\label{fig:suppMat-single_full_earh_vs_moon}
\end{figure}

\begin{figure}[h]
	\centering
	\begin{subfigure}{0.49\textwidth}
		\centering
		\includegraphics[width=3.2in]{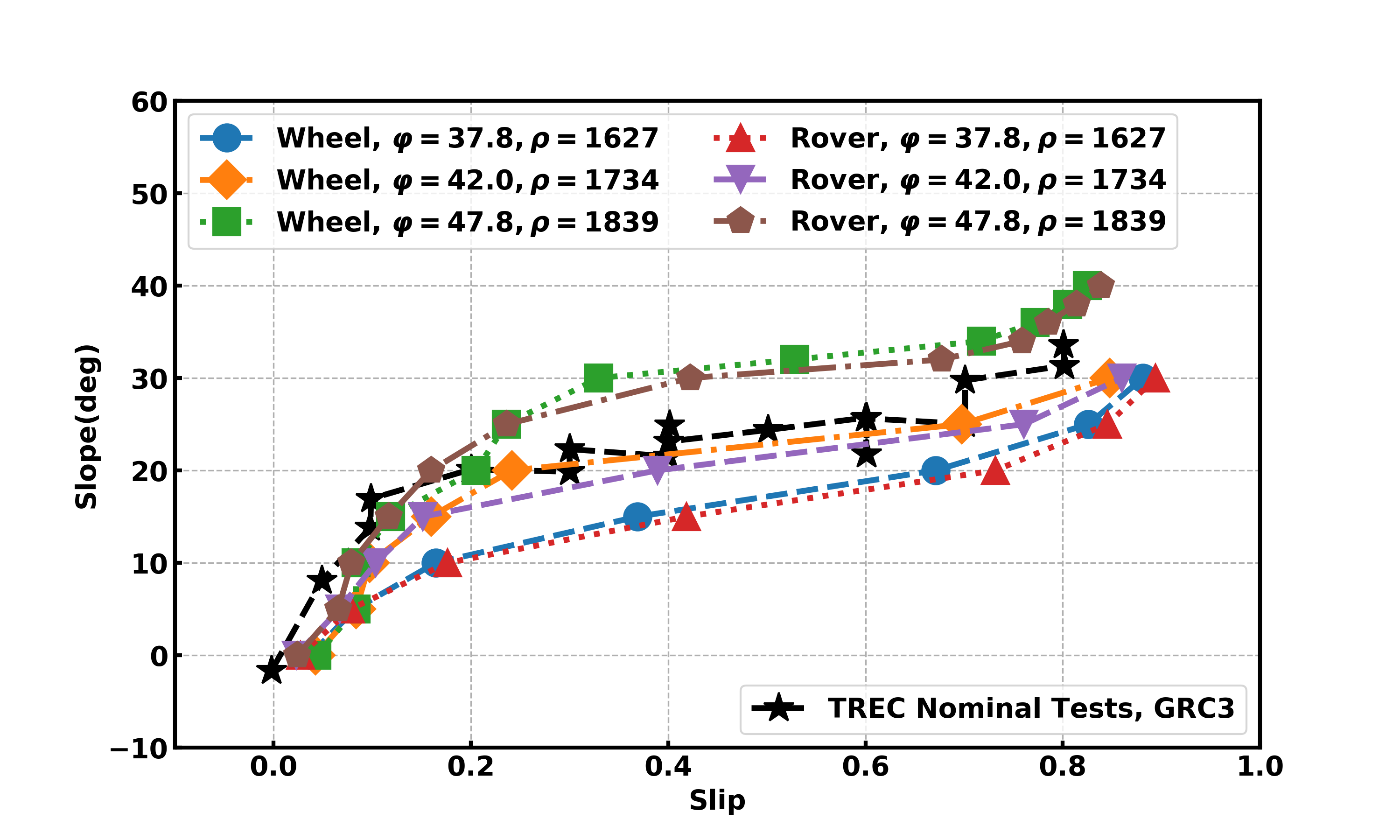}
		\caption{Earth gravity on GRC-3}
	\end{subfigure}
	\begin{subfigure}{0.49\textwidth}
		\centering
		\includegraphics[width=3.2in]{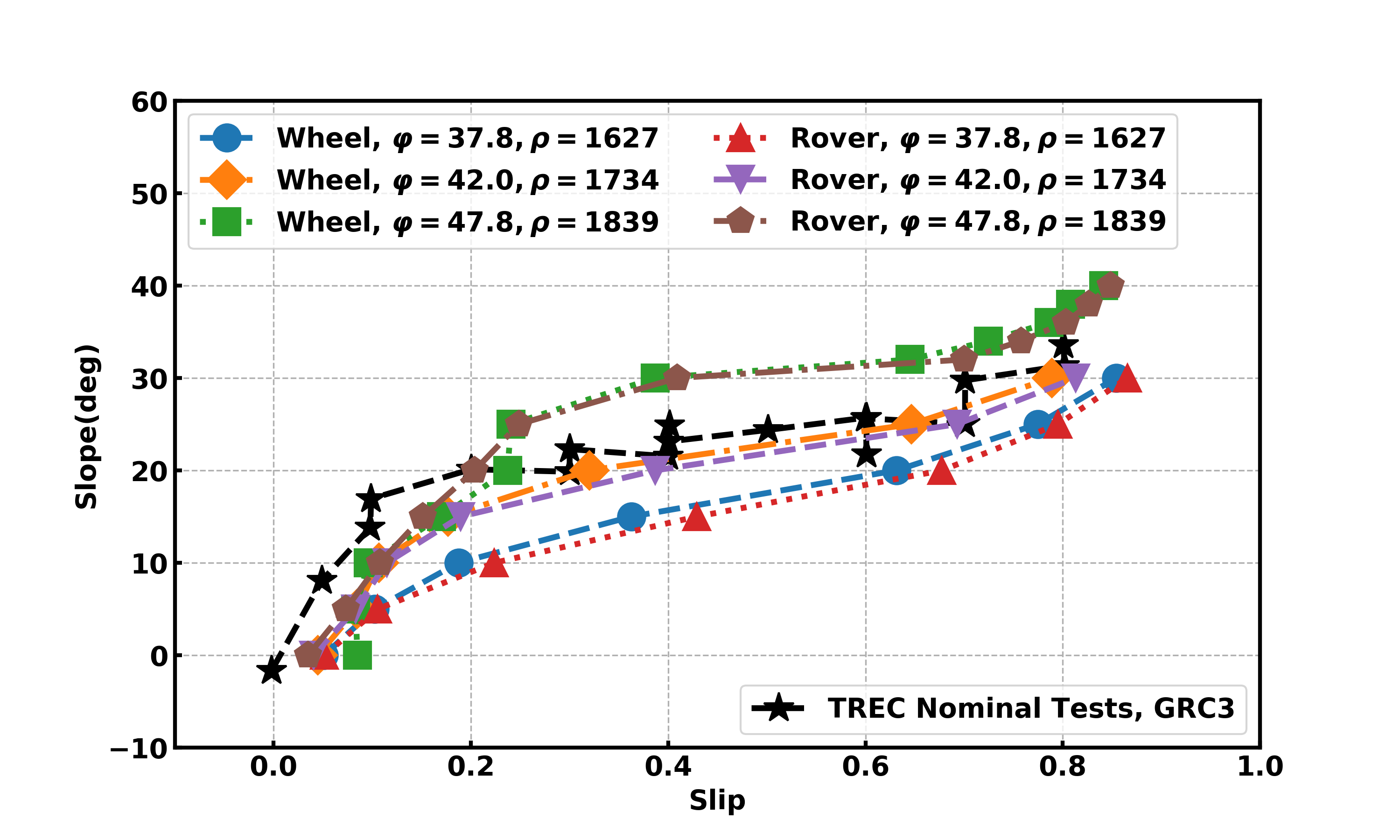}
		\caption{Moon gravity on  GRC-3} 
	\end{subfigure}
	\caption{A comparison between single wheel and MGRU3 simulations using GRC-lunar soil simulant with both Earth and Moon gravity.} 
	\label{fig:supMat-wheel_vs_rover_earth_moon}
\end{figure}

\begin{figure}[h]
	\centering
	\begin{subfigure}{0.49\textwidth}
		\centering
		\includegraphics[width=3.2in]{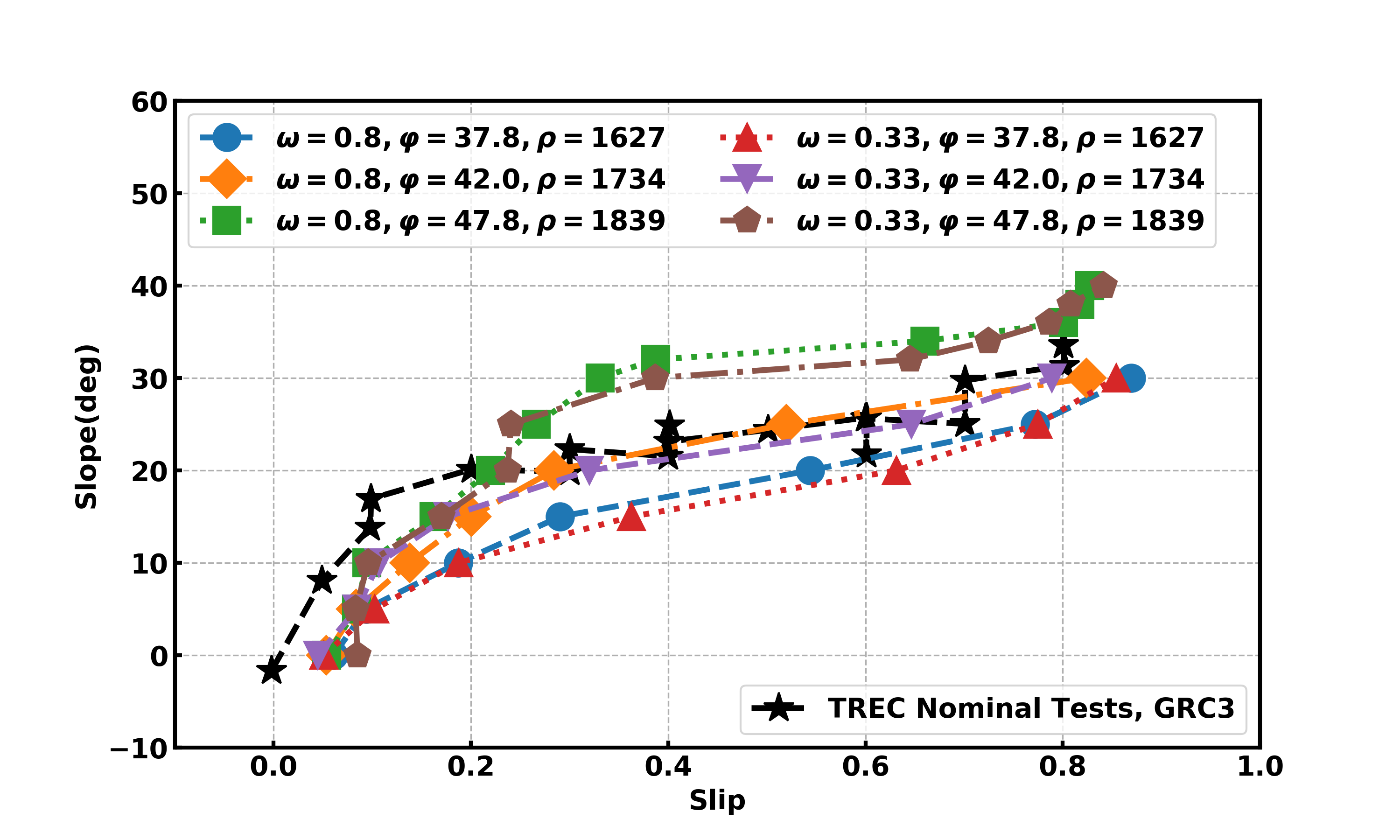}
		\caption{Single wheel on GRC-3}
	\end{subfigure}
	\begin{subfigure}{0.49\textwidth}
		\centering
		\includegraphics[width=3.2in]{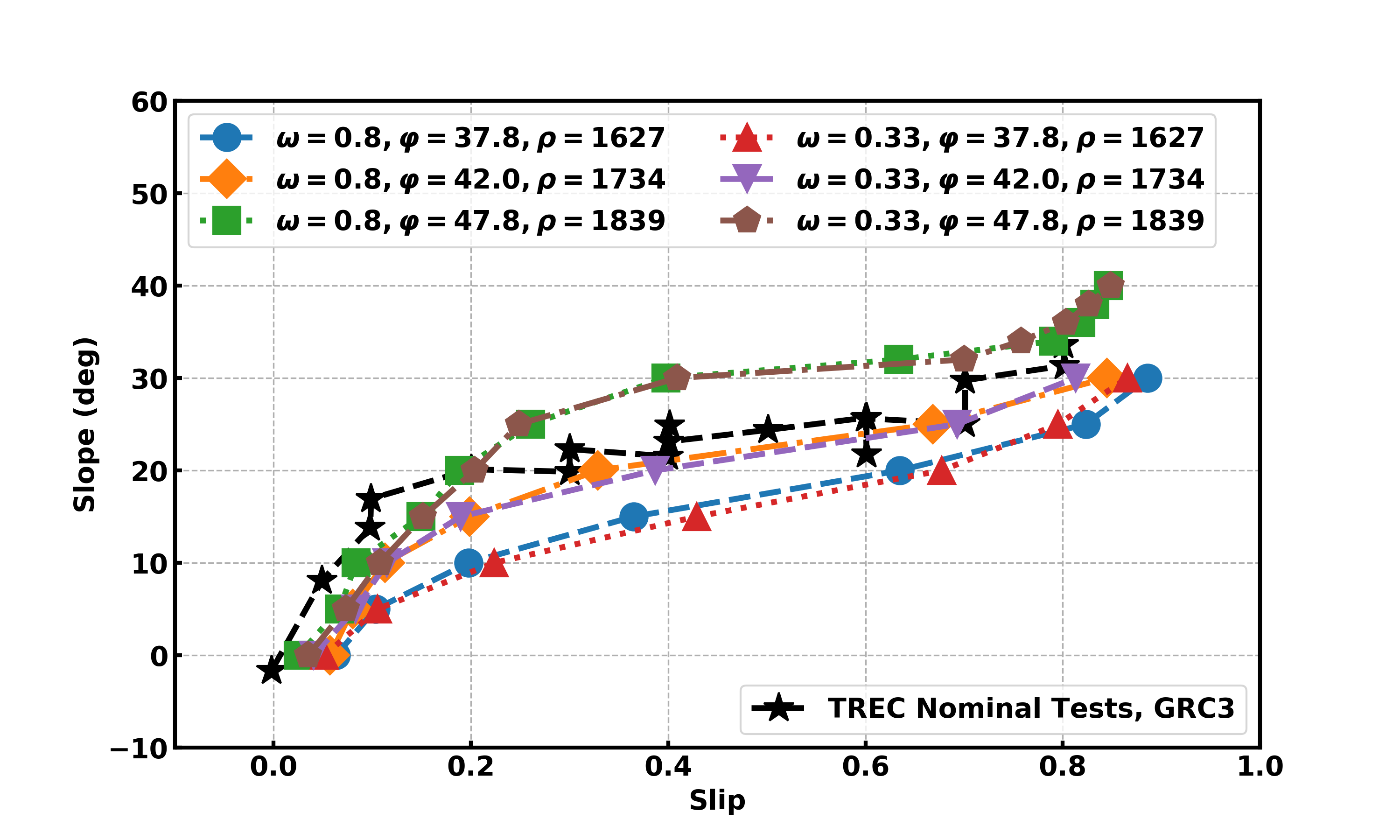}
		\caption{Full rover on GRC-3} 
	\end{subfigure}
	\caption{Single wheel and full VIPER rover simulation using on GRC-3 lunar soil simulant under Moon gravity. Two different angular velocity were used -- $0.33~\si{rad/s}$  and $0.8~\si{rad/s}$.} 
	\label{fig:supMat-diff_angW}
\end{figure}

\begin{figure}[h]
	\centering
	\begin{subfigure}{0.49\textwidth}
		\centering
		\includegraphics[width=3.2in]{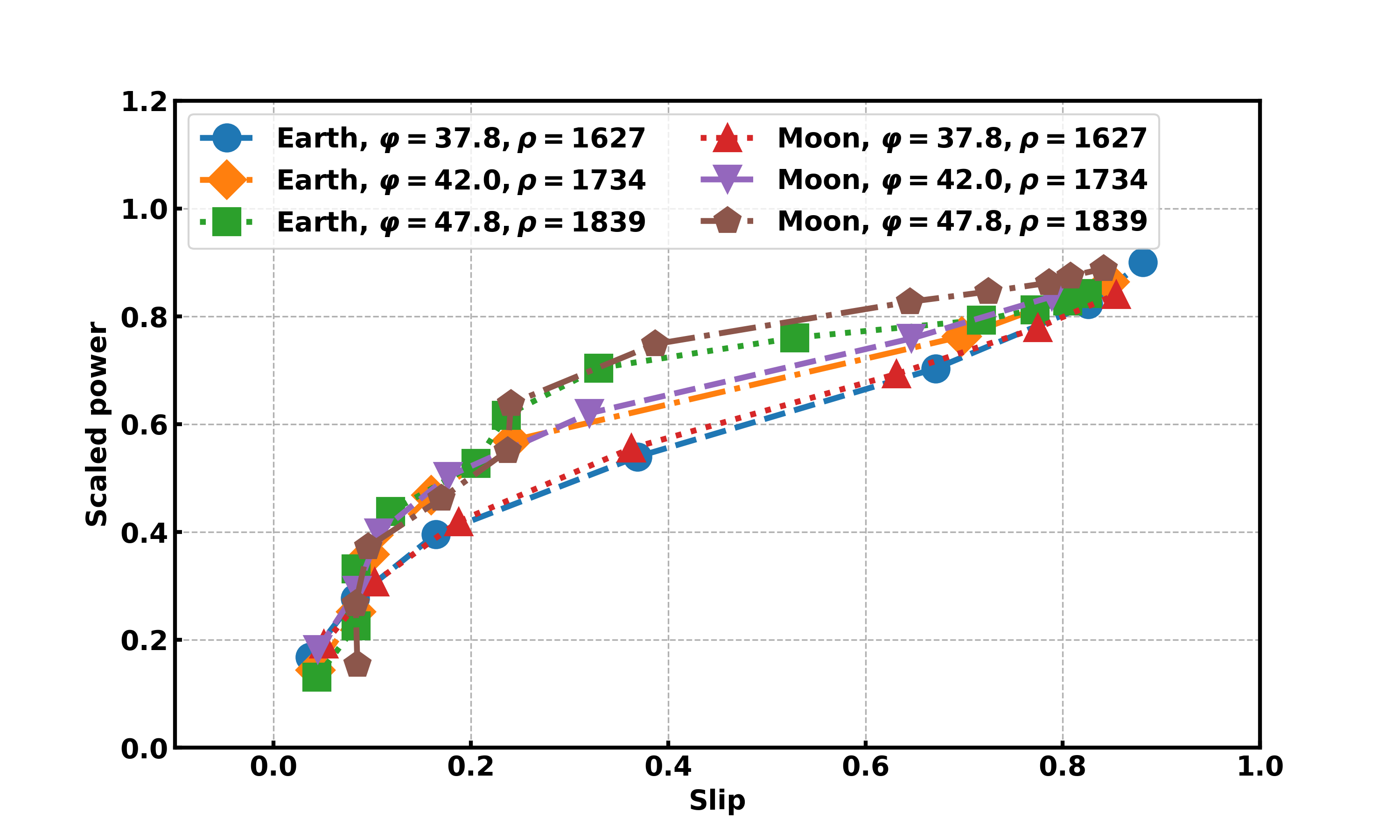}
		\caption{Single wheel: $\omega$=0.8~\si{rad/s} on Earth, $\omega$=0.33~\si{rad/s} on Moon}
	\end{subfigure}
	\begin{subfigure}{0.49\textwidth}
		\centering
		\includegraphics[width=3.2in]{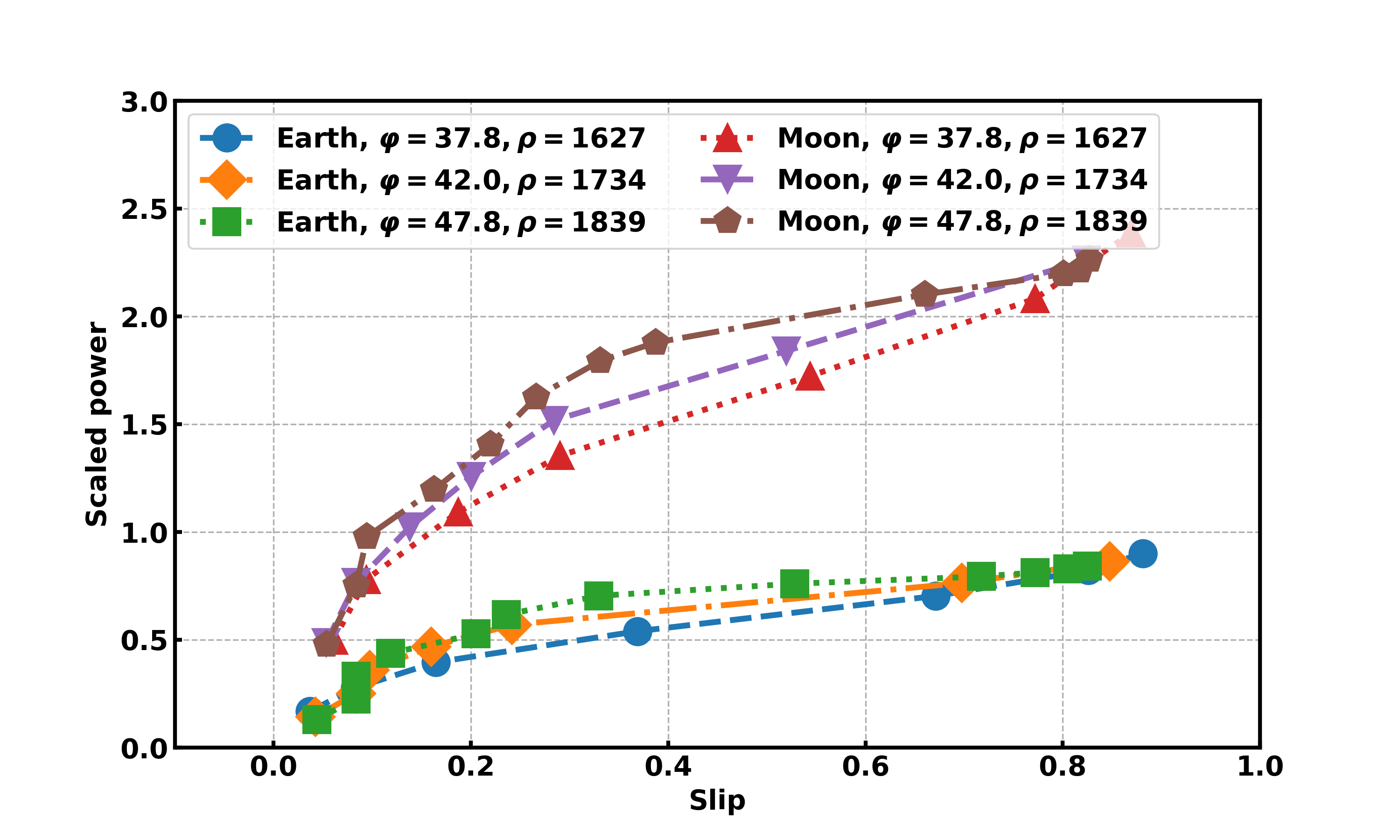}
		\caption{Single wheel: $\omega$=0.8~\si{rad/s} on Earth, $\omega$=0.8~\si{rad/s} on Moon} 
	\end{subfigure}
	\begin{subfigure}{0.49\textwidth}
		\centering
		\includegraphics[width=3.2in]{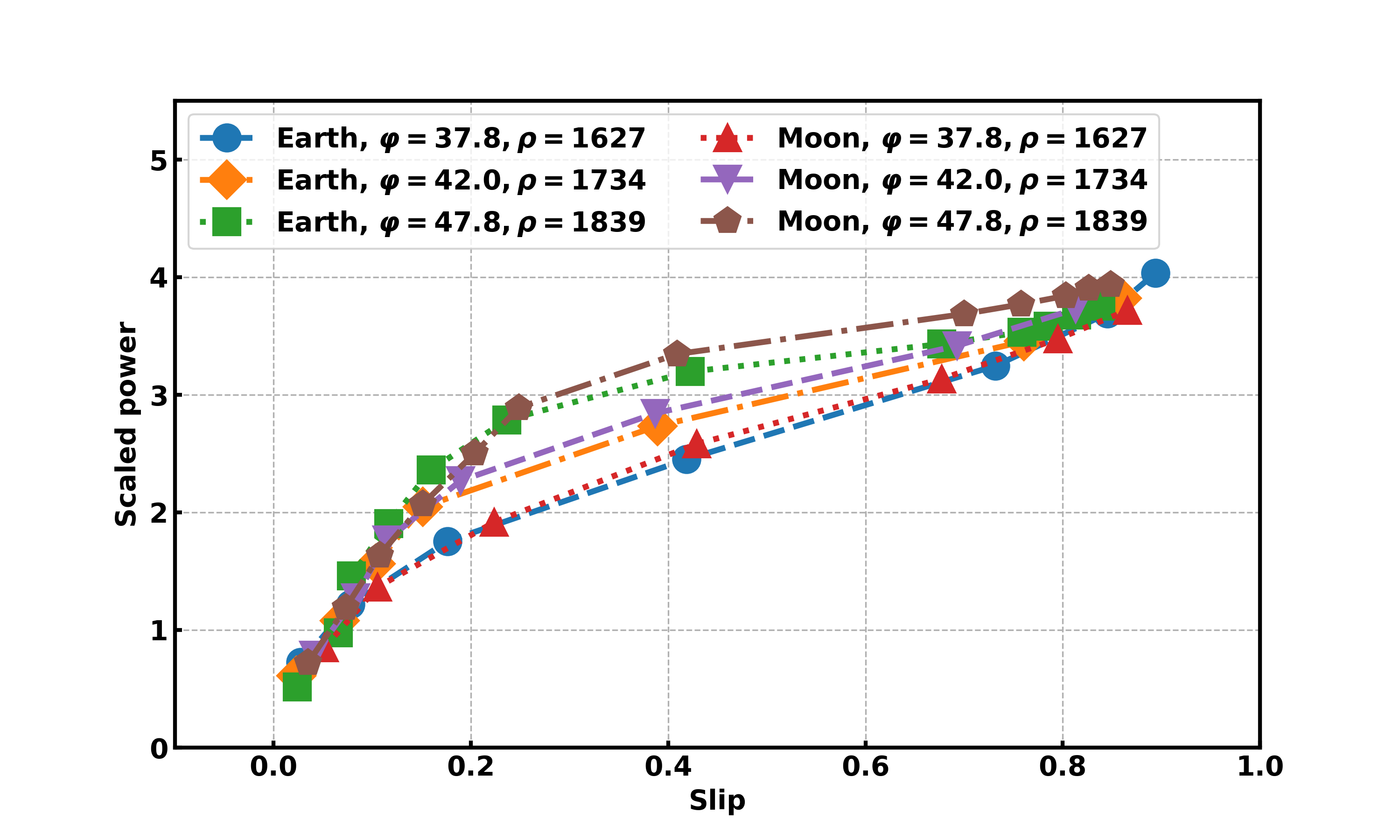}
		\caption{Full rover: $\omega$=0.8~\si{rad/s} on Earth, $\omega$=0.33~\si{rad/s} on Moon}
	\end{subfigure}
	\begin{subfigure}{0.49\textwidth}
		\centering
		\includegraphics[width=3.2in]{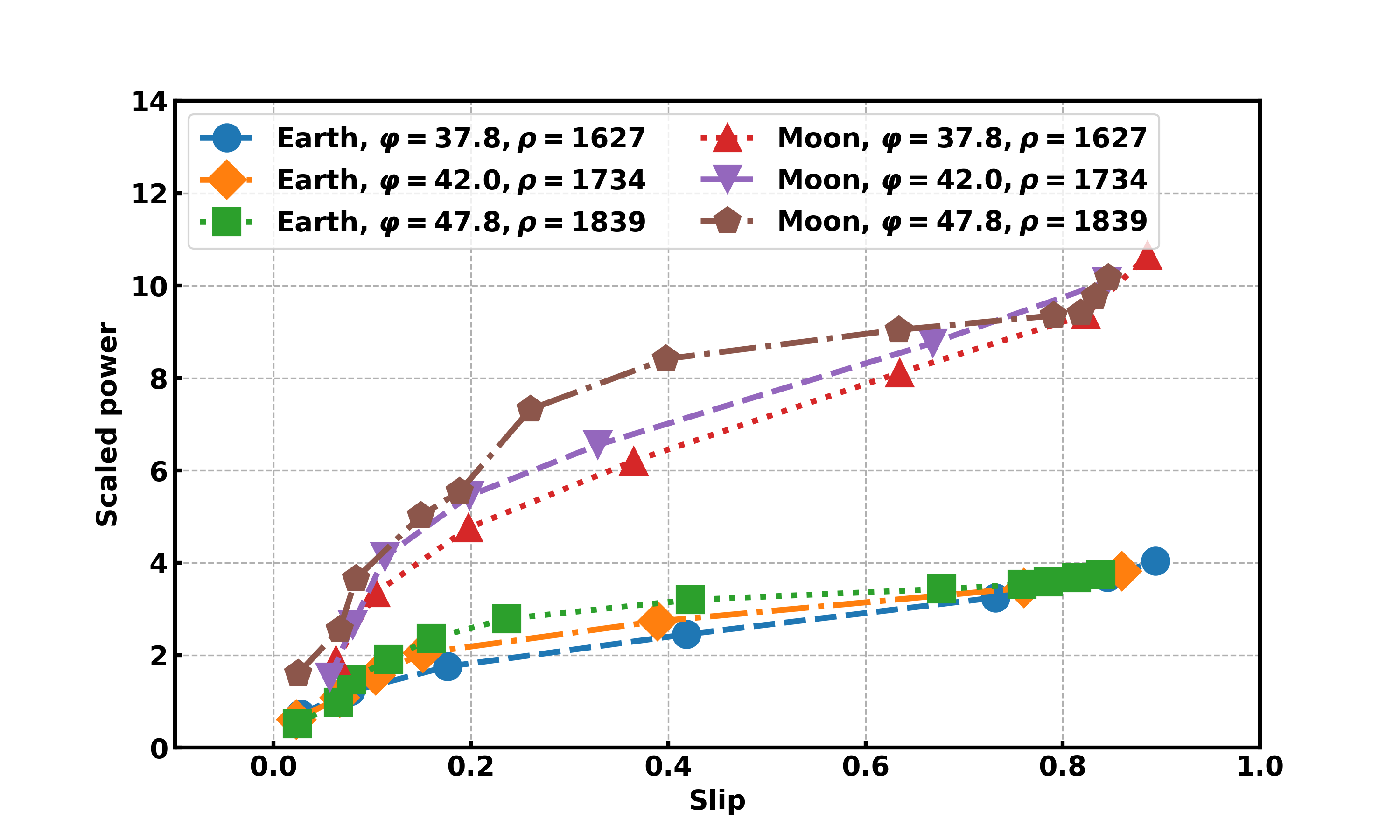}
		\caption{Full rover: $\omega$=0.8~\si{rad/s} on Earth, $\omega$=0.8~\si{rad/s} on Moon} 
	\end{subfigure}
	\caption{Scaled wheel power at steady state of single wheel/full rover simulation using CRM on GRC-3 lunar soil simulant.} 
	\label{fig:supMat-rover_power_grc3}
\end{figure}

%\begin{figure}[h]
%	\centering
%	\includegraphics[width=3.5in]{crm/vv_time/17_5kg_VV_Mode_GRC3_Omega_timeHis.png}
%	\caption{Single wheel simulation using on GRC-3 lunar soil simulant. Simulations were performed under constant angular velocity and translational velocity (VV) mode. This figure shows the time histories of the wheel's angle velocity associate with different wheel slip ratio.} 
%	\label{fig:vv_omega_time}
%\end{figure}

%========================== GRC-3
\begin{figure}[h]
	\centering
	\begin{subfigure}{0.49\textwidth}
			\centering
			\includegraphics[width=3.2in]{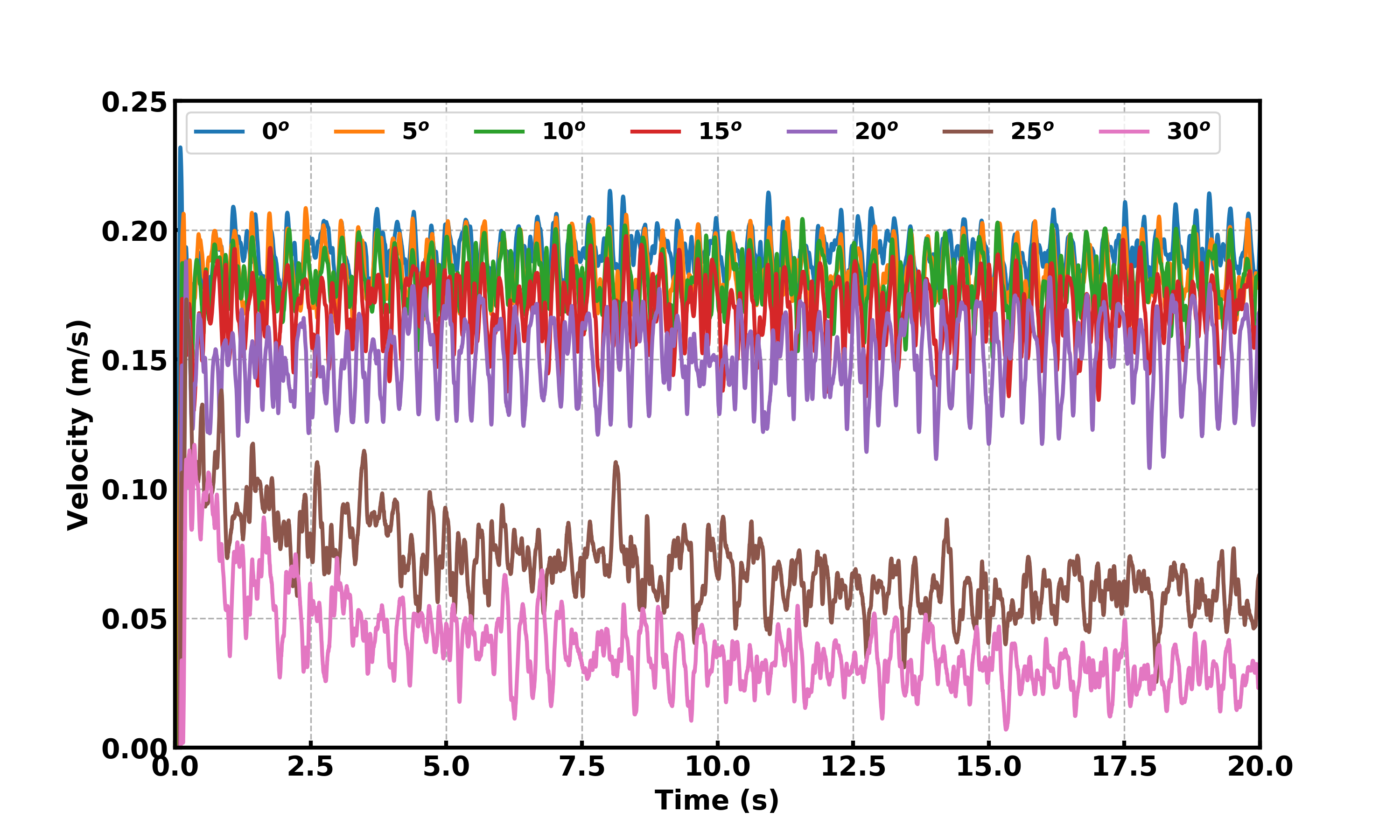}
			\caption{Earth gravity with $\omega=0.8~\si{rad/s}$}
		\end{subfigure}
	\begin{subfigure}{0.49\textwidth}
			\centering
			\includegraphics[width=3.2in]{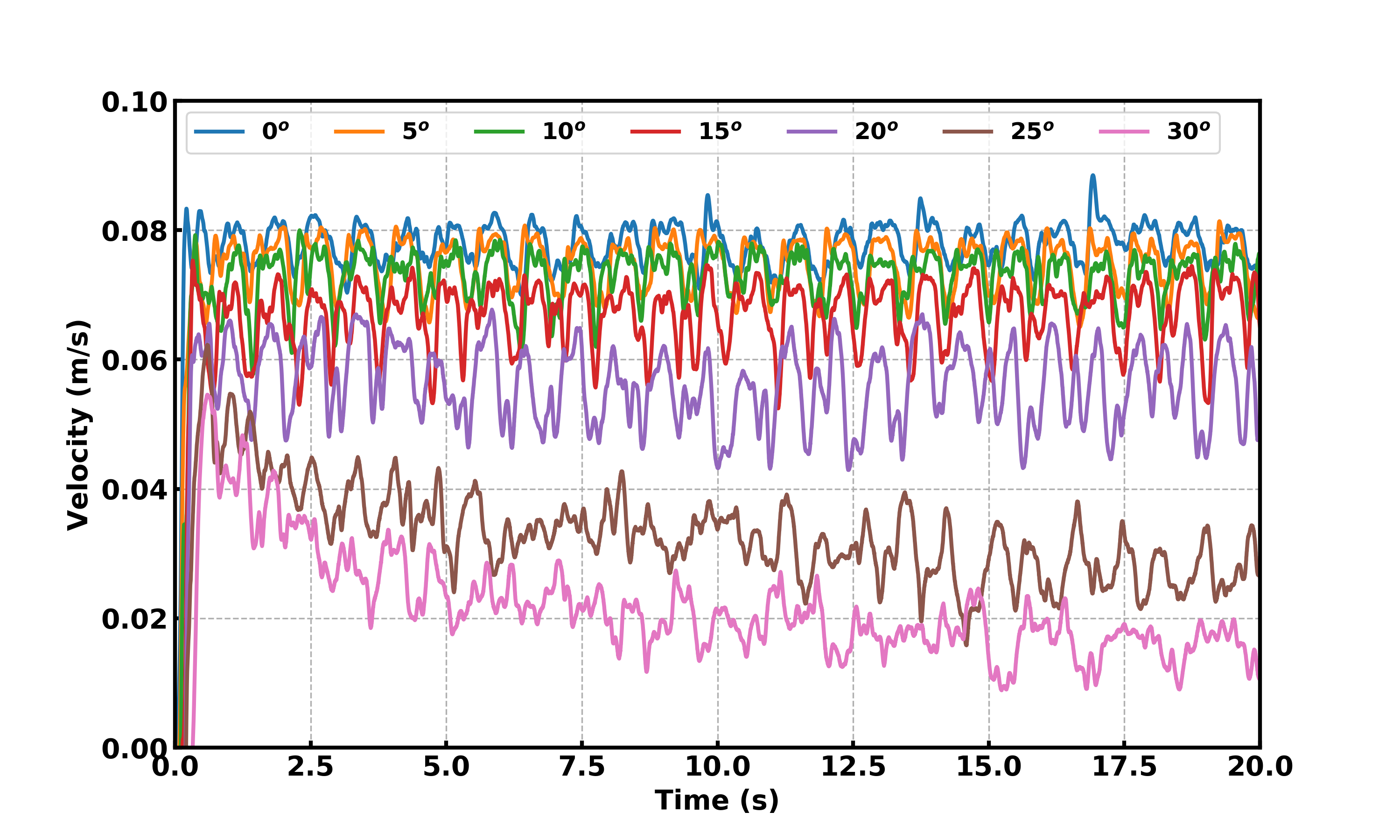}
			\caption{Moon gravity with $\omega=0.33~\si{rad/s}$} 
		\end{subfigure}
	\caption{Time history of translational velocity of single wheel simulation on GRC-3. Tests were done for $\theta$ between $0^{\circ}$ and $30^{\circ}$ in increments of $5^{\circ}$ with bulk density of 1734 $\si{kg/m^3}$ and friction angle $42.0^{\circ}$. The information in these two plots was used to generate the yellow and purple curves in Fig.~\ref{fig:suppMat-single_full_earh_vs_moon}a.} 
	\label{fig:wheel_vel_his_grc3}
\end{figure}

\begin{figure}[h]
	\centering
	\begin{subfigure}{0.49\textwidth}
			\centering
			\includegraphics[width=3.2in]{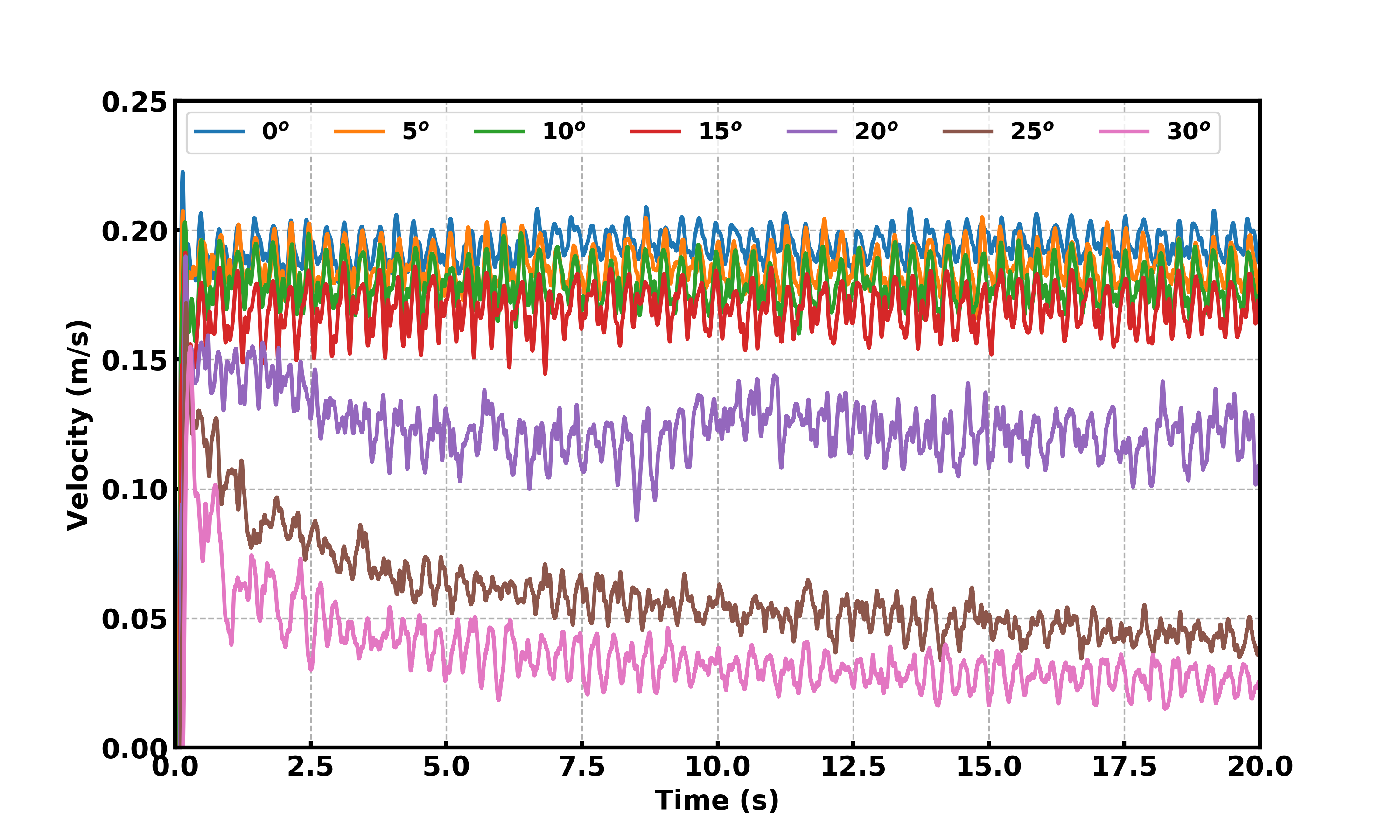}
			\caption{Earth gravity with $\omega=0.8~\si{rad/s}$}
		\end{subfigure}
	\begin{subfigure}{0.49\textwidth}
			\centering
			\includegraphics[width=3.2in]{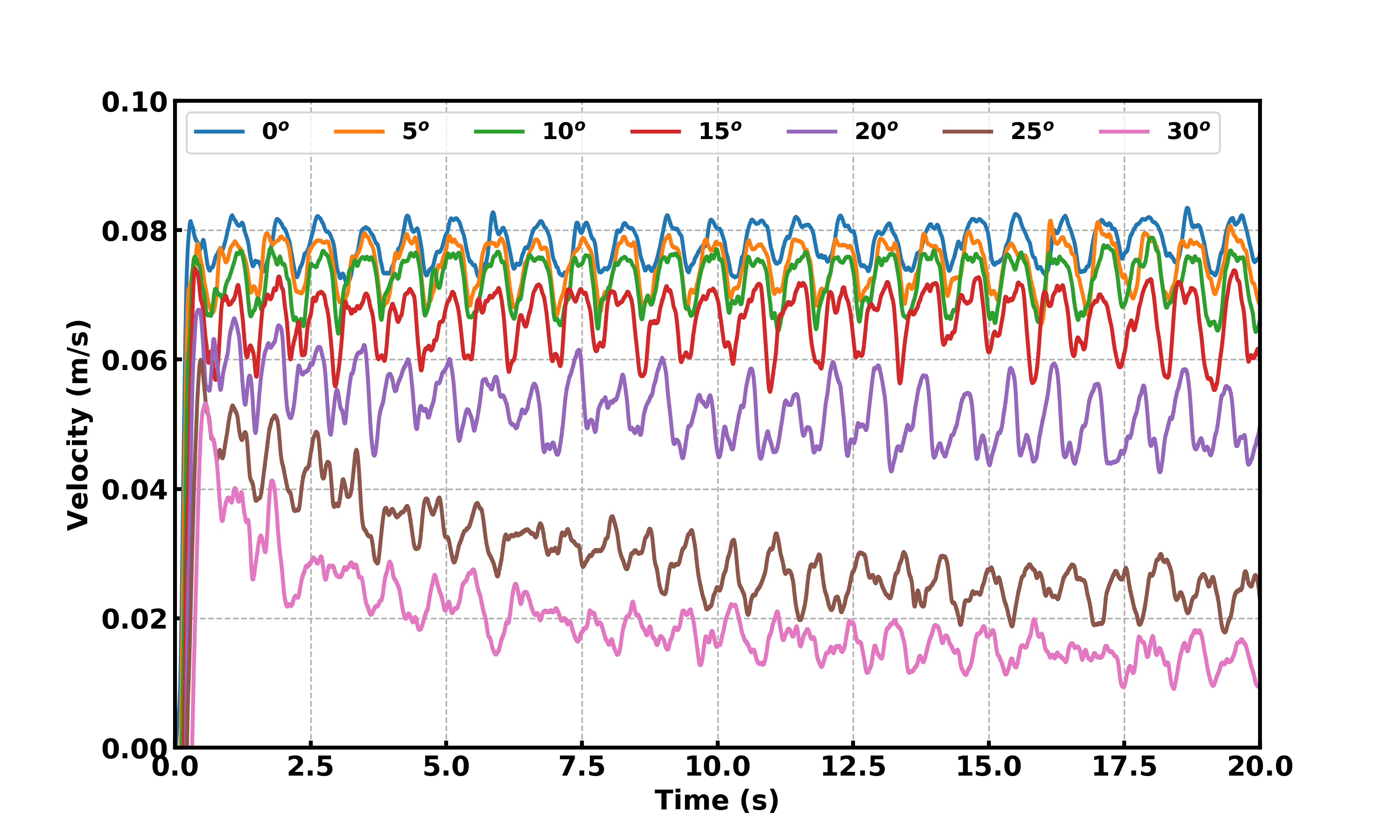}
			\caption{Moon gravity with $\omega=0.33~\si{rad/s}$} 
		\end{subfigure}
	\caption{Time history of translational velocity of full rover on GRC-3. Tests were done for $\theta$ between $0^{\circ}$ and $30^{\circ}$ in increments of $5^{\circ}$ with bulk density of 1734 $\si{kg/m^3}$ and friction angle $42.0^{\circ}$. The information in these two plots was used to generate the yellow and purple curves in Fig.~\ref{fig:suppMat-single_full_earh_vs_moon}b.} 
	\label{fig:rover_vel_his_grc3}
\end{figure}

\begin{figure}[h]
	\centering
	\begin{subfigure}{0.49\textwidth}
			\centering
			\includegraphics[width=3.2in]{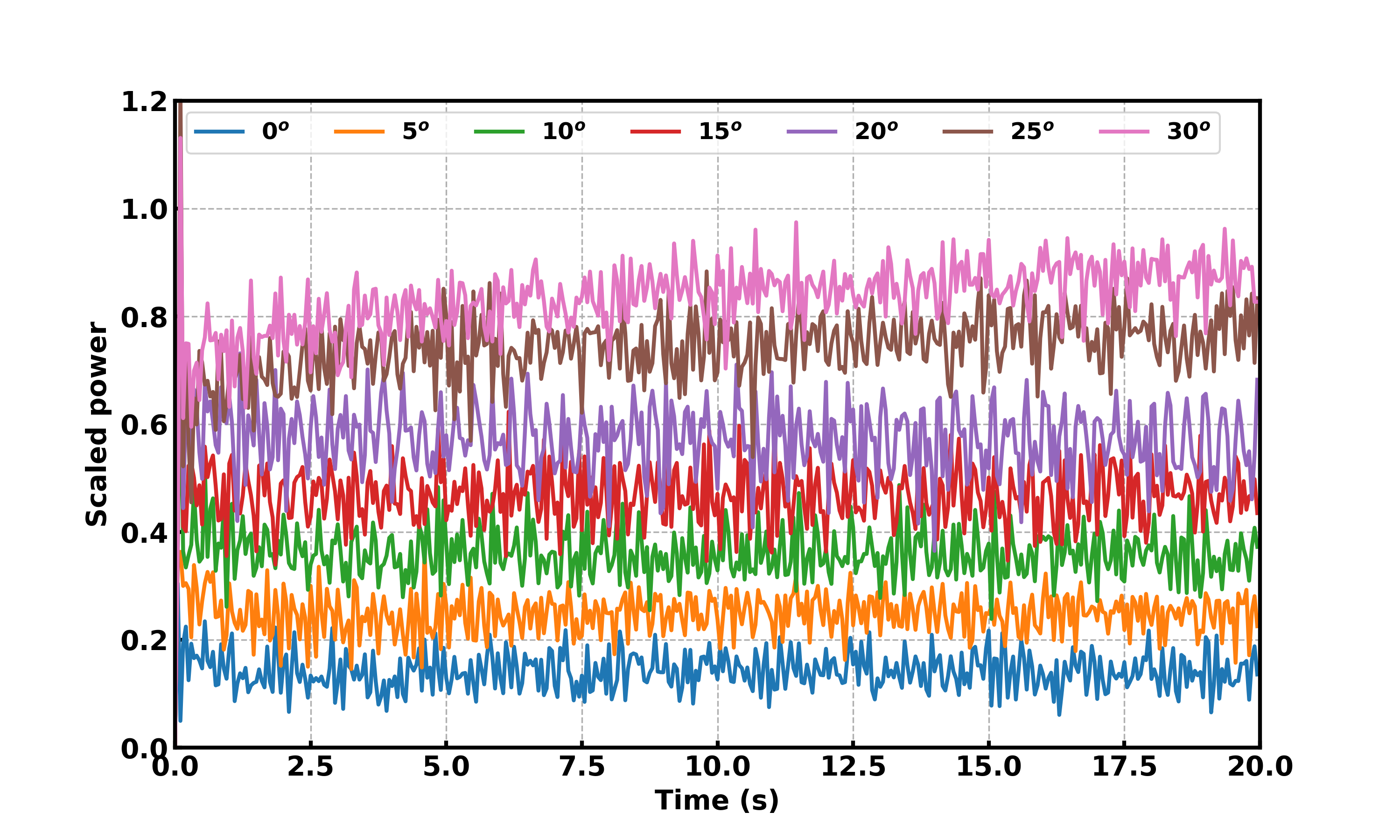}
			\caption{Earth gravity with $\omega=0.8~\si{rad/s}$}
		\end{subfigure}
	\begin{subfigure}{0.49\textwidth}
			\centering
			\includegraphics[width=3.2in]{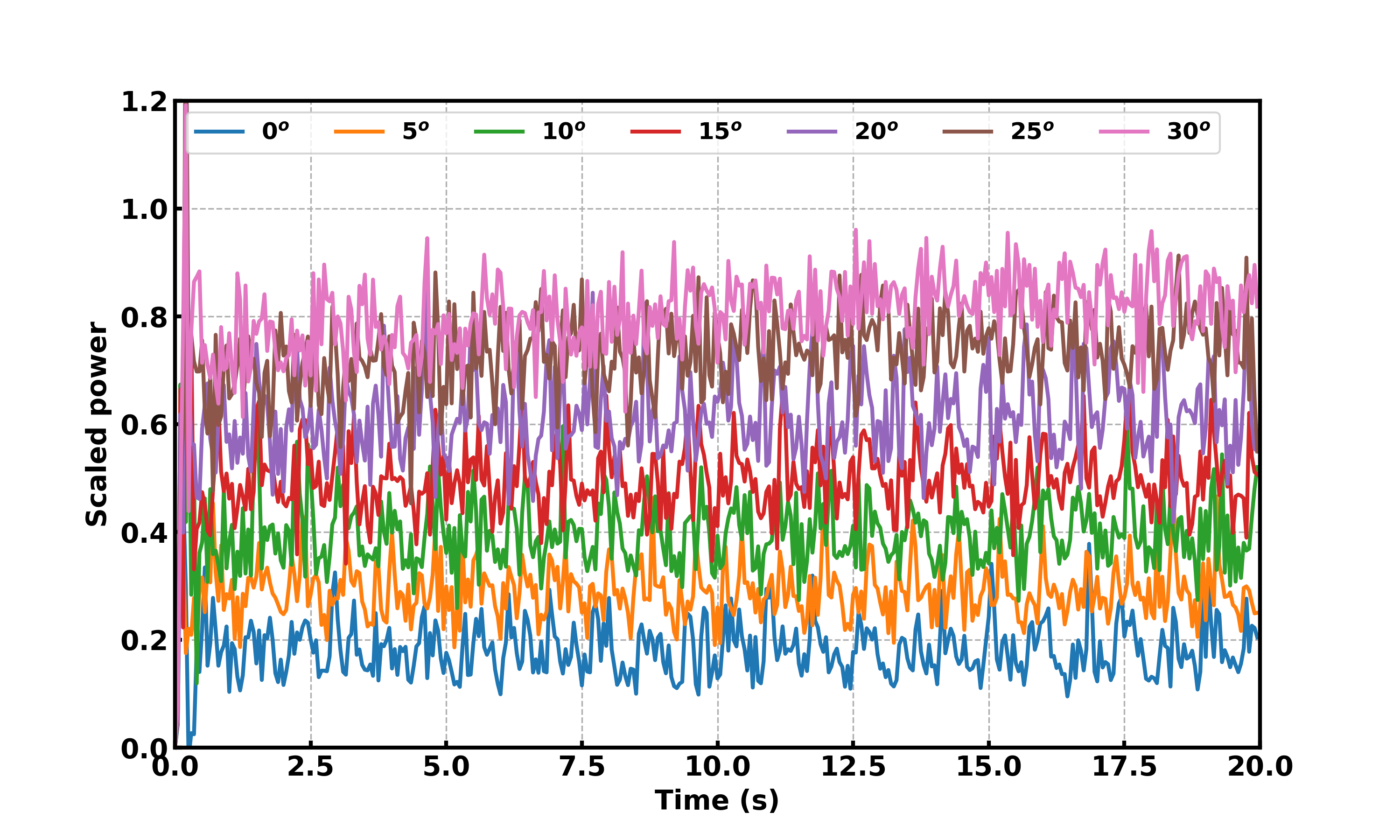}
			\caption{Moon gravity with $\omega=0.33~\si{rad/s}$} 
		\end{subfigure}
	\caption{Time history of scaled wheel power of single wheel simulation on GRC-3. Tests were done for $\theta$ between $0^{\circ}$ and $30^{\circ}$ in increments of $5^{\circ}$ with bulk density of 1734 $\si{kg/m^3}$ and friction angle $42.0^{\circ}$. The information in these two plots was used to generate the yellow and purple curves in Fig.~\ref{fig:supMat-rover_power_grc3}a.} 
	\label{fig:wheel_power_his_grc3}
\end{figure}

\begin{figure}[h]
	\centering
	\begin{subfigure}{0.49\textwidth}
			\centering
			\includegraphics[width=3.2in]{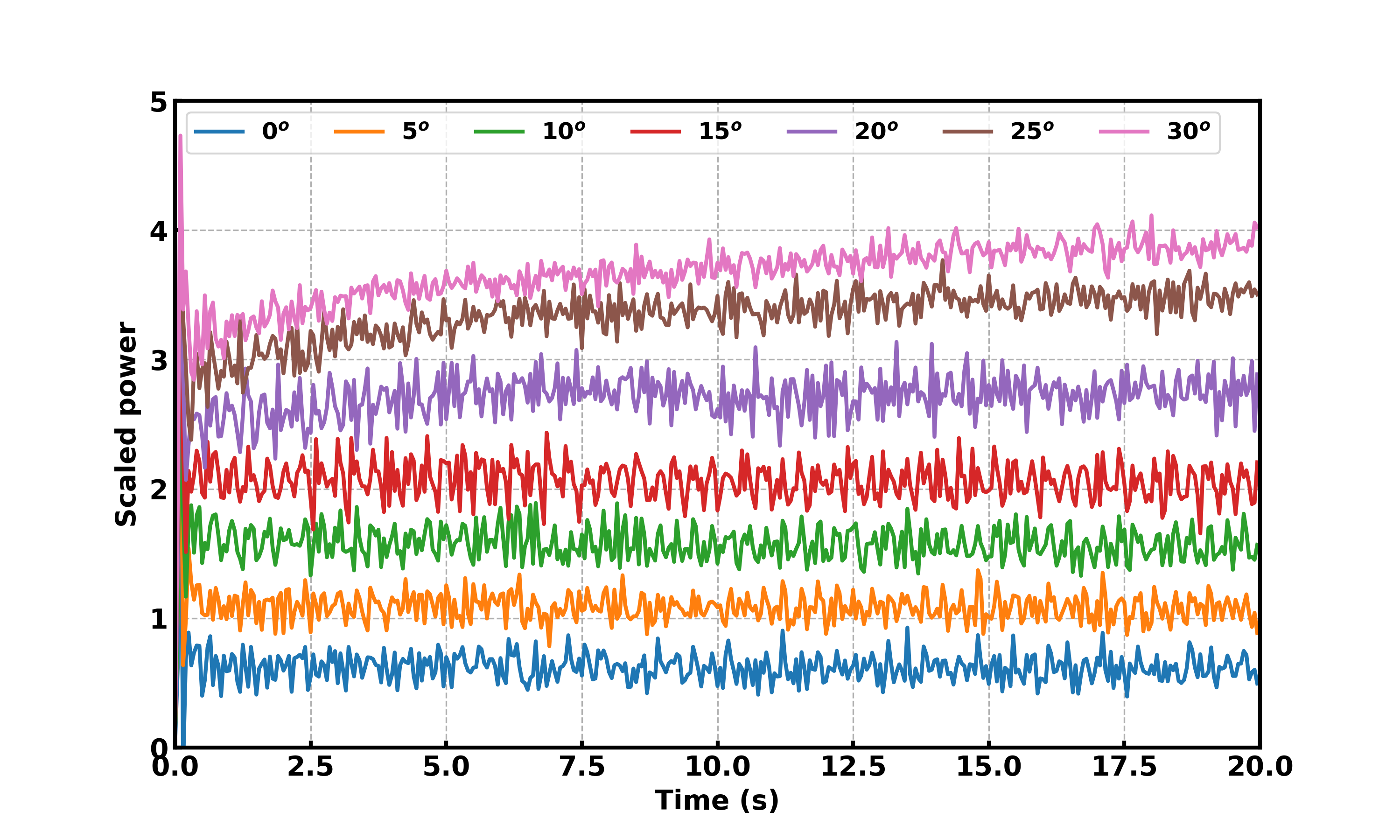}
			\caption{Earth gravity with $\omega=0.8~\si{rad/s}$}
		\end{subfigure}
	\begin{subfigure}{0.49\textwidth}
			\centering
			\includegraphics[width=3.2in]{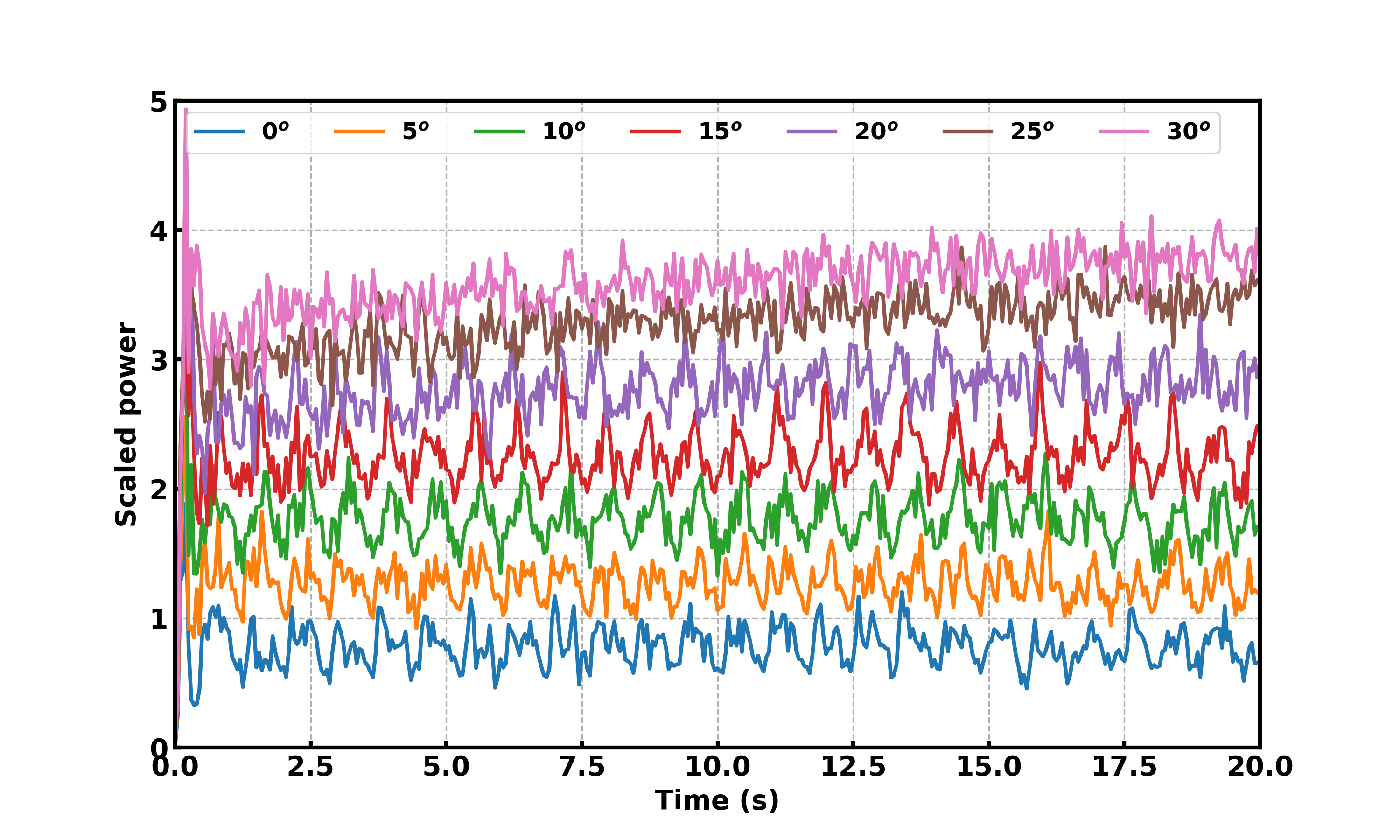}
			\caption{Moon gravity with $\omega=0.33~\si{rad/s}$} 
		\end{subfigure}
	\caption{Time history of scaled wheel power of full rover simulation on GRC-3. Tests were done for $\theta$ between $0^{\circ}$ and $30^{\circ}$ in increments of $5^{\circ}$ with bulk density of 1734 $\si{kg/m^3}$ and friction angle $42.0^{\circ}$. The information in these two plots was used to generate the yellow and purple curves in Fig.~\ref{fig:supMat-rover_power_grc3}c} 
	\label{fig:rover_power_his_grc3}
\end{figure}

%========================== GRC-1
\begin{figure}[h]
	\centering
	\begin{subfigure}{0.49\textwidth}
			\centering
			\includegraphics[width=3.2in]{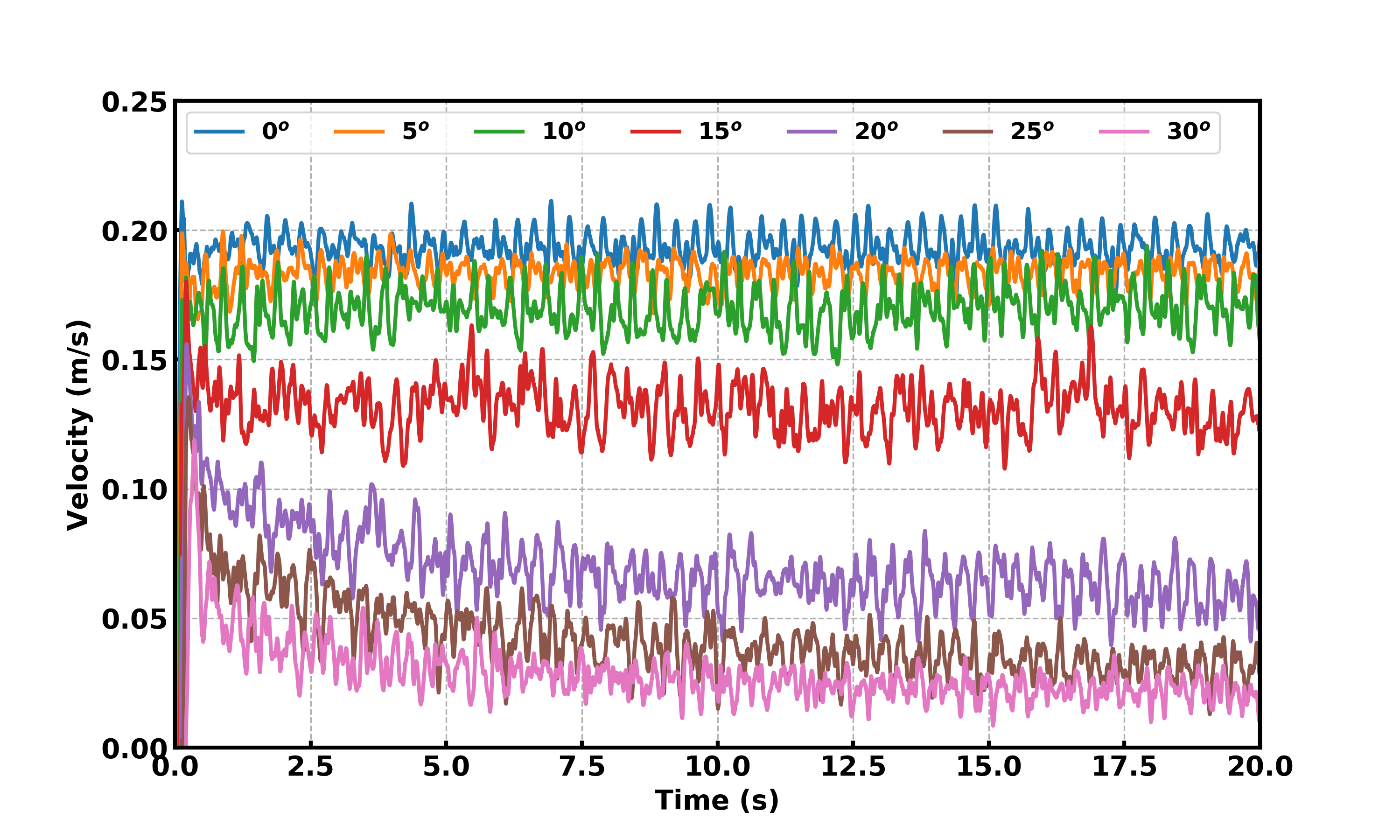}
			\caption{Earth gravity with $\omega=0.8~\si{rad/s}$}
		\end{subfigure}
	\begin{subfigure}{0.49\textwidth}
			\centering
			\includegraphics[width=3.2in]{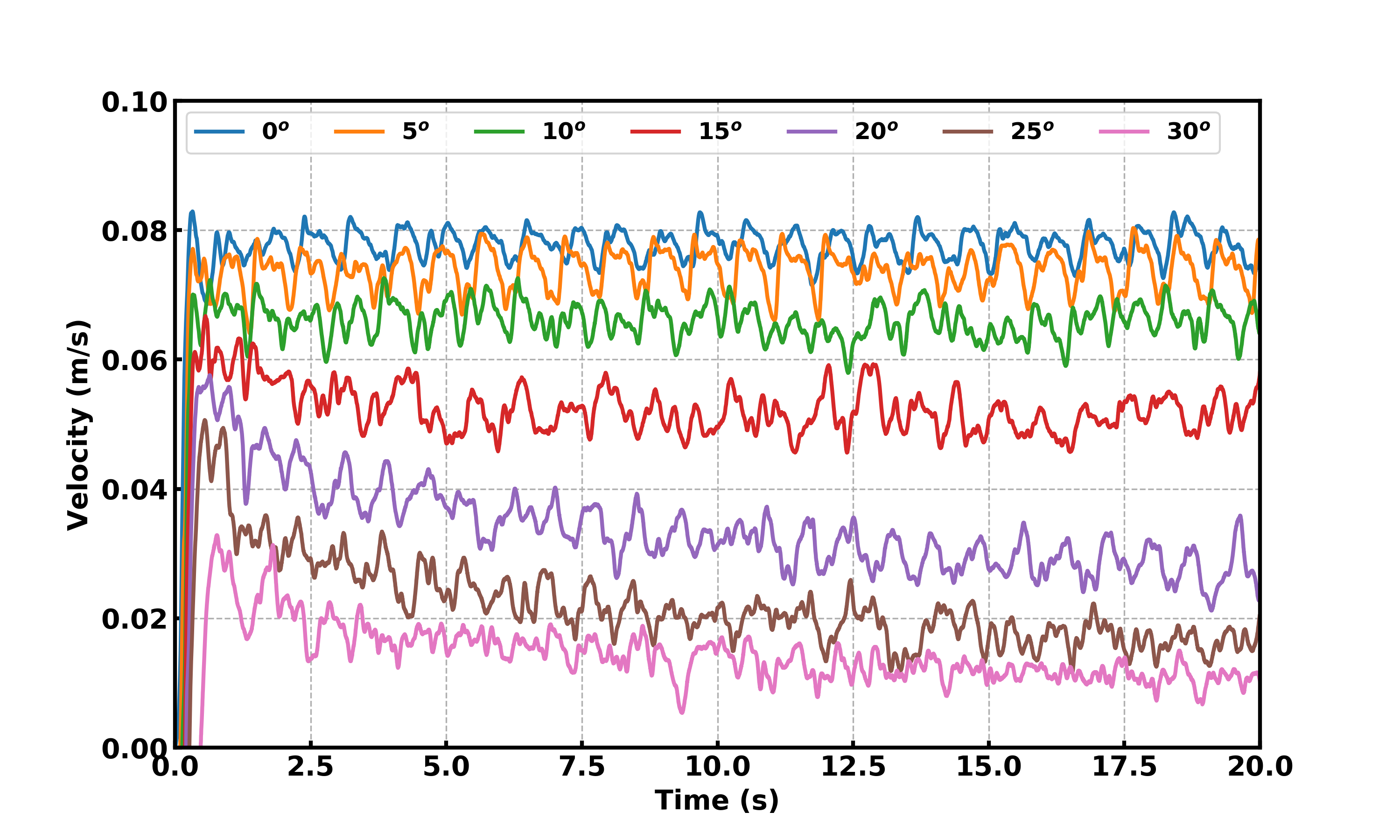}
			\caption{Moon gravity with $\omega=0.33~\si{rad/s}$} 
		\end{subfigure}
	\caption{Time history of translational velocity of single wheel simulation on GRC-1. Tests were done for $\theta$ between $0^{\circ}$ and $30^{\circ}$ in increments of $5^{\circ}$ with bulk density of 1760 $\si{kg/m^3}$ and friction angle $38.4^{\circ}$. The information in these two plots was used to generate the green and brown curves in Fig.~\ref{fig:single_full_earh_vs_moon}a} 
	\label{fig:wheel_vel_his_grc1}
\end{figure}

\begin{figure}[h]
	\centering
	\begin{subfigure}{0.49\textwidth}
			\centering
			\includegraphics[width=3.2in]{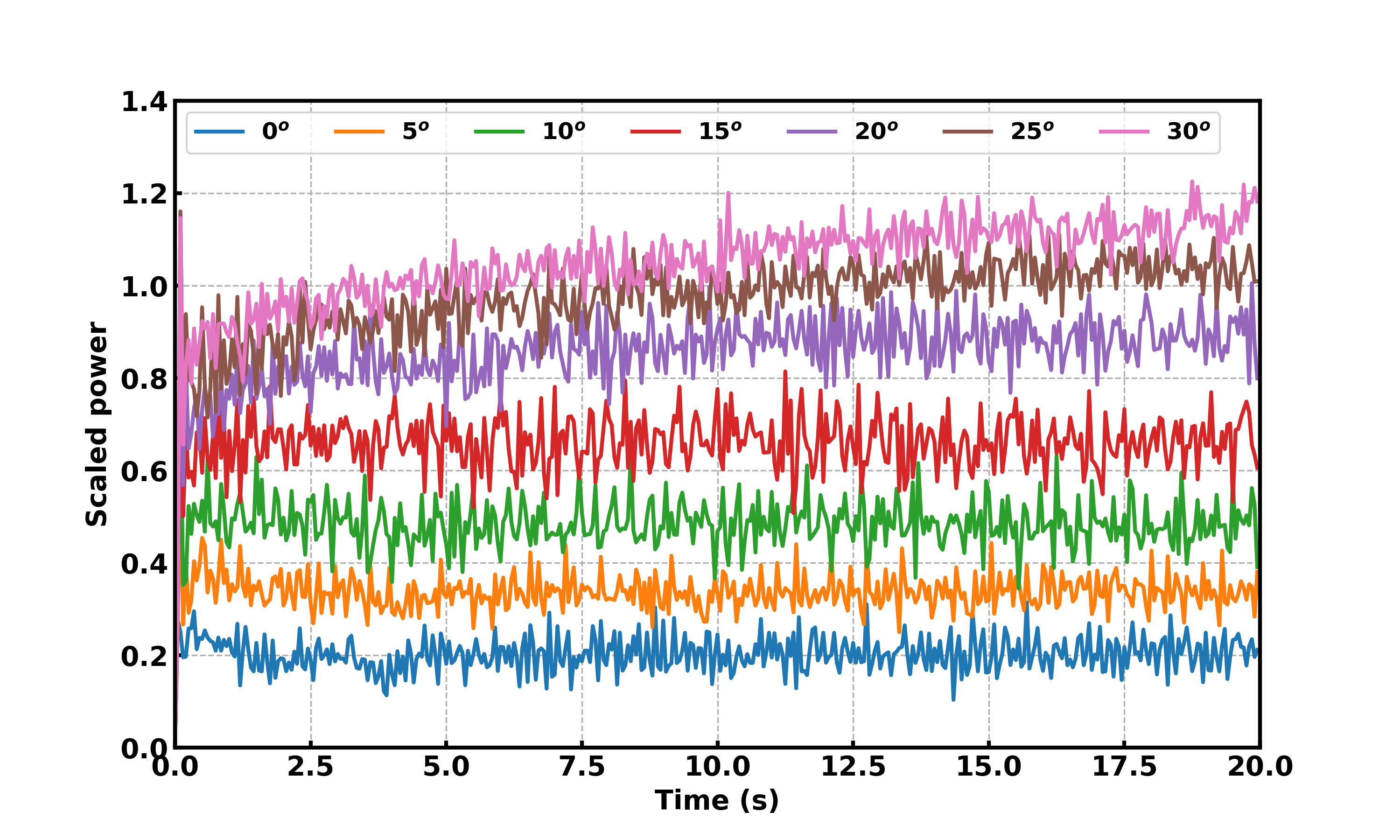}
			\caption{Earth gravity with $\omega=0.8~\si{rad/s}$}
		\end{subfigure}
	\begin{subfigure}{0.49\textwidth}
			\centering
			\includegraphics[width=3.2in]{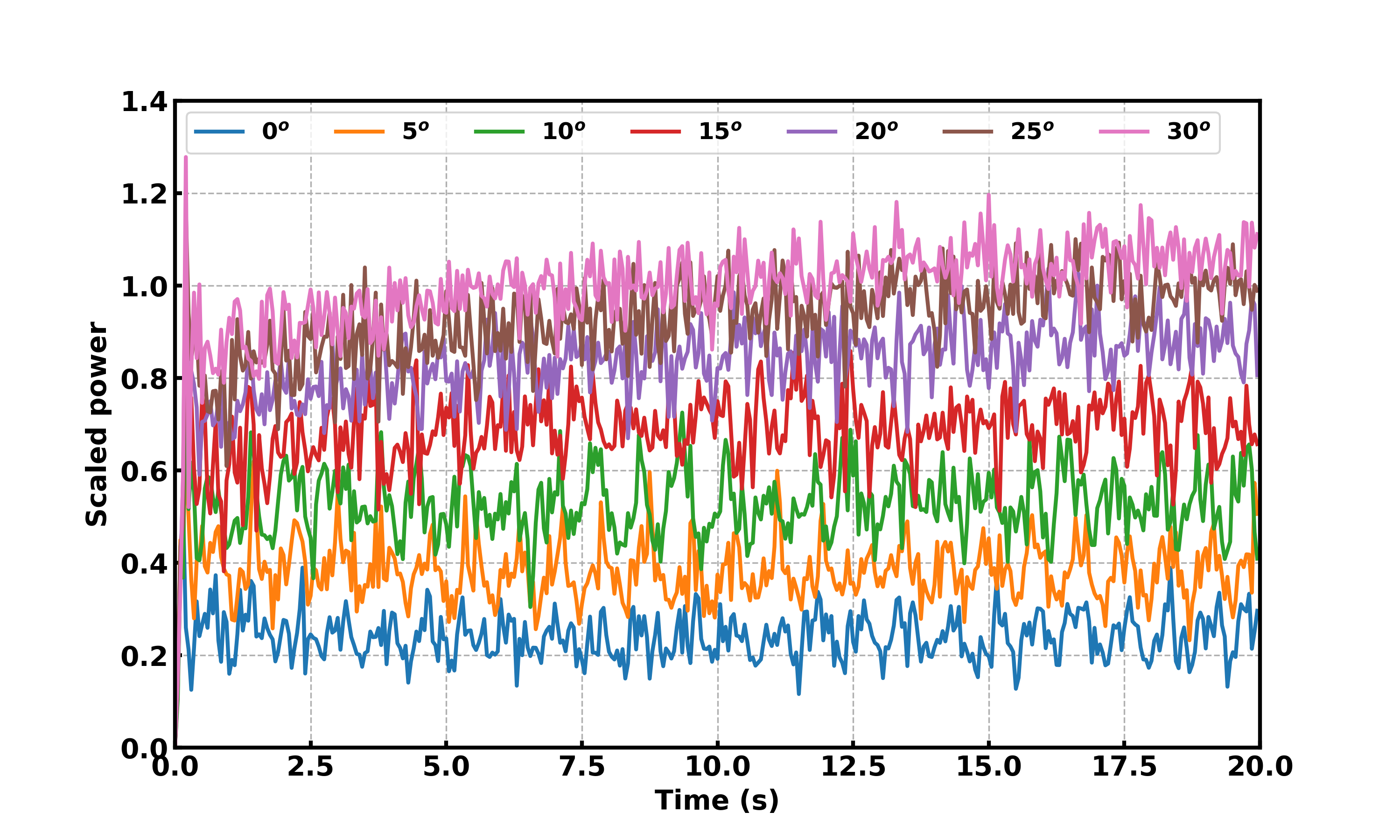}
			\caption{Moon gravity with $\omega=0.33~\si{rad/s}$} 
		\end{subfigure}
	\caption{Time history of scaled wheel power of single wheel simulation on GRC-1. Tests were done for $\theta$ between $0^{\circ}$ and $30^{\circ}$ in increments of $5^{\circ}$ with bulk density of 1760 $\si{kg/m^3}$ and friction angle $38.4^{\circ}$. The information in these two plots was used to generate the green and brown curves in Fig.~\ref{fig:rover_power_grc1}a} 
	\label{fig:wheel_power_his_grc1}
\end{figure}

\begin{figure}[h]
	\centering
	\begin{subfigure}{0.49\textwidth}
			\centering
			\includegraphics[width=3.2in]{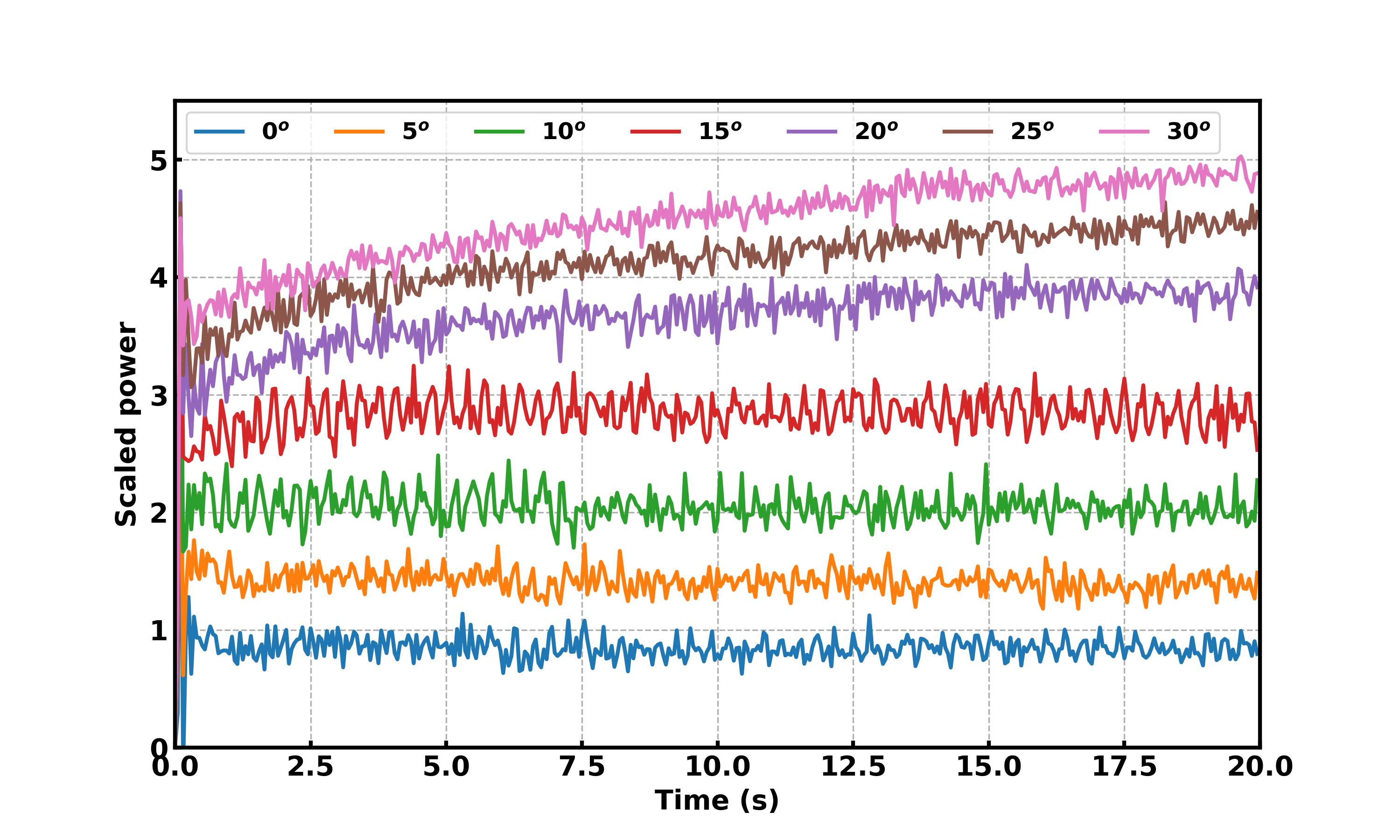}
			\caption{Earth gravity with $\omega=0.8~\si{rad/s}$}
		\end{subfigure}
	\begin{subfigure}{0.49\textwidth}
			\centering
			\includegraphics[width=3.2in]{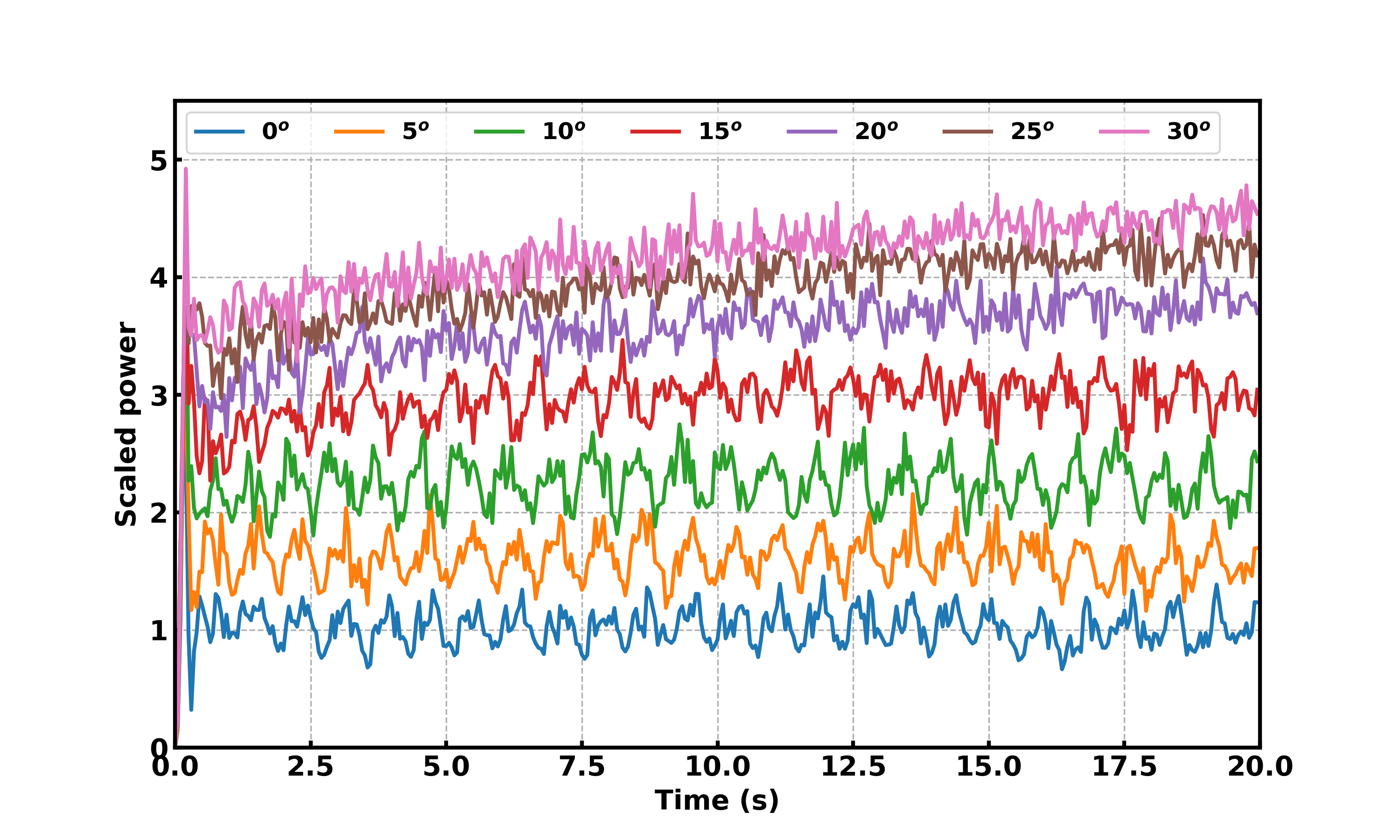}
			\caption{Moon gravity with $\omega=0.33~\si{rad/s}$} 
		\end{subfigure}
	\caption{Time history of scaled wheel power of full rover simulation on GRC-1. Tests were done for $\theta$ between $0^{\circ}$ and $30^{\circ}$ in increments of $5^{\circ}$ with bulk density of 1760 $\si{kg/m^3}$ and friction angle $38.4^{\circ}$. The information in these two plots was used to generate the green and brown curves in Fig.~\ref{fig:rover_power_grc1}c.} 
	\label{fig:rover_power_his_grc1}
\end{figure}

\end{document}